# On the complexity of spinels:
# Magnetic, electronic, and polar ground states

Vladimir Tsurkan,[1,2] Hans-Albrecht Krug von Nidda,[1] Joachim Deisenhofer,[1] Peter Lunkenheimer,[1] and Alois Loidl[1]

[1]*Center for Electronic Correlations and Magnetism, Institute of Physics, University of Augsburg, 86135 Augsburg. Germany*
[2]*Institute of Applied Physics, MD 2028, Chisinau, R. Moldova*

**Abstract**

This review aims to summarize more than 100 years of research on spinel compounds, mainly focusing on the progress in understanding their magnetic, electronic, and polar properties during the last two decades. Over the years, more than 200 different spinels, with the general formula $AB_2X_4$, were identified or synthesized in polycrystalline or single-crystalline form. Many spinel compounds are magnetic insulators or semiconductors; however, a number of spinel-type metals exists including superconductors and some rare examples of *d*-derived heavy-fermion compounds. In the early days, they gained importance as ferrimagnetic or even ferromagnetic insulators with relatively high saturation magnetization and high ordering temperatures, with magnetite being the first magnetic mineral known to mankind. From a technological point of view, spinel-type ferrites with the combination of high electrical resistance, large magnetization, and high magnetic ordering temperature made them promising candidates for many applications. However, spinels are also known as beautiful gemstones, with the famous "Black Prince's Ruby" in the front centre of the Imperial State Crown. In addition, spinels are important for the earth tectonics, and the detection of magnetite in a Martian meteorite even led to the speculation of life on Mars. However, most importantly in the perspective of this review, spinels played an outstanding role in the development of concepts of magnetism, in testing and verifying the fundamentals of magnetic exchange, in understanding orbital-ordering and charge-ordering phenomena including metal-to-insulator transitions, in developing the concepts of magnetic frustration, in establishing the importance of spin-lattice coupling, and in many other aspects. The still mysterious Verwey transition in magnetite was one of the very first illuminating examples of this complexity, which results from the fact that some ions can exist in different valence states in spinels, even at a given sublattice. In addition, the *A*- site as well as the *B*-site cations in the spinel structure form lattices prone to strong frustration effects resulting in exotic ground-state properties. The *A*-site ions are arranged in a diamond lattice. This bipartite lattice shows highly unusual ground states due to bond-order frustration, with a strength depending on the ratio of inter- to intra-sublattice exchange interactions of the two interpenetrating face-centred cubic lattices. The occurrence of a spiral spin-liquid state in some spinels is an enlightening example. Very recently, even a meron (half-skyrmion) spin structure was identified in $MnSc_2S_4$ at moderate external magnetic fields. In case the *A*-site cation is Jahn-Teller active, additional entanglements of spin and orbital degrees of freedom appear, which can give rise to a spin-orbital liquid or an orbital glass state. In systems with such a strong entanglement, the occurrence of a new class of excitations - spin-orbitons - has been reported. The *B*-site cations form a pyrochlore lattice, one of the strongest contenders of frustration in three dimensions. A highly degenerate ground state with residual zero-point entropy and short-range spin ordering according to the ice rules is one of the fascinating consequences, which is known already for more than 50 years. At low temperatures, in *B*-site spinels the occurrence of spin molecules has been reported, strongly coupled spin entities, e.g., hexamers, with accompanying exotic excitations. A spin-driven Jahn-Teller effect is a further possibility to release magnetic frustration. This phenomenon has been tested in detail in a variety of spinel compounds. In addition, in spinels with both cation lattices carrying magnetic moments, competing magnetic exchange interactions become important, yielding ground states like the time-honoured triangular Yafet-Kittel structure. Very recently, it was found that under external magnetic fields this triangular structure evolves into very complex spin orders, which can be mapped on spin super-liquid and spin super-solid phases. In addition, due to magnetic frustration, competing interactions, and coupling to the lattice, very robust magnetization plateaus appear in a variety of spinel compounds as function of an external magnetic field. Furthermore, spinels gained considerable importance in elucidating the complex physics driven by the interplay of spin, charge, orbital, and lattice degrees of freedom in materials with partly filled *d* shells. This entanglement of the internal degrees of freedom supports an exceptionally rich variety of phase transitions and complex ground states, in many cases with emerging functionalities. It also makes these materials extremely susceptible to temperature, pressure, or external magnetic and electric fields, an important prerequisite to realize technological applications. Finally, yet importantly, there exists a long-standing dispute about the possibility of a polar ground state in spinels, despite their reported overall cubic symmetry. Indeed, recently a number of multiferroic spinels were identified, including multiferroic spin super-liquid and spin super-solid phases. The spinels also belong to the rare examples of multiferroics, where vector chirality alone drives long-range ferroelectric order. In addition, a variety of spinel compounds were investigated up to very high pressures up to 40 GPa and in high magnetic fields up to 100 T, revealing complex ($p,T$) and ($H,T$)-phase diagrams.

Corresponding author:
alois.loidl@physik.uni-augsburg.de

**CONTENT**

# 1. Introduction

Spinels are a large group of compounds having the general formula $AB_2X_4$ with the same crystal structure as the mineral $MgAl_2O_4$. The anions $X$ often are $O^{2-}$, $S^{2-}$, $Se^{2-}$, or $Te^{2-}$. However, spinels with the anion being fluorine $F^-$, nitrogen $N^{3-}$ or cyanide $(CN)^{1-}$ were also reported to form stable compounds. $A$-site cations can be the alkali ion $Li^+$, the earth-alkaline ions $Mg^{2+}$, $Ca^{2+}$, and $Ba^{2+}$, the heavier 3d transition-metal ions Mn, Fe, Co, Ni, Cu, and Zn, as well as Ge, Cd, and Hg. $B$-site cations favourably are Al, Sc, Ti, Cr, Mn, Fe, Co, and Ni, as well as Ga, In, and Tl. In addition, spinels with $B$ = Mo, Rh, or In were synthesized. In connection with the anions $X$ = $F^-$ or $(CN)^-$, the $B$ sites can also be occupied by the alkali ions $Li^+$, $Na^+$, or $K^+$. More than 200 spinel compounds, which were identified or synthesized over the years, are listed in Refs. [1,2]. Mineral species with a spinel-type structure and with their original names given by mineralogists or crystallographers are described in Ref. [3]. Like a great variety of oxides, spinel compounds also consist of a face-centred cubic (fcc) lattice of the $X^{2-}$ anions, which is stabilized by interstitial cations.

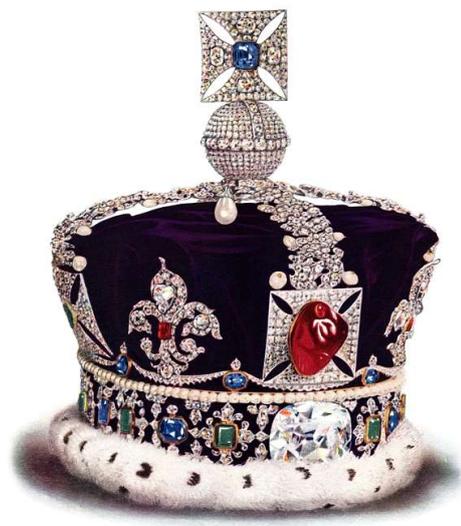

Fig. 1. Spinels as gemstones. "Black Prince's Ruby" in the front center of the Imperial State Crown. This "ruby" later was identified as chromium-doped magnesium-iron-oxide spinel.
From Wikimedia Commons.

There are two types of interstices in such an anion lattice: Tetrahedral interstices with four nearest neighbours and octahedral interstices with six nearest neighbours. If all tetrahedral sites are empty and all octahedral sites are full, the crystal has rock-salt structure. If only tetrahedral sites are occupied, the crystal has zinc-blende structure. In the spinel structure, twice as many octahedral as tetrahedral sites are occupied and the spinel structure can be thought as an ordered mixture of zinc-blende and rock-salt unit cells. In spinel compounds there exist two general classes having this lattice, the 2-4 spinels with the formula $2B^{2+}X^{2-} \bullet A^{4+}X_2^{-2}$ and the 2-3 spinels with $A^{2+}X^{2-} \bullet B_2^{3+}X_3^{2-}$ resulting in the general formulas $AB_2X_4$ or $B(AB)X_4$, where the ions within the parenthesis occupy octahedral sites only. The third possibility, the realization of 1-6 spinels is very rare. If there is only one type of cation at the octahedral sites, the spinel is called normal. If there are equal amounts of both kinds of cations at the octahedral sites, the spinel is inverse. Spinels with unequal numbers of each type of cation in the octahedral site are called mixed [4]. Another common way to describe the dense anion and cation packing of spinels assumes to subdivide the cubic unit cell into eight octants. The spinel structure then is a face centred compilation of two types of cubes, $A_2X_4$ ($A$ in four-fold coordination) and $B_4X_4$ ($B$ in six fold coordination) [5].

In nature, some spinels are found as beautifully coloured and transparent gemstones, with a variety of colours due to different dopants. In early times, they were confused with rubies, the most prominent example being the Black Prince's Ruby in the front centre of the British Imperial State crown, which later was identified as spinel (Fig. 1). In 1779 they were named by Jean Demeste, derived from the Latin word "spinella", which means little thorn and refers to their sharp octahedral crystal shapes. The determination of the crystal structure of spinels by Bragg [6] and independently by Nishikawa [7] was one of the very early accomplishments of crystal-structure analysis. The spinel structure is cubic with a large unit cell containing 8 formula units, e.g., 8 $A$, 16 $B$ and 32 $X$ atoms. In most cases the anions $X$ can be approximated by cubic close packing, while the cations fill certain interstices as described above. Later it was found that in addition to the normal variant of the spinel structure $AB_2X_4$ and inverted structure of $B(AB)X_4$ can exist, which was termed inverse spinel [8].

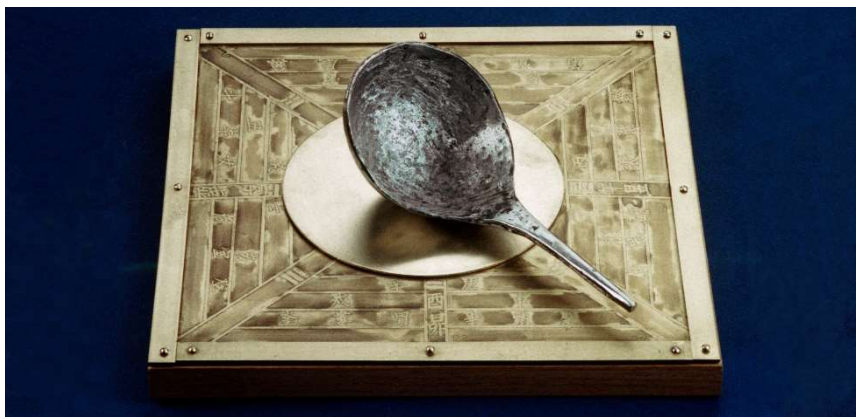

Fig. 2. South pointing spoon (Si Nan spoon compass).
Model of a magnetic compass used in ancient China. An iron spoon manufactured from naturally occurring magnetite is placed on a non-magnetic plate. When it comes to rest after turning, the handle points south and hence, the name "Si Nan" means South seeker.
Fig. reproduced from Ref. [10].

Due to their remarkable electrical and magnetic properties, spinels are of considerable technological interest, specifically due to the need of insulating or semiconducting ferromagnets. One should have in mind that permanent magnetism was first observed in lodestone, naturally magnetized pieces of the mineral magnetite, $Fe_3O_4$, which was also called the Magnesian stone, resulting in the nomenclature of the word magnetism. The development of models and theories of magnetism are closely related to magnetit, starting more than 2500 years ago. Early on it was recognized that lodestone can attract pieces of iron and the Greek philosopher Thales of Milet (625 – 545 BC) probably provided the first scientific discussion of lodestone. About 585 B.C. he stated, as witnessed by Aristotle [9] that loadstone attracts iron because it has a soul. Since then, this material fascinated countless scientists. In ancient China a variety of compass needles were invented, either using south-pointing chariots, two-wheeled vehicles that carried a movable pointer to indicate the south, no matter how the chariot turned, or south-pointing spoons (see Fig. 2). Each time the spoon manufactured from naturally occurring magnetite is spun on a non-magnetic plate, it comes to rest with its neck pointing south [10].

Even nowadays non-metallic ferromagnetic (FM) materials are rare and interestingly, many spinels are insulating ferrimagnets, in many cases with relatively high saturation magnetization and high magnetic-ordering temperatures. Hence, an early fundamental characterization of structural, electrical and magnetic properties of various spinels, which will be discussed later in this review, has been performed in the Natuurkundig Laboratorium of the N. V. Philips Gloeilampenfabrieken in the Netherlands. Before World War II, Verwey and coworkers performed the most pertinent work on spinel compounds, later on, in the early fifties, this work was continued by Gorter, Lotgering, Romeijn, and coworkers to name a few.

Spinels, in addition of being beautiful gemstones and utilized in early compass needles, probably are one of the most important classes of minerals concerning technological applications and an amazing variety of other aspects. Because of their multiple compositions, electron configurations, and valence states, spinels have demonstrated remarkable magnetic, optical, electrical, and catalytic properties. Potential applications of magnetic properties of spinels containing $3d$ magnetic ions like Fe, Co, Cr, or Ni, can be found in information technology, in biotechnology, and in applications for spintronic devices. Ferrites, which mainly are ceramic and composite spinel compounds, already play an important technological role because of their interesting electrical, magnetic, and dielectric properties. The combinations of high electrical resistivity, low eddy currents, low dielectric losses, high saturation magnetization, and high permeabilities, together with high Curie temperatures, good chemical stability, and mechanical compactness makes them extremely useful for applications such as rod antennas, electronic devices, sensors, memory devices, data storage, or tele-communication. The optical properties of transparent or semi-transparent spinels were utilized for photo-luminescence as well as for magneto-optical recording. Furthermore, the electrical characteristics of spinels have allowed their application in the field of energy-storage materials, such as super-capacitors. In addition, they also have been widely used as electrodes in Li-ion batteries, and last not least, their controllable composition, structure, valence, and morphology have made them suitable as catalysts in a variety of chemical reactions. A detailed overview over possible applications of spinel compounds is given in Ref. [2].

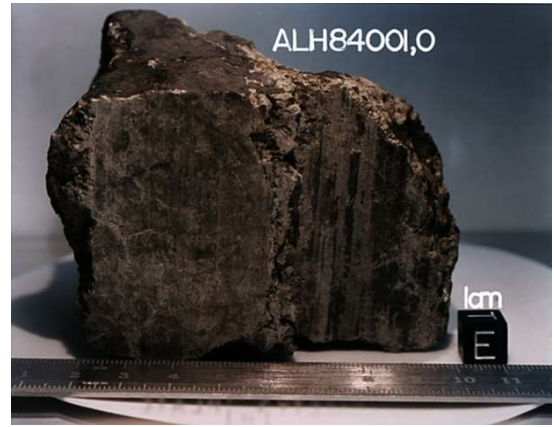

Fig. 3. Life on Mars.
Martian meteorite ALH84001 containing magnetite microcrystals. This meteorite was found in the Alan Hill icefield (Antarctica). It was speculated that these magnetite crystals were produced in ancient times by magneto-receptive Martian bacteria (see text).
Image credit: NASA/JSC/Stanford University

Quite general, magnetic insulators are promising materials for spintronics, spin-caloritronics, nonvolatile memories, and microwave applications. Magnetic insulating heterostructures in form of single crystalline thin films, grown on suitable substrates, are important and even critical for realizing quality interfaces for these applications. However, magnetic and insulating single crystalline materials with a Curie temperature well above room temperature are rare. In these introductory remarks, we tried to document that spinel ferrites constitute a class of materials that were recognized having significant potential for these applications and FM as well as ferrimagnetic (FiM) insulating spinel oxides were synthesized as thin films or heterostructures and were carefully characterized. To document this broad and important field of applications, we include a chapter dealing with synthesis and possible applications of spinel-derived thin films (Chapter 3.5.). An informative review on epitaxially-grown spinel ferrite thin films has been provided earlier by Y. Suzuki [11].

Spinels are also of utmost importance for the earth tectonics: spinel compounds form the so-called transition region between the lower and the outer mantle of the earth structure, extending roughly from 450 to 600 km below the earth's surface, forming an important zone for seismic activities of the earth. In this regime of the earth's mantle, the spinel-olivine transition is important in driving the plate tectonics.

Finally, spinels in form of magnetite or greigite are believed to act as magnetic receptors in magneto-receptive species, including honeybees, birds, sea turtles, salmons and a variety of phytoplankton and bacteria [12]. The identification of magnetite in the Martian meteorite ALH84001 (Fig. 3), which was collected in the Antarctica in Alan Hills ice field, even led to the conclusion of life on mars [13] by correlating the found magnetite micro crystals with the activity of magneto-receptive bacteria. Regarding this Mars meteorite discovery, the possibility of life on mars was even mentioned in a press-release satement by US president Clinton on August 7, 1996. It should be noticed that scientifically this claim lateron was strongly challenged.

During the past decades, magnetic nanoparticles have attracted continuous interest and were intensively explored and studied for both, basic research phenomena and potential applications in a variety of different fields ranging from data storage, energy storage, spintronics, and magneto-optical devices, to contrast agents for magnetic resonance imaging. Quite generally, they also are of interest in biomedicine in areas like medical diagnostics, drug delivery, and magneto-hyperthermia, in environmental purification, and in catalysis, to name a few. In first respect, nano-structured materials exhibit unique magnetic features depending on nanoscale size modulation and grain-boundary engineering. When compared to the bulk, nanomaterials are strongly dependent on the microcrystalline grain size and on distribution, thickness, and chemical constitution of the grain boundaries. In addition to grain size and size dispersion, materials engineering regarding core-shell design, shape, morphology, crystallinity, and surface decoration opens up a variety of prospects for potential application and innovation. Over the years, spinel-type ferrite nanoparticles played an important role, specifically with respect to cobalt ferrites and magnetite. The physics and chemistry of magnetic nanoparticles constitute an entire new research field on its own and it is impossible to review their synthesis, structural and magnetic characterization, and possible applications in this review, which mainly focuses on the basic principles of magnetism of mostly single-crystalline spinel compounds. The interested reader is referred to reviews on the importance of magnetic nanoparticles in general [14,15,16] and specifically on work dealing with spinel-type nanoparticles and nanocomposites [17,18,19,20], which only provide a very first look into this recent universe of nanoscience.

Here we mainly will focus on magnetic, electric, and multiferroic (MF) properties of spinel compounds, where in the context of this work, multiferroicity denotes the coexistence of magnetic and polar order. As will be documented in the course of this work, the complex and partly exotic magnetism of many spinel compounds results from magnetic frustration and competing interactions. As mentioned above, the pyrochlore network formed by the $B$ sites can give rise to strong magnetic frustration in three dimensions (3D) [21]. In the bipartite diamond lattice of the $A$ sites, depending on the magnetic exchange interactions within or in-between the two sublattices, again strong spin-frustration effects can occur [22]. The importance of frustration and competing interactions was recognized early on in studying the magnetic ground-state properties of spinels. In $B$-site spinels the magnetic moments occupy the corners of a pyrochlore lattice and are coupled by antiferromagnetic (AFM) exchange. In the seminal work on antiferromagnetism in spinels, Anderson [23] noticed that in such a lattice nearest-neighbour forces alone can never lead to long-range order. However, it is possible to achieve essentially perfect short-range order while maintaining a finite entropy. Down to the lowest temperatures, there will be residual zero-point entropy and Anderson noticed the closeness of this problem to the ice-rules developed by Pauling for hydrogen bonds in water. In spinels were both, $A$ and $B$ sites are occupied by magnetic moments exhibiting dominant AFM exchange between the $A$ and $B$ cations, Yafet and Kittel [24] predicted a resulting triangular spin structure, the famous Yafet-Kittel (YK) spin structure, which later indeed was identified in $MnCr_2S_4$ [25]. In the more recent past, it has been recognized that as function of an external magnetic field, the spin structure of this compound exhibits unusual magnetic states, like an ultra-robust magnetization plateau or a spin supersolid phase [26]. In a number of spinel compounds a variety of complex spin configurations driven by magnetic frustration, e.g., composite spin degrees of freedom forming spin loops in $ZnCr_2O_4$ [27] or the formation of a spin-spiral liquid state in $MnSc_2S_4$ [22,28], were identified. As will be documented in Chapter 3.2., also the orbital degrees of freedom play an important role in the physics of spinel compounds, ranging from the cooperative Jahn-Teller effect to an orbital-ordering driven dimensional reduction in spinel systems like $MgTi_2O_4$ or $CuIr_2S_4$. Such orbital effects are described in detail in a recent review by Khomskii and Streltsov [29].

In the late fifties and early sixties, the magnetism of spinels was important for developing theories predicting the ground state properties of Néel type, YK like, or spiral spin structures. In addition, spinel compounds gained considerable importance in elucidating the complex physics driven by the interplay of spin, charge, orbital, and lattice degrees of freedom in materials with partly filled $d$ or $f$ shells. This entanglement of the internal degrees of freedom supports an exceptionally rich variety of phase transitions and complex ground states, in many cases with emerging functionalities. It also makes these materials extremely susceptible to temperature, pressure, or external magnetic and electric fields an important prerequisite to achieve technological functionality. The first illuminating example of a strong entanglement of spin, charge, and lattice degrees of freedom was given by Verwey [30] by detecting a "mysterious" transition in magnetite close to 120 K, where the resistivity increases by two orders of magnitude. Despite enormous attempts in recent times, the microscopic nature of this metal-to-insulator transition (MIT), nowadays called Verwey transition, so far has not been completely unravelled. Other examples of the coupling of internal degrees of freedom in spinel compounds are, e.g., colossal magneto-resistance (CMR) effects in $FeCr_2S_4$ [31], the occurrence of multiferroicity in $CdCr_2S_4$ [32], $CoCr_2S_4$ [33], $FeCr_2S_4$ [34], and $MnCr_2S_4$ [35], structural phase transitions induced via a spin-Jahn-Teller effect [36,37,38] in a number of chromite spinels, and the formation of iridium octamers and spin dimerization at the charge ordering MIT in $CuIr_2S_4$ [39]. The complexity of the orbital physics in some of the spinel compounds is best documented by the observation of a spin-orbital-liquid (SOL) state in $FeSc_2S_4$ [40] and the report on the observation of an orbital glass in $FeCr_2S_4$ [41].

In this review, we mainly will summarize 20 years of research guided by the authors of this work. Of course, we always will refer to important results in the field: Since the crystallographic studies of Bragg [6] more than 100 years ago, outstanding solid-state researchers including a number of Nobel-prize winners have contributed to this fascinating spinel research important for physics, chemistry, material science, geology, and biology. There exist a number of reviews on spinel compounds. First, one has to mention the overview over structural, magnetic, and electrical properties of different classes of spinel compounds mainly published as Philips Research Reports and summarizing the detailed work performed in the Philips Research Laboratories, mainly by Romeijn [42,43], Gorter [44,45,46,47,48], Lotgering [49,50] and Blasse [51]. A review focusing on structural details of spinels was provided by Hill and coworkers [1]. Grimes [52] gave a short review with a specific look on their potential industrial applications. Takagi and Niitaka [53] described magnetic properties of spinels, mainly with respect to magnetic frustration effects. A review focusing on physico-chemical aspects of spinels was published by Zhao et al. [2] and Liu et al. [54] summarized the possible use of spinel compounds as multivalent battery cathodes.

## 2. General properties of spinel compounds

### 2.1. Synthesis and single-crystal growth

High-purity spinels can be synthesized in ceramic form and as single crystals by a variety of methods. Many of the preparation routes are given in the original literature cited in this review. A detailed summary of synthesis methods of spinel compounds is given in [2]. Here we only want to mention the extreme sensitivity of many physical properties on minor changes of impurities or defects introduced by the different synthesis routes. One illuminating example is $FeSc_2S_4$: There exist contrasting experimental results on spin and orbital order (OO) in this compound [40,55,56]. It was unambiguously shown that the magnetic order in some samples result from excess Fe forming a second phase corresponding to a vacancy-ordered iron sulphide with composition close to the 5C polytype of pyrrhotite ($Fe_9S_{10}$) [55]. It is also now well established that the growth of thio-spinels by chemical transport reaction using chlorine as transport agent, by minor substitution of chlorine ions for sulphur, results in an enhanced defect-induced conductivity and in distinct differences of magnetocapacitive effects. This has been documented in full detail comparing single crystals grown by chemical transport using chlorine as well as bromine as transport agents and comparing the results to stoichiometric ceramic samples [57].

### 2.2. Structural properties

Compounds with the general formula $AB_2X_4$ having the same structure as the mineral $MgAl_2O_4$, which originally was referred to as red gemstone, are named normal spinels. Starting with the mineral spinel $MgAl_2O_4$ one easily can substitute $Mg^{2+}$ and $Al^{3+}$ with a wide range of cations with different valences. Hence, the general chemical formula for spinel compounds can be written as $A^{2+}B_2^{3+}X_4^{2-}$ or $A^{4+}B_2^{2+}X_4^{2-}$ where the $A$ sites in most cases include ions with valence 2+ and 4+, while the $B$ sites host divalent or trivalent cations. The anions $X$ are oxygen or chalcogenide ions. Spinels with $X$ = S usually are called thiospinels.

Bragg [6] and Nishikawa [7] determined the spinel structure, which belongs to the space group $O_h^7$ ($Fd\bar{3}m$) or number 227 in the International Tables. The "normal" spinel structure, which is reproduced from Ref. [58] and shown in Fig. 4, is based on a cubic close-packed fcc structure of $O^{2-}$ anions introducing two types of interstitial positions. There are 64 tetrahedral and 32 octahedral interstices, of which 8 respectively 16 are occupied by cations. The former is the $A$ site, which is tetrahedrally coordinated, the latter correspond to the octahedral $B$ sites. Each anion in the spinel structure is surrounded by one $A$ and one $B$ cation. The distribution of the cations, divalent and trivalent, among the two sites $A$, $B$ may vary between normal spinel (e.g., divalent on $A$ site, trivalent on $B$ site) and inverse spinel (e.g., divalent and trivalent on $B$ site, trivalent on $A$ site). In the spinel structure, the cations occupy the special positions $8a$ and $16d$. The anions occupy the general positions $32e$, which for their complete description require an additional parameter, generally designated as fractional coordinate $x$ (please note that in many publications the fractional coordinate is called $u$ parameter, but in this review the symbol $u$ is used as generalized exchange parameter), in oxide spinels sometimes known as the oxygen and in thiospinels known as the sulphur parameter. If the origin of the unit cell is taken at the centre of symmetry, then for $x = 0.250$ the anions form an exactly cubic close-packed array and define a regular tetrahedral coordination polyhedron about the $8a$ sites (point symmetry $\bar{4}3m$) and a regular octahedron about the $16d$ sites ($m3m$).

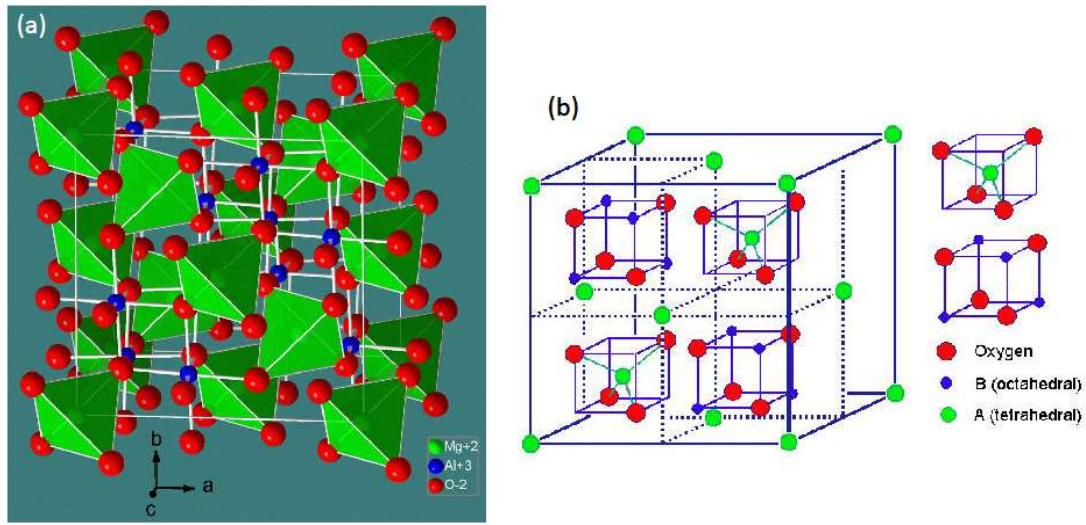

Fig. 4: Crystal structure of spinels. Structure and arrangement of octahedral and tetrahedral sites in the normal spinel (Figure taken from Ref. [58]).
(a) Normal spinel structure of $MgAl_2O_4$. Mg ions (light green) are embedded within oxygen tetrahedra (green). Al ions (blue) have an octahedral oxygen environment.
(b) Arrangement of the octahedral ($B$ sites) and tetrahedral ($A$ sites) lattice sites in the normal spinel structure. The structure is subdivided into eight cubes and can be described by a face centred compilation of two types of cubes, $A_2X_4$ ($A$ in four-fold coordination) and $B_4X_4$ ($B$ in six-fold coordination). The cubic elementary cell contains eight formula units.

However, there also exist a number of spinels with partial inversion and a lot of experimental and theoretical work has been performed to model the distribution of cations on $A$ or $B$ sites. The cation distribution is influenced by electrostatic energy and size effects as outlined by Sickafus et al. [59]: i) anions and cations exhibit repulsion and attraction due to electrostatic energy. These Coulomb terms imply that the $A$ site is occupied by either large ions with low charge or small ions with high charge, ii) within one sublattice, there exists an energy for ordering phenomena between different ions. This energy term becomes important when two transition metals can occupy the same site and iii) concerning the site-preference energy of certain ions: Some ions have a distinct preference for distinct coordination numbers, e.g., $Fe^{2+}$ for tetrahedral and $Cr^{3+}$ for octahedral coordination. The site preference of the spinel structure is given by the difference between the stabilization of this ion on $B$ sites and that on $A$ sites. Dunitz and Orgel [60] and independently McClure [61] have deduced site preferences in spinels from an electrostatic model, which can be summarized as follows: There is a distinct octahedral site preference of ions with $d^3$ configuration, there is no site preference of ions with $d^0$, $d^1$, $d^2$, $d^4$, and $d^8$ configuration. Ions with $d^5$, $d^6$, $d^7$, $d^9$, and $d^{10}$ configuration prefer tetrahedral sites. From these rules it is clear that a number of spinels can be synthesized in pure normal or pure inverse form, but always reveal partial inversion. A representative result is provided by ternary aluminium oxides $AAl_2O_4$ where the $A$-site ions are Mn, Fe or Co. Despite the fact that in these compounds, the transition metals prefer the $A$ site and aluminium the $B$ site, minor site inversion with a small percentage of the transition metals on the $B$ site cannot be excluded [62], which might explain differences in published results and the spin-glass like ground states in the Fe and Co compounds.

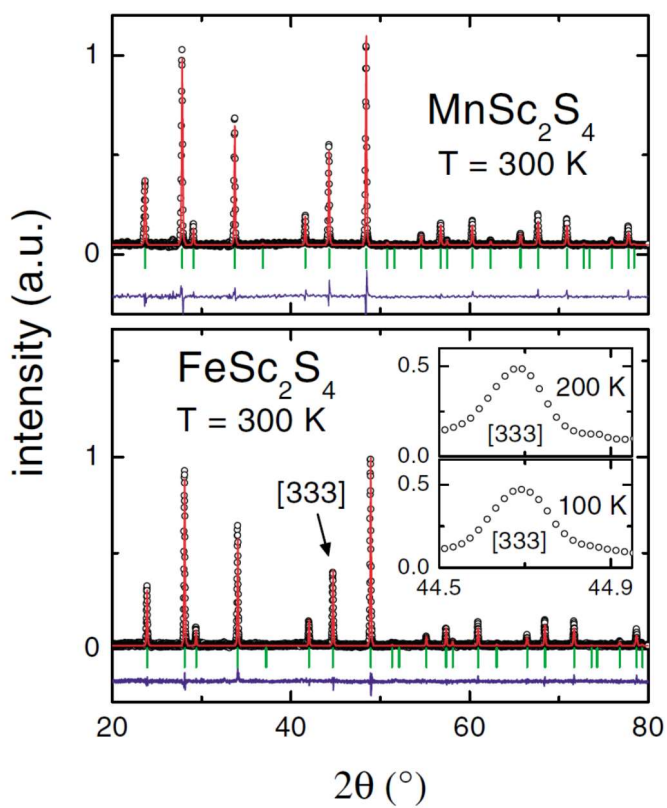

Fig. 5. Neutron diffraction profiles. Diffraction pattern of MnSc$_2$S$_4$ (upper frame) and FeSc$_2$S$_4$ (lower frame). The solid lines represent fits utilizing a Rietveld analysis. The difference spectra shown below both diffraction profiles document the absence of any impurity phases.
Inset: [333] reflection at 100 and 200 K, respectively showing no indications of line splitting or broadening.

Reprinted figure with permission from Fritsch et al. [40]. Copyright (2004) by the American Physical Society.

The applicability of the hard-sphere ionic model for the spinel structure has been proven by Hill et al. [1]. These authors have shown that the lattice constants of binary and ternary oxide spinels are a simple and approximately equal function of the effective radii of the octahedral and tetrahedral cations, but are essentially independent of the electro-negativities of the respective ions. They also documented that for thiospinels the cell edge is primarily a function of the radius of the octahedral cation alone, with small additional contributions from the tetrahedral cation radius and from the octahedral cation electronegativity. As outlined above, the anions in the spinel structure are located on equipoint 32$e$, with their detailed position determined by one parameter $x$, called the fractional coordinate or the anion positional parameter. Hill et al. [1] also showed that there is a simple relationship between this positional parameter and the ratio of the octahedral and tetrahedral bond lengths. In case the bond lengths are equal, the positional parameter $x = 0.2625$.

A detailed review on the physical and crystallographic properties of spinels has been given by Romeijn [42,43]. Spinel compounds can be synthesized in high quality and purity, in ceramic form, and in many cases as single crystals. An example is given in Fig. 5 showing the neutron-diffraction pattern of two polycrystalline thiospinels, namely MnSc$_2$S$_4$ and FeSc$_2$S$_4$ [40]. The room-temperature spectra were fitted using Rietveld refinement. Bragg peaks with resolution-limited width and the good agreement of experimentally observed and calculated intensities document high-quality and strain-free samples. The difference spectra, shown in the lower parts of each frame, document the absence of any impurity phases. From the detailed Rietveld refinement (solid lines in Fig. 5) the lattice constant and the fractional coordinate of the sulphur atom were determined as $a = 10.621$ Å and $x = 0.2574$ Å for the Mn compound, and $a = 10.519$ Å and $x = 0.2546$ Å for the Fe compound, respectively. The deviation of the sulphur parameter $x$ from the ideal value 1/4 indicates a slight trigonal distortion of the octahedra around the $B$ sites, while the tetrahedra remain undistorted. These results obtained in Sc spinels are typical values observed in spinel compounds and document the large unit cell with eight formula units per cell. In this work the [333] Bragg reflection of the iron compound was followed as function of temperature (inset of Fig. 5) and the absence of any broadening or splitting allowed to exclude any structural phase transition in the temperature range investigated.

### 2.3. Crystal-electric field and Jahn-Teller effect

The magnetic and electronic properties of transition-metal compounds are strongly influenced by the crystal-electric field (CEF) splitting of the energy levels of the $d$ electrons in a crystalline lattice. Since the ground-breaking work of Bethe [63] and van Vleck [64], there exists an enormous amount of literature covering this field, which certainly nowadays belongs to textbook knowledge. There are two limiting descriptions of the atomic outer electrons after the atoms have been brought together to form a crystal: Crystal-field theory rests on the assumption that the outer electrons are localized at discrete atomic positions, whereas band theory assumes that each electron belongs collectively to all the atoms of the periodic crystalline array [65]. The crystal-field model assumes that in a crystalline lattice a group of ligands, which can be represented by negative point charges, surrounds a transition-metal ion. This CEF will lift the degeneracy of the $d$ electrons and the resulting CEF splitting strongly depends on the symmetry of the surrounding electric field. A prototypical example of the CEF splitting of $d$ electrons, most representative for spinel compounds, is shown in Fig. 6, for a tetrahedral and an octahedral environment of the transition-metal ion. The angular parts of the lobes of the $d$ orbitals are plotted in the left frame of Fig. 6. The energies of the electrons within the five $d$ orbitals separate into two groups, namely into $e_g$ and $t_{2g}$ orbitals. In octahedral symmetry, the $e_g$ orbitals directly point towards the neighbouring negative charges, while the $t_{2g}$ orbitals point in between these charges. The Coulomb energies of these different ionic constellations yield a splitting of the involved energy levels. In tetrahedral symmetry where the orbital lobes do not point directly to or in between the neighbouring ions, these Coulomb-energy considerations result in a significantly smaller splitting of the $d$ levels. In both symmetries, the splitting is always barycentric, following a centre-of-gravity-type rule. Depending on the valence of the transition-metal ion, the five $d$ levels can now be occupied by up to 10 electrons, strictly following Pauli's exclusion principle. If the crystal-field splitting is moderate or small, the distribution of electrons within these levels strictly follow Hund's rules, i.e. each electronic level first is populated with one spin species before it is doubly occupied (high-spin case). If the CEF splitting is strong, the low levels are fully occupied, before the excited levels become occupied and a low-spin state is realized. The Jahn-Teller (JT) theorem [66] states, "any non-linear molecular system in a degenerate electronic state will be unstable and will undergo a distortion to form a system of lower symmetry and lower energy thereby removing the degeneracy". Notably, spinel compounds played an important role in realizing the importance of the JT effect in transition-metal oxides. A transition-metal ion and its immediate anion neighbours can be considered as the molecule in this case, and the occurrence of degeneracy in the ground state of the "molecule" may be predicted from the optical spectrum and the crystal-field theory [61,67]. The degeneracy of an electronic state makes a symmetric configuration of atoms unstable, causes a structural phase transition, thereby lowering one of the electronic energy levels and hence the total electronic energy of the atom embedded in a deformable crystal lattice. Via the structural phase transition and the lowering of one energy level, one specific type of orbital orientation is favoured, yielding long-range OO. Hence, the JT transition always is directly connected with orbital-ordering phenomena. Staying with the example of $d$ electrons in octahedral or tetrahedral environment as given in Fig. 6, the corresponding ions with the electronic configurations $d^4$ ($Cr^{2+}$, $Mn^{3+}$) and $d^9$ ($Cu^{2+}$, $Ag^{2+}$) in an octahedral CEF or $d^1$ ($Ti^{3+}$, $V^{4+}$) and $d^6$ ($Fe^{2+}$, $Ni^{3+}$) in tetrahedral surroundings, usually are called "Jahn-Teller ions" and are prone to orbital-ordering effects. Of course, there are also many ions that are not JT active, e.g., for $Mn^{2+}$ ($d^5$) each orbital would be occupied by just one electron, preventing any energy gain by a lattice distortion. Compounds, which contain these ions having an orbital degeneracy, are found among magnetic insulators of essentially all material classes. Their properties are significantly different from those of the corresponding materials containing non-JT ions: Their crystal structure is distorted, they exhibit structural phase transitions and in most cases, the magnetic structure seems to be more complicated with an anomalously pronounced magnetic anisotropy and magnetostriction. Goodenough [68], Gehring and Gehring [69], and Kugel and Khomskii [70] provided detailed reviews on JT phenomena in solids. In electronically strongly correlated systems, OO may also be established by the Kugel-Khomskii mechanism [70] where the orbital degeneracy is removed via a purely electronic interaction. The beauty, structural, and electronic importance of orbital physics is described in detail in a recent review by Khomskii and Streltsov [29].

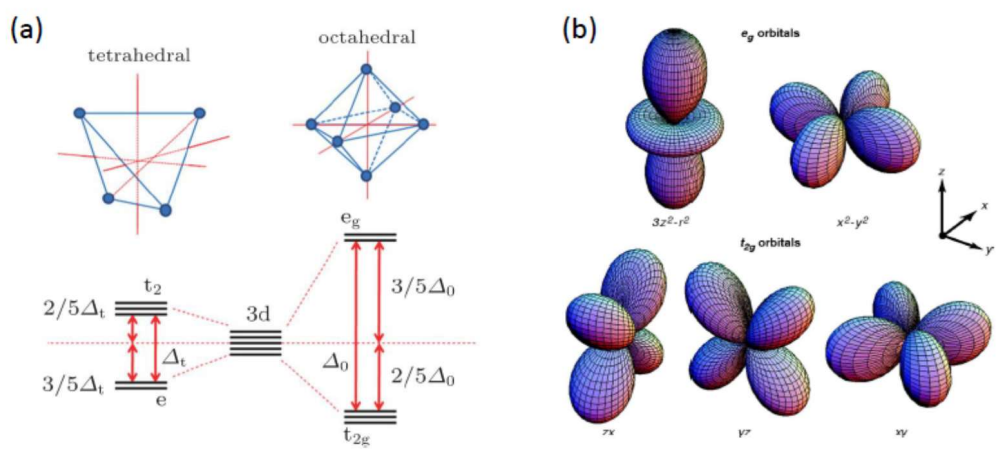

Fig. 6. Crystal-electric field splitting and the symmetry of 3$d$ orbitals.
(a) CEF splitting at tetrahedral and octahedral lattice sites. Due to the lattice symmetry and shape of the orbitals, the splitting at the octahedral sites is significantly larger. In tetrahedral symmetry the ground state is labelled $e$. The missing of the subscript $g$ (gerade) comes from the fact that the tetrahedral environment lacks a center of inversion. In both symmetries, the CEF splitting of the orbitals is barycentric (see text). (b) Shape (angular part of the $d$ orbitals) of $e_g$ (upper row) and $t_{2g}$ orbitals (lower row) orbitals.

With respect to the spinel compounds reviewed in this work, $Fe^{2+}$ at the tetrahedral $A$ site is of prime importance. As outlined above, the energy-gain of a structural transition driven by orbital-ordering phenomena is small and hence the transition energies are low. In $FeCr_2S_4$ the JT transition appears at 10 K and can easily be identified via heat-capacity experiments or via ultrasound investigations [71]. In $FeSc_2S_4$, the JT energy is too low, OO is completely suppressed and this $A$-site spinel belongs to the rare examples of a SOL, where spin and orbital degrees of freedom are strongly entangled and do not show any type of long-range order down to the lowest temperatures [40]. These aspects of orbital physics will be discussed later in full detail.

### 2.4. Magnetic exchange interactions

Paramagnetic (PM) materials contain "magnetic" atoms with partially filled shells giving rise to localized effective magnetic moments containing spin and orbital contributions. At high temperatures, these moments are subject to thermal excitation and exhibit a net magnetic moment only in external magnetic fields. This gives a magnetic susceptibility $\chi = M/H$ where the magnetization $M$, in not too high magnetic fields $H$, follows the well-known Curie law

$$\chi = C/T$$

with the Curie constant

$$C = \frac{Ng^2 S(S+1)\mu_B^2}{3k_B}$$

Here $N$ is the number density of magnetic moments with spin $S$, $g$ corresponds to the Landé factor, $\mu_B$ is the Bohr magneton and $k_B$ the Boltzmann constant. In case of quenched orbital moments, as is the case in many transition-metal compounds, $g = 2$ corresponds to the spin moment only. Residual SOC becomes apparent via small deviations from the spin-only moment. To understand magnetic-ordering phenomena, it is assumed that each magnetic moment (spin) feels an effective internal field of all the other spins. This effective average internal magnetic field is called a molecular field or the Weiss field $H_i = \lambda M$ where $\lambda$ is the molecular-field coefficient. The magnetic susceptibility including magnetic interactions via this molecular field is given by the Curie-Weiss (CW) law

$$\chi = \frac{C}{T - \Theta_{CW}}$$

with $\Theta_{CW} = \lambda C$, the CW temperature, which is positive in the case of FM interactions tending to align magnetic moments parallel and which is negative in the case of AFM interactions leading to antiparallel alignment. One should be aware that in the late forties and early fifties it was unclear whether an AFM spin pattern can be the true ground state [72]. Concerning these critical remarks it is highly elucidating to read Anderson's historical memories about the early concepts of magnetism [73]. At that time, the best proof of antiferromagnetism was the experimental neutron-scattering result on a series of transition-metal oxides by Shull et al. [74], documenting that this state indeed does exist in nature. In complex materials with more than one spin species, there exists the possibility that their magnetic moments align antiparallel, while not being equal. These materials are called ferrimagnets. The first prediction and theoretical description of ferrimagnetism was given by Néel [75] and one of the first neutron-scattering studies on the magnetic structure of the spinel compound magnetite strongly supported Néel's proposed FiM structure [76].

However, from the very beginning, it was clear that magnetism and the alignment of spins with relatively high ordering temperatures could not be explained utilizing classical dipolar magnetic interactions. The origin of magnetic exchange can only be described by quantum mechanics and these quantum-mechanical exchange-interaction effects were independently discovered in ground-breaking works by Heisenberg [77] and Dirac [78]. Extending Heitler and London´s work on the bonding of the hydrogen molecule [79], Heisenberg interpreted the "Weiss field" as quantum mechanical exchange [80]. Assuming that the electrons carrying magnetism are localized on the atoms, the exchange Hamiltonian can be given by the following sum

$$H = -2\sum_{i,j} J_{ij}\, S_i\, S_j.$$

Here $J_{ij}$ is the exchange integral between neighbouring spins $S$ at sites $i$ and $j$. This is precisely what Heisenberg assumed in his theory of ferromagnetism and this direct exchange results from a finite overlap of neighbouring wave functions. Depending on the interatomic distances, it can be positive or negative. Negative values imply AFM exchange.

Heisenberg´s model can easily be compared with the molecular mean-field theory as considered above. Applying the Weiss theory, the internal magnetic field is given by

$$g\mu_B H_i = -2 \sum_j J_{ij} S_j$$

Then the molecular-field coefficient is determined by

$$\lambda = 2zJ/Ng^2\mu_B^2$$

This equality is derived with the assumption that the exchange integral is nonzero only for $z$ nearest neighbours and always is equal to $J$. From this equation the CW temperature can be calculated as

$$\Theta_{CW} = 2zJS(S+1)/3k_B$$

.

This formula provides a direct relation of the molecular mean-field theory to the quantum mechanical Heisenberg-type exchange.

In the course of numerous studies of antiferromagnets it became clear that in many compounds the magnetically interacting atoms are far apart and often separated by intervening non-magnetic ions, with MnO - at that time- being the most illuminating example. Kramers was the first to point out that it should be possible to have exchange coupling mediated by non-magnetic ions [81] and later on this so-called superexchange (SE) mechanism in magnetic insulators was detailed by Anderson [82]. It was Kanamori [83] who documented the dependence of sign and strength of the SE interaction on the orbital state within the cations surrounding crystalline field. Independently Goodenough and coworkers [84,85,86,87] established a complex set of semi-empirical rules to describe the SE interactions in magnetic insulators with different cations and different bond angles covering both kinetic and correlation interactions. Nowadays these rules are known as Kanamori-Goodenough rules and were successfully applied in interpreting magnetic exchange and spin configurations in a large variety of magnetic compounds. The two most prominent rules are i) 180° exchange between half-filled orbitals is strong and AFM and ii) 90° exchange between half-filled orbitals is FM and rather weak.

### 2.5. Magnetic frustration

A large number of compounds with localized magnetic moments does not show magnetic long-range order at low temperatures despite strong magnetic interactions. Long-range magnetic order can be suppressed by substitutional or structural disorder, by dimensionality or by frustration effects. Geometric frustration in spin systems can be found in a variety of crystallographic structures: Ising spins on a triangular lattice, spins on a Kagome or honeycomb lattice, or spins on a pyrochlore lattice. The ground state in frustrated spin systems either can be a spin liquid, a dynamically disordered quantum state, or as experimentally observed in the vast majority of cases, the ground state is non-generic and is driven by some non-leading marginal magnetic interactions. The most prominent examples of quantum spin liquids, are the resonating valence-bond state introduced by Anderson [88] and the Kitaev-type spin liquid on a honeycomb lattice [89]. Reviews on frustrated magnets or spin liquids can be found in Refs. [21] as well as in [90,91].

While from a theoretical point of view very different and complex spin-liquid phases can evolve, from an experimental point of view the identification of strong frustration effects in magnetic systems is much easier. From an analysis of the PM susceptibility, the magnitude of frustration can easily be identified, as indicated in Fig. 7. In systems with localized magnetic moments, the high-temperature magnetic susceptibility follows a CW behaviour, defining a CW temperature $\Theta_{CW}$, which in antiferromagnets lies on the negative temperature axis. The AFM phase transition is expected to occur at the modulus of this temperature, at the AFM ordering temperature $T_N \sim |\Theta|$. In frustrated magnets however, conventional magnetic order is suppressed and a spin-liquid ground state with zero ordering temperature or the onset of non-generic magnetic order at a temperature $T_f$ will be observed instead. From a plot of the inverse susceptibility vs. temperature, a frustration parameter can be defined via $f = |\Theta_{CW}|/T_N$. Following the proposal of Ref. [90] we define strong frustration by values $f > 10$.

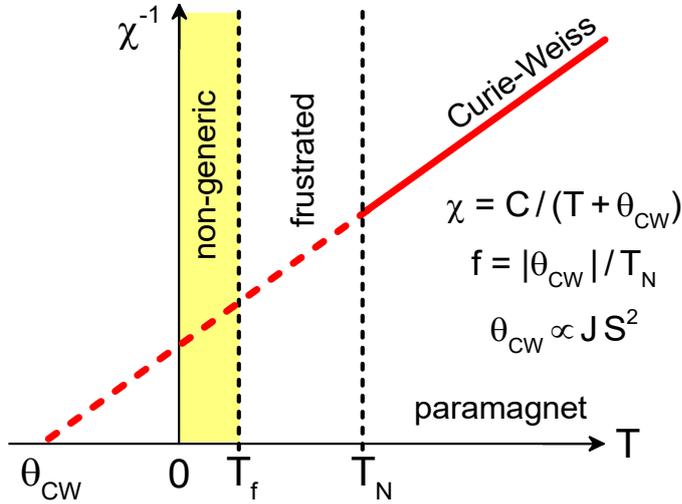

Fig. 7. Magnetic susceptibility in frustrated magnets.
At high temperatures, the inverse susceptibility follows a CW law with a given negative CW temperature $\Theta_{CW}$ (bold and dashed red line). In non-frustrated magnets, magnetic order is expected approximately at $T_N \sim |\Theta_{CW}|$. By frustration effects magnetic order is suppressed and PM susceptibility can be found deep below $|\Theta_{CW}|$. At low temperatures the system realizes a spin-liquid phase or non-leading interactions drive a non-generic spin order below $T_f$.

## 3. The fascinating physics of spinel compounds

### 3.1. Complex magnetism: Néel, Yafet-Kittel, and spin-spiral ground states

Spin ordering in magnetic compounds including spinel ferrites, like magnetite $Fe_3O_4$, was described in Néel's work [75], which is based on the Weiss molecular-field approximation within the framework of the Heisenberg model: the ground state is that of the Heisenberg model with the spin operators replaced by classical spins of fixed lengths, equal to the respective quantum numbers. Néel assumed the dominant exchange interactions between nearest-neighbour $A$-$B$ pairs of spins of the form $H_{AB} = J_{AB} S_A S_B$, with $J_{AB} > 0$. This assumption led to $A$-site spins being antiparallel to the $B$-site moments, which is called Néel configuration. Shortly after Néel's proposal of a two-sublattice model of uncompensated antiparallel spins for ferrimagnets [75], Anderson put forth a microscopic model utilizing an indirect exchange mechanism [82], while Kanamori [83] and independently Goodenough and coworkers [84,85,86,87] proposed a number of coupling rules between neighbouring and overlapping electron orbitals. Depending on geometrical constraints and electron orbitals involved, these Kanamori-Goodenough rules provided a realistic estimate of FM or AFM exchange mechanisms important for establishing long-range ordered magnetic ground states. From the very beginning, spinel-type compounds served as a scientific playground and as paradigmatic examples to test these ideas and to describe a variety of complex magnetic phases. Néel's picture of FiM ordering was confirmed in the early fifties by neutron diffraction experiments on magnetic material including magnetite [76].

The fact that spin configurations can be much more complex was proposed by Yafet and Kittel [24] assuming further nearest neighbour magnetic $A$-$A$ and $B$-$B$ interactions, which are AFM and not negligible. As the most significant result, these authors found a triangular spin structure of the magnetic moments located at two magnetic sublattices, which nowadays is known as the YK spin structure. To our knowledge, this was the first proposal of non-linear spin configurations in ordered magnets. Experimental evidence for the existence of this triangular YK structure has been provided by Lotgering [49] for some normal chromium-spinel compounds. This analysis of magnetic-susceptibility data has later been applied in full detail also for $MnCr_2S_4$ [25], arriving on similar conclusions about the stability of the YK spin structure. Jacobs [92] carried out high-field susceptibility measurements on $MnCr_2O_4$. His experimental results indicated that the magnetic moments were not collinear, and he described the results in terms of the YK theory. However, triggered by a critical remark of Anderson [82] on the stability of the YK structure and by the proposal of the existence of spiral spin structures by Villain [93] and by Yoshimori [94], an enormous activity concerning magnetic exchange interactions in spinel magnets started.

For spinel compounds, in a series of manuscripts a theory of the ground-state spin configuration has been presented by Lyons, Kaplan, Dwight, and Menyuk (LKDM) [95,96,97,98] almost 60 years ago. This theory has been later reviewed and summarized by Kaplan and Menyuk [99]. Using a model of classical Heisenberg spins and considering only $BB$ and $AB$ nearest-neighbour interactions, LKDM were able to document that in this case the ground-state magnetic structure is determined by a single parameter $u$ only, given by

$$u = \frac{4 J_{BB} S_B}{3 J_{AB} S_A}.$$

This effective exchange parameter $u$ represents the relative strength of the two nearest-neighbour $BB$ and $AB$ interactions multiplied by the appropriate spin values $S$. For $u \leq u_0 = 8/9$, the collinear Néel configuration, where all $A$-site spins are parallel to each other and antiparallel to the $B$-site moments, is the stable ground state. For $u > u_0$, it was shown that a FiM spiral configuration has the lowest energy out of a large set of possible spin configurations and that it is locally stable for $u_0 < u < u'' = 1.298$. For $u > u''$, this spiral configuration is unstable. Therefore, it was suggested that the FiM spiral is very likely the ground state for $u_0 < u < u''$, but can definitely not be the ground state for $u > u''$.

Subsequently, magnetic ordering of the normal spinel ferrimagnets $CoCr_2O_4$ and $MnCr_2O_4$ has been investigated by detailed neutron-scattering experiments in order to prove these LKDM predictions. In these spinel compounds the magnetic $Co^{2+}$ or $Mn^{2+}$ ions occupy the $A$ sites, while the magnetic $Cr^{3+}$ ions occupy the $B$ sites. The FiM ordering temperatures of the two compounds are of order ~ 90 K and ~ 50 K and their $u$ values as defined above where determined as ~ 2.0 and 1.6 respectively. According to the LKDM theory, these $u$ values correspond to the locally unstable regime. For $MnCr_2O_4$, Corliss and Hastings [100] found that the FiM spiral model proposed by LKDM does indeed give an approximately qualitative account of the intensities of the satellite lines. They also were able to show that neither a simple Néel model nor the YK model account for the major qualitative features of the diffraction pattern even at the lowest temperatures. Similar results with analogous conclusions were obtained by Menyuk et al. [101] for $CoCr_2O_4$. Their results indicated that the true configuration determined experimentally closely approximates the model predictions. However, they concluded that the range of exchange-interaction ratios obviously well extends beyond the onset of the theoretically predicted instability. 40 years later, Tomiyasu et al. [102] reinvestigated both chromite spinels by magnetization and neutron scattering experiments on single crystalline compounds. The use of single crystals and the enormous experimental achievements in neutron scattering techniques allowed performing the experiments with much better resolution. They analysed their results using a FiM spin-spiral structure similar to the one proposed by LKDM, but with the cone angles of the individual magnetic sublattices not restricted to the LKDM theory. The main and new result of their work is that the spin structure of both compounds is essentially the LKDM FiM spiral, but with long-range order (LRO) of all the spin components replaced by LRO of the longitudinal spin components and short-range order (SRO) of the transverse, or spiral, components.

Later on, Ederer and Komelj [103] presented first-principle LSDA+U calculations of the magnetic coupling constants in the spinel magnets $CoCr_2O_4$ and $MnCr_2O_4$. One of the main conclusions of their work is that the coupling between the $A$-site cations, neglected in the classical LKDM theory, is of appreciable size in both compounds and definitely not negligible. As a summary of these investigations of chromium spinel oxides, Fig. 8 provides an overview of predictions of spin states as function of the $u$ parameter of the LKDM theory including some spinel compounds, which are discussed in the course of this work. With increasing values of $u$, magnetic frustration increases. Only magnetite has $u < 8/9$ and exhibits the predicted Néel-type spin structure. $MnCr_2O_4$ and $CoCr_2O_4$ are well beyond the critical $u$ value corresponding to the locally unstable region. For these compounds, a spin-spiral ground state has been observed, however, with only SRO of the transverse (conical) spin components, while the longitudinal (FM) components exhibit LRO.

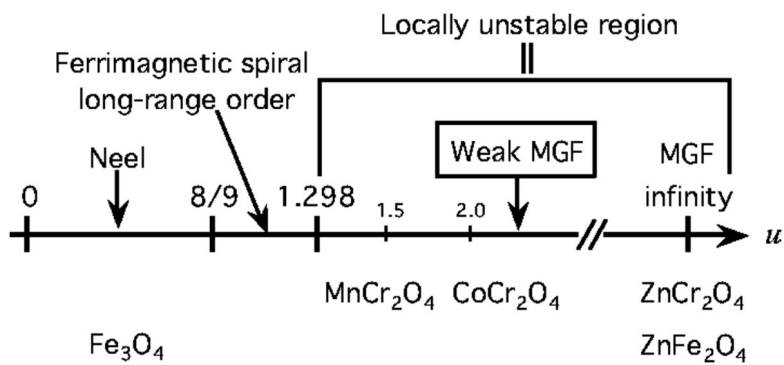

Fig. 8. Spin structure as function of frustration. Spin structure vs. exchange parameter $u$, predicted by the LKDM theory for materials with cubic phases. MGF abbreviates "magnetic geometrical frustration." Reprinted figure with permission from Tomiyasu et al., [102]. Copyright (2004) by the American Physical Society.

While the investigation of oxide spinels started in the early fifties, detailed and systematic research on thiospinels, or more generally on spinels with the anions being a chalcogenide, namely $X$ = S, Se or Te, started ten years later, despite the fact that first magnetic measurements on thiospinels were carried out by Lotgering [49,50] much earlier. However, intensive research started with the observation that the chalcogenide spinels $CuCr_2X_4$ ($X$ = S or Se) are metallic ferromagnets with Curie temperatures well above 300 K [104] and the subsequent discovery that a number of chromium spinels $MCr_2X_4$, with $M$ = Cd or Hg, and $X$ = S or Se, are semiconducting ferromagnets [105,106]. While it seemed that from a viewpoint of application oxide spinels are more important, thiospinels exhibit a much broader variety of ground states being highly interesting when viewed with the focus on basic research. The rich physics of the thiospinels will be detailed throughout this review. Interestingly, the electronic ground state of $CuCr_2S_4$, exhibiting metallic conductivity, is even today far from being completely understood. Lotgering [104] and Lotgering and van Stapele [107] assumed monovalent copper at the $A$ site and mixed valent $Cr^{3+}$ and $Cr^{4+}$ at the $B$ site. Alternatively, Goodenough proposed divalent copper at the $A$ site and conventional trivalent chromium at the $B$ site [108]. Goodenough's model gained support from detailed neutron-scattering experiments on $CuCr_2Se_4$ from Robbins et al. [109]. These neutron diffraction experiments indicate a magnetic moment of 3 $\mu_B$ on each $B$ site and zero moment at the $A$ site. The neutron results and the metallic conductivity are consistent with the ionic configuration $Cu^{2+}$ and $Cr^{3+}$ in which the cupper orbitals are delocalized and form a partially filled band. A detailed review on the physics of sulphospinels was published by van Stapele [110]. It is almost impossible to review all the experimental and theoretical work published on the magnetism of oxide spinels and thiospinels during the last 70 years or so. We decided to exemplify the complexity of magnetic behaviour and of the magnetic ground states by documenting a survey of the work performed on some selected spinel compounds, which early on were in the focus of our own experimental work.

During the last twenty years, we tried to solve related questions concerning the spin ground state in $MnCr_2S_4$ by a variety of experimental techniques including magnetic susceptibility, magnetization and heat capacity experiments as well as electron-spin resonance (ESR) measurements, all performed on single-crystalline material. The main conclusions including also results from other groups are detailed below. Here we mainly focus on the temperature dependent properties. The magnetic-field dependence including high-field experiments will be discussed in the chapter on MF compounds (3.4.2.) as well as in the chapter on spinels in high magnetic fields (3.10.1.). First reports on the magnetic properties of the cubic $MnCr_2S_4$ were reported by Lotgering [49] using magnetization experiments and by Menyuk et al. [111] utilizing magnetization and neutron-diffraction experiments. $MnCr_2S_4$ is a normal cubic spinel, with $Mn^{2+}$ and $Cr^{3+}$ occupying the $A$ and $B$ sites respectively. The FiM ordering temperature is of order ~ 65 K and one observes a pronounced maximum in the temperature-dependent magnetization close to 40 K. The magnetization results and the absence of a (200)-reflection peak in neutron-diffraction experiments [111] were interpreted in the framework of a collinear Néel model. However, this was only possible assuming large and FM $B$-$B$ interactions and a significantly reduced magnetic moment of the $Mn^{2+}$ ions. In subsequent magnetization experiments, Lotgering [25] documented a low-temperature transition at 5 K to a canted spin state, with spin canting of the manganese moments occupying the tetrahedral $A$ sites, constituting the triangular YK spin structure. This magnetic transition into a canted YK phase was further corroborated by the observation of the appearance of a (200) reflection below the transition temperature in subsequent neutron-diffraction experiments [112]. Thirty years later, Tsurkan et al. [113] started a series of experiments in single crystalline $MnCr_2S_4$. Figure 8 shows the temperature-dependent magnetization below 80 K for a series of external magnetic fields up to 2 kOe (0.2 T). These measurements were performed with the external field along

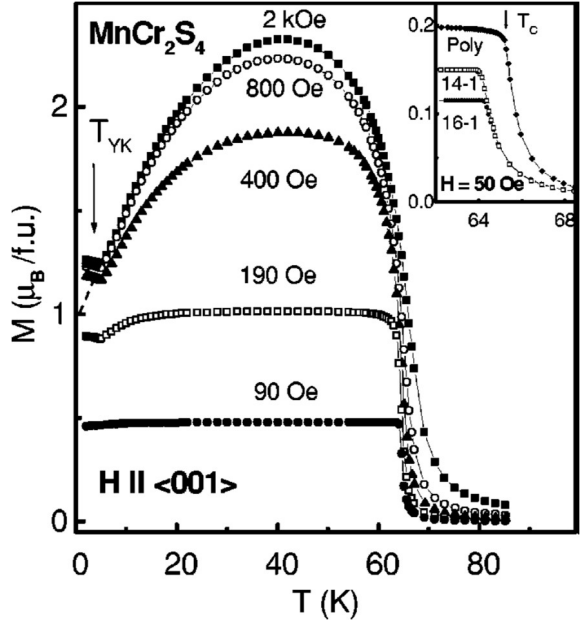

Fig. 9. Temperature dependent magnetization of MnCr$_2$S$_4$.

$M$ in $\mu_B$ per formula unit, in different magnetic fields applied along the <001> direction for a single crystalline sample. Inset: Magnetization in the region of the ferromagnetic transition for two different single crystals and for polycrystalline material in an external field of 50 Oe. Arrows indicate magnetization anomalies at the Curie temperature $T_C$ and at the Yafet-Kittel transition temperature $T_{YK}$. Solid lines are guides to the eye.
Reprinted figure with permission from Tsurkan et al. [113]. Copyright (2003) by the American Physical Society.

the crystallographic <001> direction. The significant field dependence results from demagnetizing fields. For low external fields the sample is in a multidomain state, while for high fields (> 1 kOe) a single-domain state is formed. For higher fields, the temperature dependent magnetization reveals the prominent maximum close to 40 K. A break of slope close to the transition temperature into the triangular YK state at $T_{YK} \sim 5$ K is clearly visible. The inset in Fig. 9 documents that the FiM transition temperatur $T_c \sim 65$ K is rather constant as measured in different single crystals, but is slightly enhanced in polycrystalline material.

The measurements in the single-domain phase indicate a striking non-monotonic temperature dependence of the magnetization with a maximum close to 40 K. For low temperatures, the extrapolated magnetization tentatively approaches a value of 1 $\mu_B$ (dashed line in Fig. 9), which corresponds to the expected resulting magnetic moment, if the moments of the Cr and Mn ions are assumed to be in a strictly antiparallel collinear arrangement. However, below 5 K the magnetization levels off to a nearly constant value of about 1.23 $\mu_B$ down to the lowest temperature at 1.8 K. This change of slope of the magnetization below 5 K is attributed to the transition into a non-collinear YK-like triangular spin structure. It is assumed that an additional AFM $A$-$A$ magnetic exchange interaction induces a finite angle between the two manganese spins located at the two lattice sites on the diamond lattice.

FiM ordering at $T_c$ and the phase transition into the triangular canted spin phase below $T_{YK}$ were further elucidated by detailed heat-capacity experiments in zero external magnetic field and in fields of 100 kOe (10 T). These results are plotted in Fig. 10(a) together with the results of low-field magnetization measured at 20 Oe observed at the same single crystal, Fig. 10(b). In zero magnetic field, the transition from the PM into the FiM phase is indicated by a well-defined spike-like anomaly at $T_c$. A $\lambda$-like anomaly indicates a further magnetic transition into a non-collinear spin state at $T_{YK}$. At 100 kOe (10 T) the FiM ordering transition is smeared out and shifted to higher temperatures. The YK phase transition is shifted to 10 K and reveals a broad additional hump at lower temperatures. The field-dependent measurements on MnCr$_2$S$_4$ including a detailed ($H,T$)-phase diagram will be discussed in chapter 3.4 on multiferroics. The temperature-dependent magnetization as measured at low external fields of 20 Oe is shown in Fig. 10(b). The onset of spontaneous magnetization at the FiM ordering is followed by a non-linear temperature dependence of the magnetization. The decrease is interrupted at $T_{YK}$ and is interpreted as realization of the triangular YK spin structure, which is indicated in Fig. 10(b). Here the manganese moments span a finite angle with a residual FiM moment with respect to the chromium spins. In Ref. [113] it was assumed, that just below the onset of FiM ordering at $T_c$ the manganese moments exhibit already a conical spin structure and are not strictly antiparallel to the chromium moments. However, in the temperature regime $T_C > T > T_{YK}$ with a conical-like spin structure, the transverse spin components are still paramagnetically disordered [113]. At $T_{YK}$ the manganese spins lock-in into a fixed triangular structure with AFM transverse components of the manganese spins. The residual longitudinal component exhibits ideal FiM order with respect to the chromium moments. To our knowledge, $MnCr_2S_4$ is the only compound where the triangular YK spin structure has been proven the magnetic ground state.

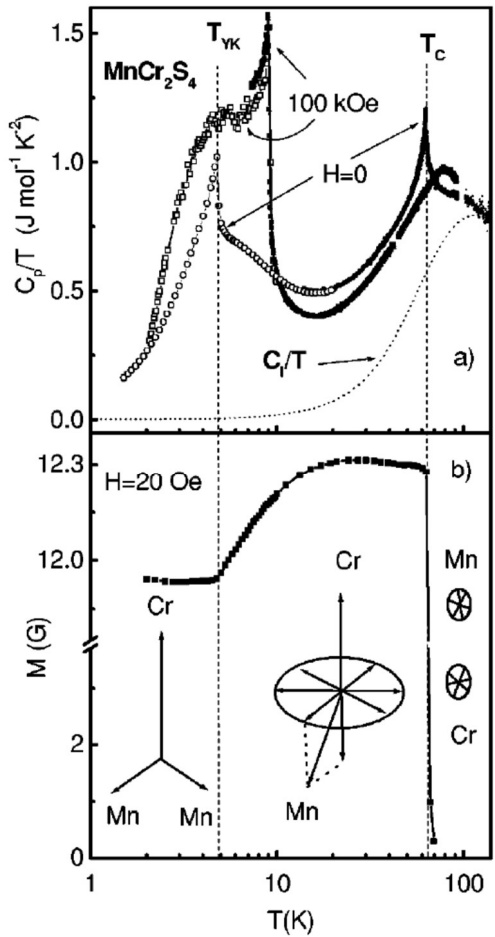

Fig. 10. Heat capacity and magnetization of $MnCr_2S_4$.
(a) Semilogarithmic presentation of the temperature dependence of the molar heat capacity $C_p/T$ for single crystalline $MnCr_2S_4$ measured by adiabatic (open squares) and ac (closed symbols) calorimetric methods in zero magnetic field and at 100 kOe. Dotted line indicates the pure phonon contribution.
(b) Semi-logarithmic plot of the temperature dependence of the magnetization of the same sample measured in low magnetic fields of 20 Oe. Dashed vertical lines indicate the temperatures $T_C$ and $T_{YK}$. The proposed spin structures (triangular, non-collinear with disordered transverse spin component, and paramagnetic) are indicated schematically. Solid lines are drawn to guide to the eye.
Reprinted figure with permission from Tsurkan et al. [113]. Copyright (2003) by the American Physical Society.

A number of spinel compounds also reveal spin spiral order at low temperatures, with $HgCr_2S_4$ probably being the most prominent example. As outlined earlier, there were early reports on FM and insulating chromium spinels by Baltzer and coworkers [105,106], which gained considerable attention. All these FM chromium spinels, e.g., $CdCr_2S_4$, $CdCr_2Se_4$, $HgCr_2S_4$, and $HgCr_2Se_4$, are characterized by a sulphur or selenium anion at the $X$ position of the spinel lattice. It was proposed that ferromagnetism is induced by an increasing importance of the indirect nearest neighbour (NN) 90 ° FM $B$-$X$-$B$ super-exchange compared to the direct AFM NN $B$-$B$ exchange, which is dominating in the oxide spinels. These exchange paths are competing in the oxides and are responsible for low AFM ordering temperatures, but the 90 ° FM $B$-$X$-$B$ super-exchange paths gain increasing importance with increasing lattice constants in the sulphides and selenides [38,106]. In chapter 3.3 on magnetic ground states governed by frustration, these competing exchange paths in the spinels will be discussed in more detail. Already Baltzer et al. [105] found that $HgCr_2S_4$, despite a FM CW temperature $\Theta \sim 140$ K, undergoes a magnetic transition at low magnetic fields close to 25 K. Soon after, utilizing neutron diffraction experiments the ground state of the mercury chromium sulphide was determined to be a spin spiral. The low-temperature spin structure was described by a helical spin order, with the propagation vector parallel to one of the cube edges and the spin-spiral plane perpendicular to the propagation vector [114]. Later on, this unconventional non-collinear spin state was investigated in detail by a variety of techniques in Ref. [115]. The magnetization of $HgCr_2S_4$ as function of an external field at temperatures between 2 and 60 K is shown in Fig. 11. At low temperatures, a clear metamagnetic transition between 1 and 2 T is documented in Fig. 11(a), while at temperatures > 20 K, the magnetization resembles that of a field-induced ferromagnet. At 2 K [Fig. 11 (b)], the magnetization seems to be even more complex with a further critical field close to 5 T and saturation magnetization beyond 8 T. The low-field transition at the critical field $H_1$ is characterized by a strong hysteresis indicative for a first-order transition.

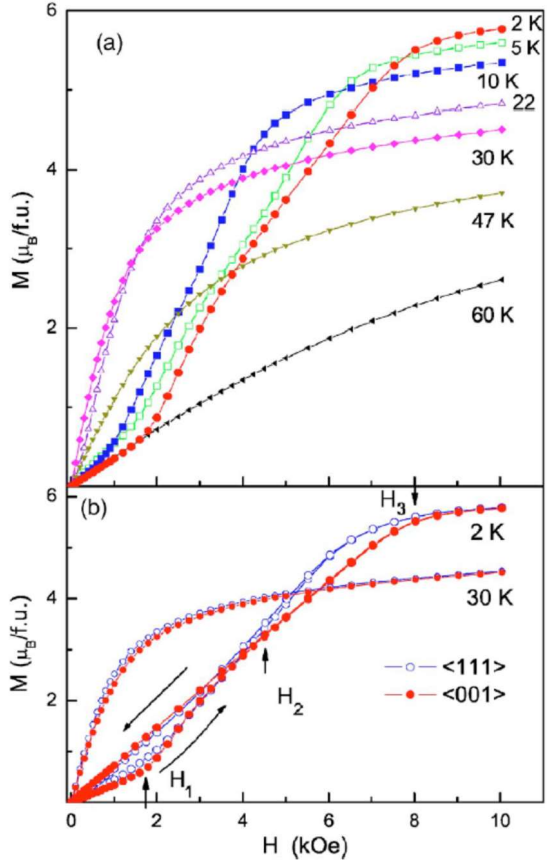

Fig. 11. Magnetization of $HgCr_2S_4$. Magnetic field-dependent magnetization at various temperatures measured in single-crystalline samples.
(a) Magnetic fields applied along the <001> direction for temperatures between 2 and 60 K.
(b) Measurements at external fields along the <111> and <001> directions; at 2 K on increasing and decreasing fields (indicated by arrows) hysteresis effects appear at low fields and close to saturation. Vertical arrows indicated critical fields. At 30 K the magnetization curves in the field-induced ferromagnetic state reveal similar demagnetization factors.
Reprinted figure with permission from Tsurkan et al. [115]. Copyright (2006) by the American Physical Society.

### 3.2. Orbital order and orbital glass

Since early 2000, orbital physics became a fascinating topic in modern solid-state physics and materials science. The spatial orientations of the electronic orbitals govern magnetic exchange and hence determine long-range spin order. If the orientational order of orbitals can be changed by external fields, e.g. strain, the magnetic order will change concomitantly. On long terms, such a tuning of electronic orbitals, so called orbitronics, could be an important ingredient of a future correlated-electron technology [116]. In normal cubic spinels JT effects giving rise to OO can arise either from the *A* or from the *B* sites, or from both simultaneously.

As outlined above, the cooperative JT effect in transition-metal compounds is a phase transition, which is driven by the interaction between localized orbital electronic states and the crystal lattice. It involves the simultaneous splitting of the degenerate electronic states with concomitant orientational ordering of the electronic orbitals and a symmetry-lowering distortion of the lattice. Already in the early fifties, structural distortions and the observation of symmetry-lowering phase transitions were interpreted in relation to the JT theorem. In 1952,

Orgel [117] suggested that the $Cu^{2+}$ compounds have unsymmetrical environments because of the JT effect. Orgel and Dunitz [118] published a survey of the stereochemistry of copper compounds and subsequently discussed the occurrence of the JT effect in numerous materials with a special focus on spinel compounds [67]. Independently, using optical absorption spectroscopy Mc Clure [61] showed that the apparent values of the crystal-field parameter $D_q$ for $Mn^{3+}$, $Cr^{2+}$ and $Cu^{2+}$ located in an octahedral environment are much larger than for other ions having the same electronic charge. This fact was explained by taking into account a JT splitting of several thousand wave numbers in addition to the cubic crystal-field splitting. Already somewhat earlier, Goodenough and Loeb [84] noted that spinels containing a sufficient concentration of octahedral-site $Mn^{3+}$ or $Cu^{2+}$ ions at room temperature were tetragonal with lattice parameters $c/a > 1$ rather than being cubic. Being unaware of the JT theorem, these authors used Pauling's time-honoured bonding concepts to show that the tetragonal distortions were due to orbital ordering that removes the orbital degeneracies of the Mn(III) and Cu(II) ions, respectively. Later on there followed detailed studies of orbital ordering transitions in a number of spinels with $Cu^{2+}$, $Fe^{2+}$, $Ni^{2+}$ or $Co^{2+}$ at the tetrahedral A site [119,120]. The authors of Ref. [120] focused on the investigation of mixed chromium spinels to elucidate the experimental fact that if either Cu(II) or Fe(II) are at the A site, the tetragonal distortion is $c/a < 1$. For Ni(II) the distortion $c/a > 1$, while for Co(II) the spinel does not distort at all. JT-active mixed chromium spinels were also investigated in detail by Kataoka and Kanamori [121]. A molecular-field treatment of the cooperative JT effect, with a specific focus on spinel compounds, was given by Englman and Halperin [122].

Since then, numerous experimental and theoretical works on the JT effect in a variety of spinel compounds have been performed and it certainly exceeds the frame of this work to provide a rigorous and complete review on the work published. We decided instead to elucidate the complexity of the JT physics and of orbital-ordering transitions by predominantly treating the normal spinel $FeCr_2S_4$. In this compound the Cr sublattice ($Cr^{3+}$: $3d^3$, $S = 3/2$) at the B sites is dominated by FM exchange via the 90° Cr-S-Cr bond angle. The low-lying $t_{2g}$ triplet in octahedral symmetry is half filled and the orbital moment is quenched. The Fe ions ($Fe^{2+}$: $3d^6$, $S = 2$) are located at the tetrahedrally coordinated A sites. In this case, the five-fold degenerate ground state is split by the crystal field into a lower orbital doublet and an excited triplet (see Fig. 6). The e ground doublet is split by SOC and additionally is weakly JT active. Goodenough [119] concluded, that in case the octahedral site cations do not contribute to the elastic anisotropy, then the elastic energy may be too small to induce a static JT distortion, but in case a distortion occurs, it should be to tetragonal symmetry ($c/a < 1$).

In $FeCr_2S_4$ magnetic order, driven by the strong FM Cr-Cr exchange, occurs close to $T_c$ = 180 K [49]. Via an AFM coupling of the Fe sublattice to the Cr moments, FiM order with antiparallel chromium and iron moments is established at the magnetic phase transition. From neutron powder diffraction at 4.2 K it has been concluded that the compound remains cubic down to the lowest temperatures and that the spin arrangement is of simple Néel type [123]. The observed magnetic intensities have well been described assuming magnetic moments of $\mu_{Fe} = -4.2$ $\mu_B$ and $\mu_{Cr} = 2.9$ $\mu_B$ (here the minus sign indicates antiparallel alignment of the moments). The former was thought to be enhanced due to moderate SOC ($Fe^{2+}$: $g = 2.2$, $\mu = 4.0$ $\mu_B$), the latter is close to the spin-only value of ($Cr^{3+}$: $g = 2$, $\mu = 3.0$ $\mu_B$). These values seem to be slightly enhanced compared to subsequent neutron-diffraction work: At 10 K, Kim et al. [124] determined ordered moments of − 3.52 (Fe) and 2.72 $\mu_B$ (Cr). In recent high-resolution neutron-diffraction experiments, Bertinshaw et al. [34] measured the temperature dependence of the iron and chromium magnetic moments. At 4 K and in zero external magnetic fields, these authors determined moments of − 3.69(6) and 2.64(3) $\mu_B$ for iron and chromium, respectively. The reduced magnetic moments in zero-field were explained taking a transverse quasi-PM component of the magnetic moment of the $Fe^{2+}$ ions into account, an effect, which will be discussed later in more detail.

Early on there have been a number of Mössbauer studies in $FeCr_2S_4$ reporting the observation of non-zero quadrupole shifts of the hyperfine pattern, usually indicating substantial deviations from cubic symmetry [125,126,127,128,129]. It has to be noted that the spatial symmetry of $Fe^{2+}$ at the tetrahedral site, according to neutron scattering [123] is cubic down to the lowest temperatures and that the observation of a nonzero quadrupole shift is not expected. From these early Mössbauer experiments there arose a substantial controversy how to explain the experimental results. In reference [126] the quadrupole shift was explained by a model utilizing the concept of a magnetically induced electric-field gradient (EFG). In Ref. [127] this interpretation was discarded in favour of magnetostriction effects, and in [128] it was concluded that the EFG might be magnetically induced for $T > 40$ K, but was implausible at lower temperatures. It was by the authors of Ref. [129] to identify an orbital-order transition at $T_{oo} \sim 10$ K: At this temperature, they identified abrupt changes of hyperfine field and EFG and explained these observations in terms of a static distortion of the tetrahedral sites with $c/a < 1$ and attributed this distortion to a JT stabilization of the $^5E_g$ ground state of $Fe^{2+}$ in tetrahedral symmetry. For temperatures > 10 K they proposed to explain the Mössbauer data in terms of a dynamic JT stabilization. They concluded that $FeCr_2S_4$ may be one of the rare examples, which allows studying the transition from a dynamic to a static JT effect. Later on this proposal has been substantiated by further Mössbauer [130,131,132] and low-temperature heat-capacity experiments [131]. These heat-capacity experiments in pure and doped compounds documented the extreme dependence of the appearance of a static JT transition on details of stoichiometry. The sensitivity of the onset of OO has been further elucidated by detailed experiments on single crystals and polycrystals documenting its dependence on the exact synthesis conditions [133].

Figure 12 schematically shows the electronic level splitting in $FeCr_2S_4$ in the different temperature regimes and the internal microscopic driving forces for these effects, which are experimentally observed in Mössbauer experiments. By the tetrahedral crystal field, $Fe^{2+}$ is split into an excited triplet $t_2$ and a ground-state doublet $e$. The tetrahedral crystal-field splitting $\Delta_{tet}$ is of order 0.5 eV. Below the onset of FiM order, the spin degeneracy of the spin state $S = 2$ at the iron site is removed and the electronic ground state will be split into five equidistant doublets by the magnetic exchange field. Finally, at the lowest temperatures a further splitting due to second-order spin-orbit interactions is expected. This splitting can be estimated as $2\delta \sim 12\,\lambda^2/\Delta_{tet}$, where $\lambda$ represents the spin-orbit interaction. Hence, despite the fact that the overall symmetry is almost cubic, there will be a finite magnetically-induced EFG. Below the JT transition with long-range OO, the lowest doublet will additionally split characterized by a splitting $\Delta_{oo}$. This splitting will compete with the splitting due to the spin-orbit interaction and in most cases will yield changes in the level splitting. In addition to these changes of the level splitting also the level scheme can be inverted [Fig. 12(d)]. Obviously, in $FeCr_2S_4$ the energies of spin-orbit and spin-phonon coupling compete and favour different ground states.

In the recent past, a number of detailed Mössbauer and muon spin-rotation experiments on $FeCr_2S_4$ [134,135,136] have been published, elucidating the temperature dependencies of hyperfine field and EFG. Some representative results of these new experiments on samples with highly improved quality concerning crystallinity and impurity states are shown in Fig. 13. In the left frame, a series of $^{57}Fe$ absorption spectra as observed in $FeCr_2S_4$ is documented, characteristic for the different temperature regimes revealing different magnetically or orbitally ordered phases. The shape of these spectra resembles those reported in earlier Mössbauer experiments. Above $T_c$, in the PM phase at 292 K, a dominant single Lorentzian resonance is observed as expected for an ideal $A$ site tetrahedral sulphur coordination of $Fe^{2+}$. Below $T_c$, in the FiM phase at 136.1 K, a symmetrical six-line pattern is observed, typical for a collinear FiM structure, yet with a small quadrupole interaction due to an axially symmetric EFG. The EFG is increasing with increasing magnetic hyperfine field. The observation of a finite quadrupole interaction at the iron site below the Curie temperature, despite the fact that neutron data [123] have identified a cubic structure even in the magnetically ordered phase, was discussed controversially [125,126,128,130,131,132,137,138,139]. As first proposed by [126] this quadrupole interaction is related to a magnetically induced EFG, which occurs for the electronic ground state of $Fe^{2+}$ in a tetrahedral crystalline electric field, at least for certain orientations of the local molecular magnetic field.

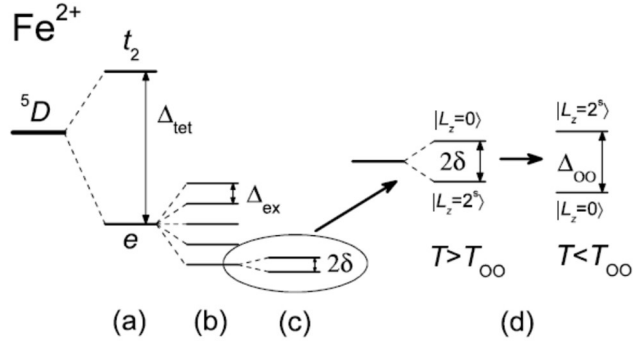

Fig. 12. Electronic level structure of tetrahedrally coordinated $Fe^{2+}$.
(a): The cubic crystal field at the tetrahedral site splits the $^5D$ free-ion ground state into a lower $e$ doublet and an excited $t_2$ orbital triplet. (b): At $T < T_c$, spin degeneracy is removed by the exchange-field splitting of the $e$ states into five orbital doublets separated by $\Delta_{ex}$.
(c): Spin-spin and second-order spin-orbit interactions finally induce a further level splitting of the $e$ ground state into five doublets separated by $2\delta$. (d): Splitting of the lowest doublet with a separation of $\Delta_{oo}$ below the JT transition in the presence of orbital order (OO). In addition to changes in the level splitting, the level scheme is inverted at the onset of OO.
Data for (d) taken from Engelke et al. [134] and Sadrollahi [136].

The temperature dependence of the magnetic hyperfine field $B_{hf}$ is shown in the upper right frame of Fig. 13. Its temperature dependence below $T_c$ is typical for a second-order magnetic phase transition and smoothly follows the sublattice magnetization down to ~ 50 K. Below 50 K, the spectra cannot be described by a single magnetic site (see absorption spectra at 24.8 K). Three sets of hyperfine interactions with equal spectral weight, characteristic for three magnetically and structurally non-equivalent sites were necessary to fit the data [134]. At this temperature, the local axial symmetry is lost and an asymmetry parameter was necessary to describe the field-gradient tensor, which is in agreement with transmission-electron microscopy studies indicating a cubic-to-triclinic phase transition [140]. The non-equivalent magnetic sites found in these Mössbauer data [134] provide further support for the conclusions drawn from muon-spin rotation regarding the change of the magnetic structure from a collinear spin alignment to a spin-spiral phase [135]. Whether this is indeed helical with accompanying changes of the angles between $B_{hf}$ and the electrical-field gradient cannot reliably be resolved by the Mössbauer experiments [134], where this spin-reorientation transition was correlated with the onset of short-range orbital ordering. The gradual decrease of $B_{hf}$ below 50 K (upper right frame of Fig. 13) is caused by continuous changes in the crystalline electric field. Finally, below 10 K a significant change in quadrupole interaction was identified with a change of sign and re-orientation of the EFG (see absorption spectra at 4.2 K). These dramatic changes in the spectra are best documented by the temperature dependence of the quadrupole splitting, shown in the lower right frame of Fig. 13. Its major axis is now pointing perpendicular to the magnetic hyperfine field. Quantitatively similar conclusions have been drawn already in Ref. [128]. The change of sign of the EFG indicates that the sequence of the lowest populated crystalline electric-field states at the Fe site becomes significantly modified upon orbital ordering. This inversion of the electronic levels of the ground state is indicated in Fig. 12(d). At the lowest temperatures, the splitting of the ground-state doublet amounts $\Delta_{oo}$. The opening of an orbital gap can be estimated from the temperature dependence of the quadrupolar splitting below 10 K (see Fig. 13). Utilizing a mean-field approach, this splitting of the ground-state doublet in the orbitally ordered phase has been determined as $\Delta_{oo} \sim 32$ K [136].

In what follows we will present a variety of magnetic, thermodynamic and sound-velocity data taken on FeCr$_2$S$_4$, which are representative for a variety of spinels with strong coupling of the internal degrees of freedom, e.g., of orbital, spin, charge, and lattice degrees of freedom. Tsurkan et al. [141] published results of temperature-dependent measurements on high-resolution x-ray synchrotron powder-diffraction, magnetic-susceptibility, sound-velocity, thermal-expansion, and heat-capacity experiments in FeCr$_2$S$_4$. In this work, the focus was on polycrystalline samples, which were synthesized close to ideal stoichiometry in reasonable quantities for the different experimental methods. Of course, single crystals can be grown by chemical transport. However, large single crystals, which can be prepared using chlorine gas as transport agent, turned out to be slightly electronically doped, because monovalent chlorine can substitute divalent sulphur ions at the $X$ sites. Impurity-free single crystals can be prepared using bromine gas as transport agent. However, these crystals are rather limited in size and for many experimental techniques the sample size is too small. For the preparation of dense ceramic samples suitable for ultrasound and optical experiments, the authors of Ref. [141] used spark plasma-sintering techniques.

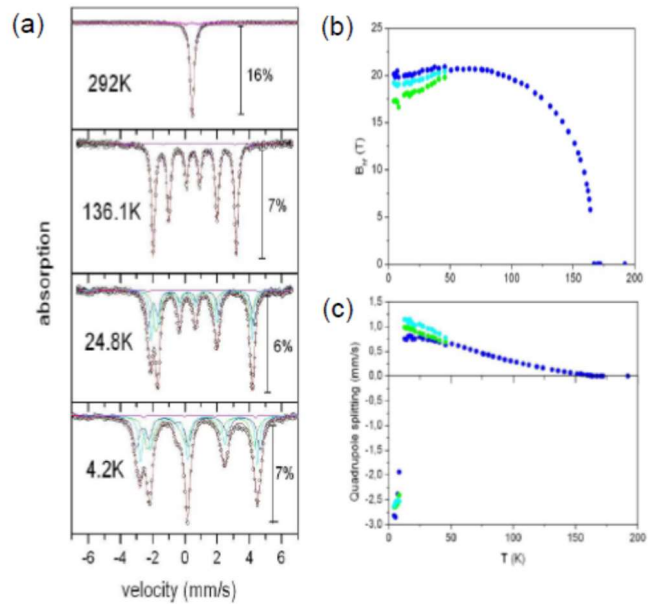

Fig. 13. Mössbauer spectroscopy on FeCr$_2$S$_4$.
(a) Temperature dependence of $^{57}$Fe absorption spectra characteristic for the different phases. At 292 K in the PM phase, a single Lorentzian line is observed. At 136 K, well below the onset of FiM order, a six-line pattern shows up typical for the collinear FiM structure with a magnetically induced axial EFG. At 24 K, three superimposed sextets with axially symmetric EFGs indicate a more complex spin structure. Below 10 K, in the orbitally ordered phase, a clear rotation of the EFG by $\Theta = 90°$ is observed. Its major component becomes negative, doubles, and has a symmetry lower than axial.
(b) Temperature dependence of hyperfine field $B_{hf}$ and (c) of the quadrupolar splitting. For temperatures < 50 K a clear splitting of the hyperfine field is observed and below the onset of OO, the quadrupolar splitting changes sign. Data for this figure taken from Engelke et al. [134] and Sadrollahi [136].

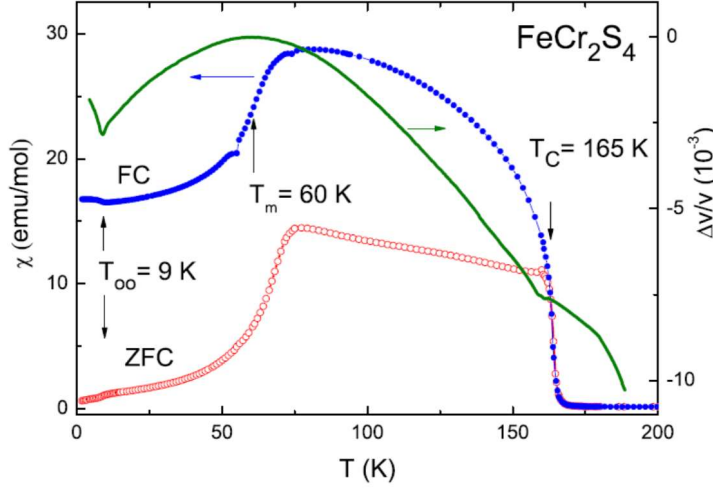

Fig. 14. Temperature dependence of the magnetic susceptibility and of the sound velocity measured in polycrystalline $FeCr_2S_4$.
Left scale: temperature dependence of zero-field-cooled (ZFC) (open red circles) and field-cooled (FC) (closed blue circles) susceptibilities $\chi = M/H$ as measured in an external magnetic field $H$ = 100 Oe.
Right scale: temperature dependence of the relative change of the longitudinal sound velocity (solid green line). The characteristic temperatures are indicated by vertical arrows. Reprinted figure with permission from Tsurkan et al. [141]. Copyright (2010) by the American Physical Society.

Field-cooled (FC) and zero-field cooled (ZFC) magnetic susceptibilities as measured in external magnetic fields of 100 Oe (10 mT) in Ref. [141] are shown in Fig. 14. The well-defined step-like increase of the susceptibility close to $T_C$ = 165 K signals the onset of FiM order with strictly antiparallel alignment of chromium and iron spins. The splitting of FC and ZFC susceptibilities can be explained in terms of domain-wall dynamics. A well-defined decrease of both susceptibilities close to $T_m \sim 60$ K signals a further magnetic transition into a more complex spin state, probably a spiral spin order, which will be discussed later. Finally, a small anomaly at $T_{OO} \sim 9$ K was interpreted as onset of OO of the JT active orbitals of the $Fe^{2+}$ ions in tetrahedral environment.

Fig. 14 also shows the relative change of the longitudinal sound velocity (right ordinate) as function of temperature. The strong coupling of the iron orbitals to the lattice degrees of freedom is documented by a significant anomaly with a clear V-shaped minimum sound velocity at $T_{oo}$. The sound velocity starts to decrease below 60 K, which probably signals the onset of orbital fluctuations with a strong coupling to longitudinal sound waves. In the temperature dependence of the longitudinal sound velocity, the onset of FiM order is indicated by a small anomaly only. Before discussing the orbital ordering transition in full detail, we shortly mention the magnetic transition close to 60 K, which will be discussed later- in the chapter on multiferroics - in more detail. The critical temperature $T_m \sim 60$ K probably indicates a transition into a non-collinear spin configuration of chromium and iron moments. This anomaly has been identified earlier in ESR experiments [142] as well as in ac susceptibility measurements [143]. However, the most convincing evidence has been derived from muon spin-rotation spectroscopy [135] as well as from Mössbauer experiments as documented in Fig. 13 and outlined above [134]. This magnetic transition probably is driven by coupling of the spin degrees of freedom to orbital fluctuations gaining considerable strength close to $T_m$.

To further elucidate the orbital ordering transition in $FeCr_2S_4$, Fig. 15 shows the temperature dependence of the specific heat as taken from Ref. [141]. For presentation purposes, the specific heat is plotted as $C/T$ vs. $T$. Two well defined $\lambda$-type anomalies indicate the phase transitions into the FiM ($T_C$) and into the orbitally ordered ($T_{OO}$) phases. The pure phonon-derived contribution to the heat capacity was estimated utilizing a Debye-Einstein model with appropriate characteristic temperatures, which is indicated by the blue dashed line in Fig. 15 [141]. After subtracting the pure lattice contributions to the heat capacity, the temperature dependence of the remaining magnetic entropy can be calculated and is shown in the inset of Fig. 15. The entropy release at the orbital ordering transition is expected to be of order $R\ln2$, corresponding to the ground state doublet. Approximately half of this entropy value is reached just at $T_{OO}$ only. It is interesting to note that full entropy release, which is expected for the orbital doublet involved in the JT transition, is reached close to the critical temperature $T_m$, again indicating that this temperature denotes the onset of orbital fluctuations and via a strong spin-lattice coupling a magnetic phase transition with a more complex, probably spiral-like spin structure is induced. At the FiM phase-transition temperature again only 50% of the full spin entropy $R(2\ln4 + \ln5) = 36.4$ Jmol$^{-1}$K$^1$ are reached, indicating the importance of strong spin fluctuations in the PM regime in agreement with the large (negative) CW temperature of $-350$ K [141]. It is important to note that no specific-heat anomaly can be observed at $T_m$. It seems that the entropy release at this spin-reorientation transition is too low to be identified in calorimetric experiments.

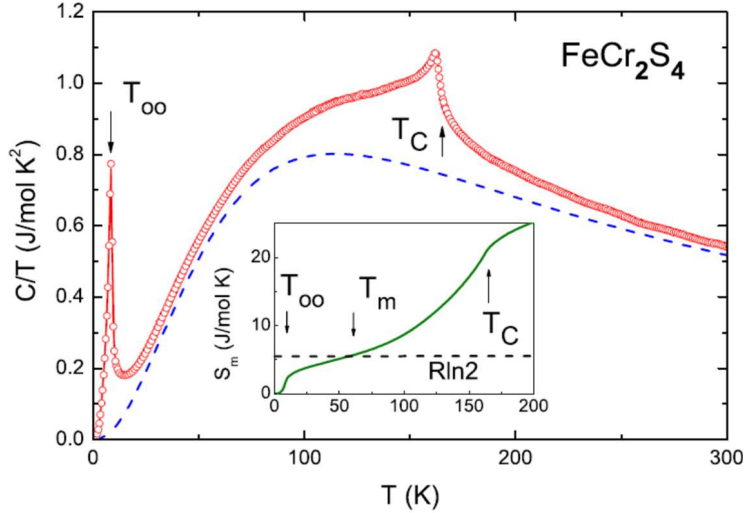

FIG. 15. Heat capacity of $FeCr_2S_4$. Temperature dependence of the heat capacity plotted as $C/T$ vs. $T$ (red open circles). Arrows indicate the characteristic temperatures $T_{OO}$, $T_m$, and $T_C$. The blue dashed line represents the calculated pure phonon contribution to the heat capacity. Inset: $T$ dependence of the magnetic entropy. The dashed horizontal line indicates the entropy $R\ln2$ characteristic of the ground-state doublet. Reprinted figure with permission from Tsurkan et al. [141]. Copyright (2010) by the American Physical Society.

The authors of Ref. [141] attempted to describe the thermodynamic anomaly as observed at the orbital ordering transition in $FeCr_2S_4$ quantitatively. The $\lambda$-like anomaly as observed at the cooperative JT transition at ~ 9 K is shown in Fig. 16 on an enlarged scale. Also shown is the temperature dependent heat capacity as determined for $Fe_{0.5}Cu_{0.5}Cr_2S_4$. In this compound, the orbital ordering transition is absent and the heat capacity at low temperatures is determined from phonon and magnon contributions only. A pure phonon-derived fit utilizing one Debye and two Einstein terms is indicated in the main frame of Fig. 16 [141]. This fit describes the experimentally observed mixed-crystal heat-capacity data reasonably well. After subtracting this lattice-derived heat capacity, the temperature dependence of the specific heat resulting from the orbital degrees of freedom and from orbital-ordering effects can be calculated. These contributions are shown in the inset of Fig. 16. Starting from low temperatures, the exponential increase of the heat capacity due to orbital ordering can be fitted utilizing a mean-field approach with a temperature-dependent mean-field gap [144].

Using this molecular-field model the temperature dependence of the opening of an orbital gap can be calculated according to $\Delta_{OO}/\Delta_0 = \tanh(\Delta_{OO} T_{OO}/\Delta_0 T)$, where $\Delta_0$ is the value of the orbital gap $\Delta_{OO}$ at $T = 0$ K. This model calculation assumes that the orbital splitting varies with temperature in the same way as the magnetization of a FM assembly of ions with spin $S = 1/2$ [144]. Using this formalism, the transition temperature is given by $T_{OO} = \Delta_0/2$ and the step-like jump at the ordering temperature in the heat capacity amounts to 1.5 R, where R is the gas constant. In the $FeCr_2S_4$, taking the orbital-ordering transition at a temperature of 9 K as given, the zero-gap value should be of order ~ 18 K. The result of a mean-field model fit is shown as red solid line in the inset of Fig. 16 and was derived assuming a zero-temperature gap $\Delta_0 \sim$ 14.8 K, in reasonable agreement with the model prediction $\Delta_0 = 2\, T_{OO}$ [144]. As documented in the inset of Fig. 16, the jump at the orbital-ordering transition is significantly smaller than the predicted value of 1.5 R. A fact that again documents that precursor phenomena due to orbital ordering are established already at significantly higher temperatures. The orbital gap as derived from these thermodynamic measurements is notably smaller than the value of the orbital gap as determined from Mössbauer spectroscopy [134], where a gap value of $\Delta_0 \sim 32$ K has been determined. To explain this difference, three things have to be stated: i) as can be seen in Fig. 12, the ground state doublet is already split, even above the orbital-ordering transition by second-order spin-orbit interactions, which compete with the spin-phonon energy, both energies obviously preferring a different ground state. At the JT transition the level splitting is dictated by the long-range spin-phonon coupling; at the phase-transition temperature the splitting is increased and reversed. In the analysis of the temperature-dependence of the specific heat, the level splitting above the onset of OO was not taken into account. ii) the heat capacity measures bulk properties averaging over possible gap anisotropies and averaging over local exchange fields, and iii) in $FeCr_2S_4$ there is significant sample dependence of the JT splitting as outlined in Ref. [141].

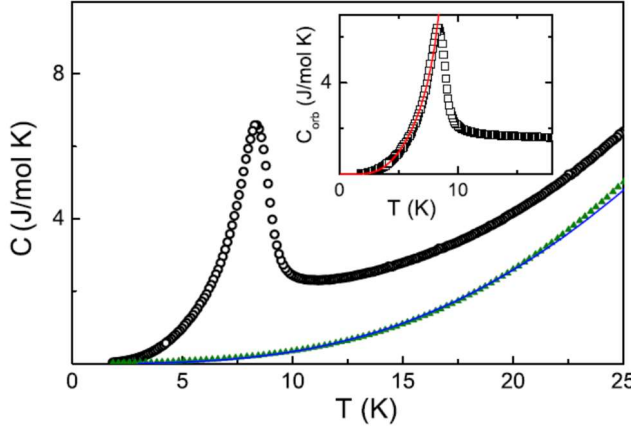

Fig. 16. Heat capacity at the onset of OO in $FeCr_2S_4$.
Heat capacity anomaly at low temperatures (open circles). The green triangles show the heat capacity of $Fe_{0.5}Cu_{0.5}Cr_2S_4$ with no JT transition. Blue solid line indicates a fit describing the pure lattice contribution. The inset shows the heat capacity with the phonon contribution subtracted solely due to orbital degrees of freedom. The red solid line is a mean-field fit of the low-temperature heat capacity. Reprinted figure with permission from Tsurkan et al. [141]. Copyright 2010 by the American Physical Society.

Symmetry-lowering phase transitions can easily be identified by infrared (IR) and Raman spectroscopy following the temperature evolution of phonon eigenfrequencies crossing structural, orbital, or magnetic phase transitions. The IR active phonon modes of $FeCr_2S_4$ have been investigated by a number of groups [145,146,147] however with no special focus on the low-temperature orbital ordering transition or on the magnetic transition at the critical temperature $T_m$. Raman active modes were reported by Choi et al. [148], only reporting anomalies in phonon frequencies and excitation linewidths around the FM ordering temperature $T_C$. In a recent far-infrared spectroscopy study, the detailed temperature dependence of all phonon eigenfrequencies was studied in detail, when crossing all critical phase transitions [149]. The temperature dependence of the phonon eigenfrequencies as reported in this work are documented in Fig. 17. The cubic high-temperature phase of $FeCr_2S_4$ is characterized by four IR-active phonon modes. These four modes exhibit a moderate and continuous temperature evolution, where the modes 3 and 4 soften, while the eigenfrequencies of modes 1 and 2 rather moderately increase on cooling. This behaviour signals the importance of spin and orbital degrees of freedom coupled to the lattice. Assuming a crystalline lattice without any further internal degrees of freedom, on decreasing temperatures the evolution of all mode-eigenfrequencies would exhibit a moderate softening due to anharmonic effects. In addition, the lowest-lying $T_{1u}(1)$ mode broadens and splits on passing the characteristic temperature $T_m$ ~ 60 K. This is further experimental evidence for symmetry lowering, associated with the appearance of an incommensurate (IC) helical magnetic order, as has been identified by muon-spin relaxation experiments [135] and as documented in Fig. 14 via magnetic-susceptibility experiments [141]. Two new modes, $R_1$ and $R_2$, appear below the onset of OO (see Fig. 17). These new modes are Raman active and have been reported in Raman experiments [148]. They are interpreted as signatures of a symmetry-lowering transition with the loss of inversion symmetry reflecting the MF and orbitally-ordered ground state (for more details see the chapter on multiferroics).

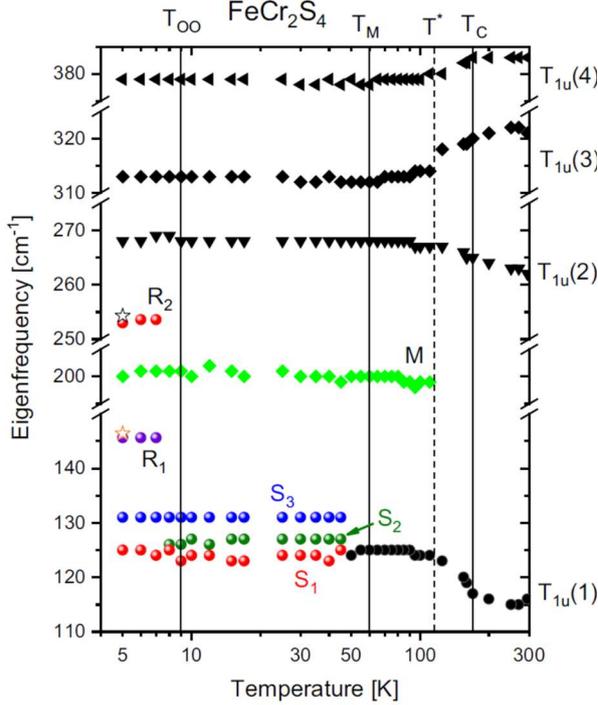

Fig. 17. Phonon eigenfrequencies in $FeCr_2S_4$. Temperature dependence of the frequencies of IR active modes (closed symbols) together with the eigenfrequencies of two Raman modes at 5 K (open stars) taken from Ref. [148]. Note that the vertical axis is broken at different wavenumbers. The four IR-active modes of $T_{1u}$ symmetry of the cubic high-temperature phase are indicated on the right axis. R modes appear in the low-temperature orbitally ordered phase, S modes below the spin-reorientation transition. A mode labelled M appears below a characteristic temperature $T^*$.
Reprinted figure with permission from Deisenhofer et al. [149]. Copyright (2019) by the American Physical Society.

Astonishingly, there appears a new mode M below the characteristic temperature $T^*$ ~ 120 K well within the ferrimagnetically ordered regime. In this temperature regime, no evidence for a structural change has been reported. However, several experimental studies noticed significant changes in this temperature regime: The magnetocrystalline anisotropy seems to change below this temperature as evidenced by ac susceptibility [143], electron spin resonance [142], as well as by magnetization and time-resolved magneto-optical Kerr spectroscopy studies [150]. Moreover, the evaluation of the Fe-Mössbauer spectra by Kalvius et al. [135] pointed towards a finite EFG in the magnetically ordered phase, which starts to increase monotonously on cooling below 120 K. The latter authors speculate about a minor lattice distortion occurring below this temperature, which does not alter the collinear spin arrangement, but influences the electric quadrupolar field at the iron nuclei. Such a scenario is consistent with an interpretation of the symmetry of mode M, which directly couples to the electronic $e$ orbitals of the Fe ions. However, another scenario should be taken into account, which will be later discussed in full detail: A number of chromium spinels reveal anomalies approximately in this temperature regime, which were thought to arise from the relaxation dynamics of off-centre chromium ions constituting a local static dipole moment. One is tempted to speculate that $FeCr_2S_4$ also belongs to this class of spinel materials with a locally broken inversion symmetry and with a tendency towards polar order.

The extreme sensitivity of the structural, electronic and polar ground-state properties of thiospinels on synthesis route and doping was investigated in detail in Refs. [133,151]. Specifically, the significant dependence of the orbital-ordering transition in $FeCr_2S_4$ on the synthesis route and on chlorine substitution is documented in Ref. [133]. In single crystals grown by chemical transport using chlorine gas as transport agent, the orbital ordering transition is suppressed. In these crystals the $\lambda$-like anomaly due to orbital ordering as observed in the temperature dependence of the specific heat becomes suppressed leaving a hump-like anomaly close to 5 K, which was interpreted as an orbital glass transition [41,152]. It was argued that at this orbital-glass transition, the orientations of the electronic orbitals freeze in into random orientations without establishing long-range OO as resulting from a cooperative JT transition. To our knowledge, $FeCr_2S_4$ is the only material where such an orbital-glass transition was identified. The dynamics and the slowing down of the orbital reorientations was studied in detail by dielectric spectroscopy by Fichtl et al. [41]. It was argued that the electron density coupled to local lattice distortions induces dipolar moments and allows following the orbital reorientational dynamics by dielectric spectroscopy. A schematic view of the orbital orientations with their local structural distortions for the different scenarios, namely a high-temperature orbital liquid (OL), OO and orbital glass is shown in Fig. 18 [153]. The OL is characterized by an orbitally degenerate ground state or by a small level splitting of the ground state, resulting in an equal thermal population of the involved energy levels. Here, on time average, the orbitals point in all symmetry-allowed crystallographic directions establishing over-all cubic symmetry. A cooperative JT transition induces long-range OO with accompanying long-range structural distortions. Due to disorder and frustration effects, long-range order can be suppressed and the electronic orbitals coupled to their local structural distortions, essentially freeze into random orientations yielding a low-temperature glass-like state. In contrast to a canonical glass transition, however, here quantum-mechanical tunnelling prevents complete freezing-in and at the lowest temperatures, orbital motion still is possible, albeit with typical time scales of the order of 0.1 s. This is exceptionally slow for electronic degrees of freedom, which reflects the involved coupled lattice distortions. In the orbital glass, the overall cubic lattice symmetry is not broken. However, locally strong deviations occur, which are practically frozen on quasi-static time scales. This orbital glass phase seems to be closely related to the freezing observed in orientational glasses and so-called plastic crystals, where molecular orientations freeze into random directions devoid of long-range orientational order [154,155].

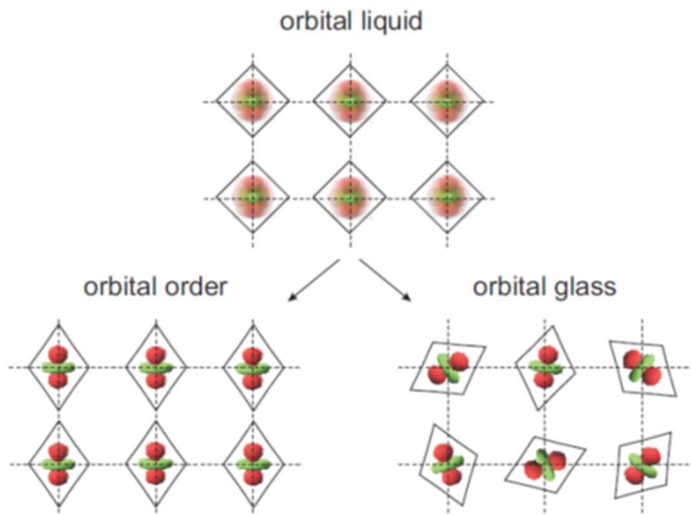

Fig. 18. Orbital degrees of freedom in solids.
Schematic representations of orbital orientations coupled to local lattice strains: the high-temperature orbital liquid characterizes a state with a degenerate orbital ground state. The arrows indicate possible transitions from the orbital liquid into an orbitally ordered phase via a cooperative JT transition or into an orbital glass, with locally varying orbital orientations coupled to local strain fields. Here the orbital orientations vary from site to site and are frozen on macroscopic time scales. Figure taken from Loidl and Lunkenheimer [153].

## 3.3. Magnetism governed by frustration and competing exchange interactions

3.3.1. Frustration in *B*-site spinels: Magnetic ions at the pyrochlore lattice

A. Exotic ground-state properties

The *B* sites in the spinel lattice constitute a pyrochlore lattice, which belongs to the strongest-frustrated lattices in three dimensions. More than 60 years ago, AFM ordering on this lattice was considered by Anderson, predicting a very high ground-state degeneracy and the absence of LRO at any temperature for Ising spins [23]. Villain reached the same conclusion for Heisenberg spins located at the pyrochlore lattice of spinel compounds, calling this type of frustrated ground-state a cooperative paramagnet [156]. Later on a number of authors has further elucidated the problem of Heisenberg spins on the pyrochlore lattice, all sharing the common conclusion that antiferromagnetically coupled Heisenberg spins on the vertices of corner-sharing tetrahedra cannot undergo long-range spin order [157,158,159]. It turned out that in the first decade of the 21$^{st}$ century, chromium spinels became the archetypal model systems to prove these theoretical predictions. There exists a vast literature documenting magnetic frustration, spin-lattice coupling and resulting exotic ground states in this class of materials.

It is known since more than 50 years that the chromium spinels $ACr_2X_4$, with $A$ = Zn, Mg, Cd, and Hg, and $X$ = O, S and Se, span an enormous range of magnetic exchange interactions, with CW temperatures ranging from – 600 K to + 200 K [105]. A summary of crystallographic data, PM moments $p_{eff}$, CW temperatures $\Theta_{CW}$, and magnetic ordering temperatures $T_m$, as reproduced from Ref. [38] is provided in Table 1. From this table it becomes immediately clear that when considering all chromium spinels, ranging from zinc to mercury at the *A* site and from oxide to selenide at the *X* site, the lattice constants strongly depend on the constituents of the lattice, increasing from oxygen to selenium and from zinc to mercury. Concomitantly the magnetic properties change considerably: the CW temperatures range from –398 K in $ZnCr_2O_4$ to + 200 K in the mercury selenide, indicating drastic changes of the magnetic exchange interactions. $CdCr_2S_4$, $CdCr_2Se_4$ and $HgCr_2Se_4$ exhibit true FM ground states and belong to the rare classes of FM insulators or FM semiconductors, while all the other compounds reveal complex AFM spin order, with ordering temperatures ranging from 10 – 20 K, irrespective of the CW temperature. Hence, also the frustration parameter changes considerably, ranging from values close to $f \sim 30$ in some oxides with the smallest lattice constants, to values of order $f \sim 1$ in the selenides with FM spin structure. At this point it has to be mentioned that the frustration parameter $f$ signals geometrical frustration effects of magnetic moments with AFM interactions. In the case of $ZnCr_2S_4$, the value of $f$ = 0.5 does not signal the complete absence of frustration effects, but rather competing FM and AFM exchange interactions of considerable but comparable strength. These competing magnetic exchange interactions result in an almost zero CW temperature, but establish a complex low-temperature spin structure, which will be elaborated in the course of this review.

The PM moments as documented in Table 1 are always close to the spin-only value of a $S$ = 3/2 magnet ($p_{eff}$ = 3.87 $\mu_B$). $Cr^{3+}$ with an electronic $d^3$ configuration is located in an octahedral environment of oxygens with a lower $t_{2g}$ triplet (see Fig. 6), which consequently is half filled. Hence a spin-only value of the magnetic moment is expected with very little spin-lattice coupling. As will be seen later, experimentally this is not at all observed. The different CW temperatures and the resulting magnetic ground states as a function of lattice constants are driven by the dominating exchange interactions: at small Cr–Cr separation, strong direct AFM exchange dominates. With increasing separation, the 90° FM Cr–$X$–Cr exchange becomes more and more important. Probably, for all lattice spacings a complex Cr–$X$–$A$–$X$–Cr SE is active. This AFM SE is weak and only of the order of 1 K, but gains importance via its high multiplicity [105]. Detailed calculations of the electronic band structure and the resulting exchange-coupling constants of this class of $A$Cr$_2X_4$ compounds have been performed by Yaresko [160]. In this work, the exchange-coupling constants have been determined up to the fourth-nearest neighbour. Using an LSDA + U approach, in these calculations the CW temperatures span the range from – 500 K for AFM ZnCr$_2$O$_4$ up to + 422 K for FM HgCr$_2$Se$_4$ utilizing a Coulomb repulsion $U$ = 2 eV. In addition, it is shown that the magnetic exchange significantly depends on the value of the Coulomb repulsion, the Hubbard $U$. Experimentally the appropriate CW temperatures range from –398 K to +184 K (see Table 1).

| Compound | $a$ (Å) | $x$ | $p_{eff}$ ($\mu_B$) | $\Theta_{CW}$ (K) | $T_m$ (K) | $f$ | spin structure |
|---|---|---|---|---|---|---|---|
| ZnCr$_2$O$_4$ | 8.317 | 0.265 | 3.85 | -398 | 12.5 | 32 | AFM |
| MgCr$_2$O$_4$ | 8.319 | 0.261 | 3.71 | -346 | 12.7 | 27 | AFM |
| CdCr$_2$O$_4$ | 8.596 | 0.265 | 4.03 | -71 | 8.2 | 8.7 | AFM |
| HgCr$_2$O$_4$ | 8.658 | 0.229 | 3.72 | -32 | 5.8 | 5.5 | spiral |
| ZnCr$_2$S$_4$ | 9.983 | 0.258 | 3.86 | 7.9 | 15/8 | 0.5 | AFM1/AFM2 |
| CdCr$_2$S$_4$ | 10.247 | 0.263 | 3.88 | 155 | 84.5 | 1.8 | FM |
| HgCr$_2$S$_4$ | 10.256 | 0.267 | 3.90 | 140 | 22 | 6.4 | spiral |
| ZnCr$_2$Se$_4$ | 10.498 | 0.260 | 4.04 | 90 | 21 | 4.3 | spiral |
| CdCr$_2$Se$_4$ | 10.740 | 0.264 | 3.82 | 184 | 130 | 1.4 | FM |
| HgCr$_2$Se$_4$ | 10.737 | 0.264 | 3.89 | 200 | 106 | 1.9 | FM |

Table 1: Structural and magnetic parameters of various chromium spinels.
Lattice constant $a$, fractional coordinate $x$, effective PM moment $p_{eff}$, CW temperature $\Theta_{CW}$, magnetic ordering temperature $T_m$, frustration parameter $f = |\Theta_{CW}|/T_m$, and spin structure. Values are taken from Rudolf et al. [38].

Figure 19 shows a schematic diagram with the characteristic ordering temperatures plotted vs. the CW temperatures. It is obvious that in this series of spinel compounds the CW temperature scales with the lattice constant (see Table 1), or to be more precise with the chromium-chromium separation. However, in the chosen type of presentation the co-action of frustration effects and competing interactions are clearly visible. ZnCr$_2$O$_4$, with the smallest lattice spacing, is dominated by strong and direct AFM exchange resulting in a large and negative CW temperature. In the selenides characterized by much larger lattice spacings, the 90° FM exchange is the dominant interaction resulting in positive CW temperatures. In the intermediate range, FM and AFM exchange compete and partly compensate, yielding characteristic temperatures close to zero. Two straight lines given by the positive, respectively negative CW temperatures define the tentative ordering temperatures of non-frustrated magnets and represent the ideal FM or AFM exchange. These idealized ordering temperatures are indicated by a yellow triangle starting at zero CW temperature.

In the AFM regime, corresponding to large and negative CW temperatures, strong magnetic frustration is active, with ZnCr$_2$O$_4$ and MgCr$_2$O$_4$ revealing the strongest effects with $f \sim 30$. These compounds seem to be prototypical examples of strongly frustrated pyrochlore lattices where no spin configuration simultaneously can satisfy all six AFM interactions among the spins located at vertices of the tetrahedral building blocks. It is theoretically documented that this so-called pyrochlore antiferromagnet has no phase transition for any spin $S$ and forms a quantum spin liquid with low-lying singlet excitations for $S = ½$ [157,158,159]. However, small perturbations away from the ideal model or further non-leading interactions can induce low-temperature phase transitions. $ZnCr_2O_4$ undergoes a first-order cubic to tetragonal phase transition concomitant with the onset of AFM order at 12.5 K [161]. The full spin entropy of this transition, $2R\ln4$, is reached well above room temperature only, consistent with strong magnetic frustration and a high CW temperature [161]. This magneto-structural transition has been interpreted as being of spin-Peierls type [162]. That the symmetry of the AFM phase may be even lower than tetragonal was concluded from a detailed AFM resonance study [163]. Several resonance modes corresponding to different structural domains were observed, but the number of domains could be reduced by field cooling the sample through the transition point. To describe the AFM resonance spectra, it was necessary to assume an orthorhombic lattice distortion in addition to the above mentioned tetragonal distortion.

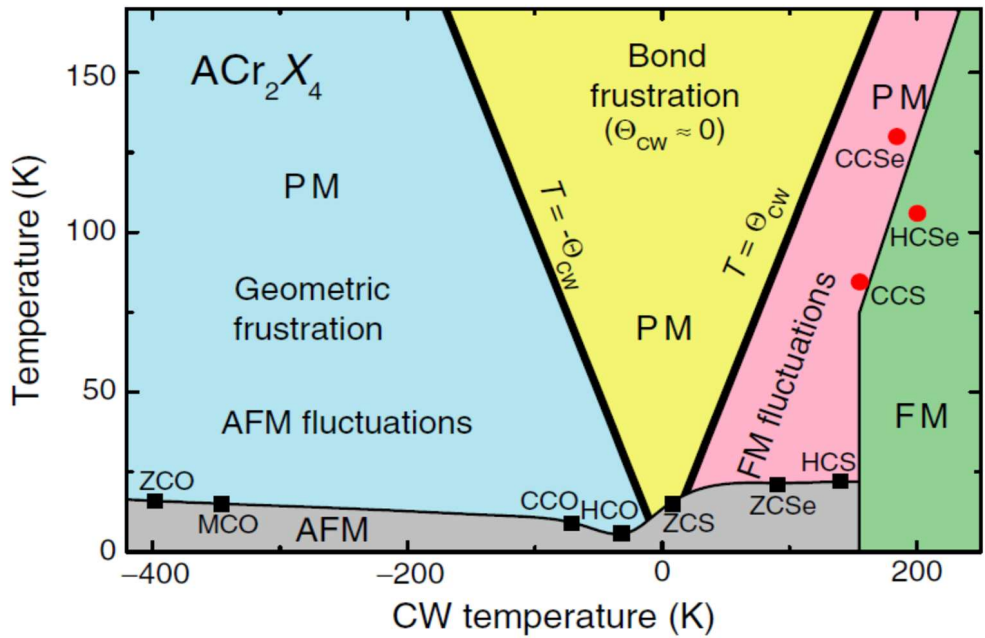

Fig. 19. Schematic magnetic phase diagram of $ACr_2X_4$ compounds: $A$ = Zn (Z), Mg (M), Cd (C), and Hg (H); $X$ = O, S, and Se. Characteristic magnetic ordering temperatures are plotted versus the CW temperature (see Table 2): FM (red circles) and AFM (black squares) ordering temperatures are indicated. Hypothetical magnetic ordering temperatures expected to follow the CW temperatures, $T = \pm \Theta_{CW}$, are indicated by thick solid lines. The CW temperatures scale with the lattice constants or – more precisely - with the separation of magnetic ions. Thin solid lines separate magnetically ordered from PM phases and are drawn to guide the eye. Reprinted figure with permission from Rudolf et al. [38]. Copyright (2007) by the Institute of Physics (IOP).

In addition, in the ordered phase a localized spin resonance located at an energy of 4.5 meV develops abruptly out of the PM spin-fluctuation spectrum. The most detailed low-temperature model is based on synchrotron x-ray single-crystal diffraction [164]. This model comprises a complex low-temperature structure with three types of tetrahedra with different bond distortions. The authors of Ref. [164] concluded that the match between the local structural distortions with different bond lengths and the moment arrangement within the tetrahedra is incomplete. Subsequently, Lee et al. [27] utilizing neutron-scattering experiments measured the form factor of the lowest-energy excitations in $ZnCr_2O_4$ and recognized that it results from AFM hexagonal spin loops. The measurements were performed in the PM regime just above the ordering temperature and the hexagonal spin loops were interpreted as local zero-energy modes of the pyrochlore Heisenberg antiferromagnet. Similar quasi-elastic and inelastic neutron scattering techniques experiments, documenting that the magnetic ground state in these strongly frustrated chromium oxides is not a simple antiferromagnet, but that the spins form complex spin molecules or spin clusters with accompanying molecular-like quasiparticle excitations, were conducted by Tomiyasu et al. [165,166] in $MgCr_2O_4$. Astonishingly, these authors identified a set of molecular spin excitation modes in the PM and in the magnetically ordered phase, an observation, which was explained using a concept of frustrated excitations [166]. A detailed report on the low-temperature magneto-structural transitions in this compound and the resulting complex crystallographic and spin order has also been given by Gao et al [167]. By neutron diffraction in the magnetically ordered phase these authors confirmed two main magnetic propagation vectors $\mathbf{k_1}$ = (½ ½ 0) and $\mathbf{k_2}$ = (1 0 ½). The corresponding reflections respond differently to external magnetic fields implying the formation of multiple domains. It is worth mentioning that a quantum mean-field model for the pyrochlore lattice has been developed in Ref. [168]. In this model, not single spins but the assumption of a set of interacting corner-sharing tetrahedra was utilized to explain the temperature dependence of the susceptibility in $ZnCr_2O_4$.

Now, returning to Fig. 19, with decreasing CW temperatures, AFM and FM exchange interactions compete, are almost of equal strength and again establish complex magnetic ground states. The oxides $CdCr_2O_4$ and $HgCr_2O_4$, as well the sulphide $ZnCr_2S_4$ are very close to this limit, where the two dominating exchange interactions almost completely compensate (see Fig. 19). $CdCr_2O_4$ also undergoes a spin-Peierls-like phase transition close to ~ 8 K [161,169,170]. Despite the fact that the cadmium and zinc chromium oxide are closely related compounds governed by similar frustration effects, in contrast to the distortion observed in $ZnCr_2O_4$, at the magneto-structural transition $CdCr_2O_4$ undergoes elongation along the $c$ axis and exhibits an IC Néel order at the lowest temperatures [169]. The difference in the structural distortions and the resulting difference in the spin structure could result from the fact the zinc compound is governed by geometrical frustration and is a prototypical example of a frustrated Heisenberg AFM, while in the cadmium compound frustration results from competing interactions with increasing importance of FM exchange. It has been shown that the lattice distortion of the cadmium compound stabilizes coplanar magnetic order and that the lack of inversion symmetry makes the crystal structure chiral [170]. Later on, model calculations and comparison to experiments revealed that a soft optical phonon triggers the magnetic transition and endows the lattice chirality [171]. Interestingly, the spin relaxation in the chromium oxide spinels $ACr_2O_4$ ($A$ = Mg, Zn, Cd) was investigated in the PM regime by ESR [172]. The temperature dependence of the ESR linewidth indicates an unconventional spin-relaxation behaviour, similar to spin-spin relaxation in the two-dimensional (2D) triangular lattice antiferromagnets. The authors of Ref. [172] were able to describe these ESR results in terms of a generalized Berezinskii-Kosterlitz-Thouless (BKT) type scenario for 2D systems with additional internal symmetries. Based on the characteristic exponents obtained from the evaluation of the ESR linewidth, short-range order with a hidden internal symmetry was suggested.

On cooling without external field, $HgCr_2O_4$ undergoes an orthorhombic distortion at the Néel ordering temperature $T_N$ = 5.8 K [173]. Its frustration parameter $f$ = 5.5 (Table 1) is not very large, but in this compound competing FM and AFM exchange interactions certainly play an important role. Like the other chromium oxide spinels, it shows a magnetic ordering with concomitant structural distortions. However, the symmetry of the lattice is orthorhombic as opposed to tetragonal of the other oxide compounds, and the volume change at the transition is substantially larger [173]. Elastic magnetic-neutron scattering data obtained in zero magnetic field documented that in the low-temperature phase the spins order long range with two characteristic magnetic-ordering wave vectors (1/2,0,1) and (1,0,0) [174]. All the chromium oxides gained considerable attention because of the observation of a sequence of fractionalized magnetization plateaus in high-field magnetization experiments, which will be discussed in the next chapter. An informative and detailed review on these novel states of matter induced by frustration effects in this class of chromium oxide spinels was given by Lee et al. [175].

Moving to sulphides and selenides, the CW temperatures become positive (see Fig. 19 and Table 1) indicating the increasing importance of 90° FM exchange. However, $ZnCr_2S_4$ [176,177,178], $ZnCr_2Se_4$ [179,180,181,182], and $HgCr_2S_4$ [115,183] still exhibit AFM ground states. The CW temperatures for the zinc selenide and the mercury sulphide are large and positive and document the predominant FM exchange, the CW temperature of the zinc sulphide is close to zero signalling that FM and AFM exchange interactions almost compensate each other. Large magnetostriction effects and negative thermal expansion were reported for $ZnCr_2Se_4$ by Hemberger et al. [182] and interpreted as fingerprints of strong magnetic frustration. The crystal and magnetic structures of the spinel compounds $ZnCr_2S_4$ and $ZnCr_2Se_4$ were reinvestigated by high-resolution synchrotron and neutron-powder diffraction studies [184] as well as by small-angle neutron scattering [185]. $ZnCr_2S_4$ shows two AFM transitions at 15 K and 8 K that are accompanied by structural phase transitions. The crystal structure transforms from the cubic spinel structure in the PM state, via a tetragonal-distorted intermediate phase, into a low-temperature orthorhombic phase. In the intermediate phase, the magnetic structure of $ZnCr_2S_4$ reveals FM layers with a spin helix formed by subsequent layers. At low temperatures, the AFM exchange becomes dominant and the spin helix transforms into a commensurate AFM structure reminiscent of $ZnCr_2O_4$ with almost identical propagation vectors.

$ZnCr_2Se_4$ exhibits a first-order phase transition at 21 K into an IC helical spin structure propagating along <100> [185]. This IC spin helix, characterized by a proper screw structure with a propagation vector (0 0 0.44), reminds on the spin-spiral structure of the sulphide in the intermediate phase. The magnetic excitation spectrum of the helimagnetic $ZnCr_2Se_4$ was investigated by neutron spectroscopy and analysed in detail by theoretical model calculations [186]. Low-energy magnetic excitations were observed in the single-domain proper-screw spiral phase and were interpreted as soft helimagnon modes with a small energy gap of ~ 0.17 meV, which emerge from two orthogonal wave vectors where no magnetic Bragg peaks are present. These Goldstone modes in $ZnCr_2Se_4$ were related to the occurrence of a magnetic-field driven quantum critical point (QCP), which was proposed and identified by Gu et al. [187]. The evolution of specific magnon modes across the QCP, which separates the spin-spiral from the field-induced FM phase, were recently investigated by a combination of neutron scattering, ultra sound velocity and dilatometry experiments by Inosov et al. [188]. These authors found that the magnon modes completely soften at the QCP restoring over-all cubic symmetry, and linearly increase with increasing magnetic field when entering the field-polarized phase. In addition, based on NMR experiments by Park et al. [189], a partial transfer of chromium moments to the neighbouring Se sites was proposed, a fact that could explain the finite spin-lattice interactions.

Neutron-diffraction studies of the normal spinel $HgCr_2S_4$ [183] show that the magnetic structure is a simple spiral, similar to the observations in $ZnCr_2Se_4$ [179]. The propagation vector is along the symmetry axis of the spiral and directed along a particular cube edge of a given domain. The ordered moment of $Cr^{3+}$ was found to be 2.73 $\mu_B$, in agreement with bulk magnetization measurements. Further magnetization, ESR, and specific-heat studies documented strong FM fluctuations close to 50 K and the occurrence of complex AFM order below $T_N$ ~ 22 K [115]. The highly unconventional magnetic behaviour resembles properties of a helical antiferromagnet and of a soft ferromagnet, dependent on temperature and magnetic field. Even weak external magnetic fields disturb the AFM order and strongly enhance the FM correlations. It is an interesting phenomenon, that similar to the true ferromagnets $CdCr_2S_4$ and $CdCr_2Se_4$, multiferroicity and concomitant strong magnetoelectric (ME) effects were identified in the low-temperature magnetic phases of $ZnCr_2Se_4$ [190] and $HgCr_2S_4$ [191].

Finally, with increasing FM exchange the ground states become truly FM with relatively high Curie temperatures (see Table 1 and Fig. 19) and minor frustration effects only. In $CdCr_2S_4$, $CdCr_2Se_4$ and $HgCr_2Se_4$, due to the large separation of the chromium ions, the 90° FM Cr–X–Cr exchange dominates [105,106]. Interestingly, all chromium spinels with a FM ground state or with dominant FM fluctuations in a broad temperature range reveal strong ME effects, which will be discussed later in the chapter 3.4 on multiferroics.

In this respect it is important to mention that the critical behaviour of $CdCr_2S_4$ close to the Curie temperature follows that expected for an ideal 3D Heisenberg system, however, in an unusual narrow window of temperature and in low magnetic fields only [192]. Moreover, in the FM phase of $CdCr_2S_4$ detailed investigations of the ferromagnetic resonance (FMR) revealed a pronounced anomalous cubic anisotropy not expected for the spin $S=3/2$ ground-state quartet of $Cr^{3+}$, which should remain degenerate in purely cubic symmetry. In Ref. [193] any extrinsic effect, which would explain the magneto-crystalline anisotropy could be ruled out. Indeed, a model with local trigonal distortions along the four <111> cube-space diagonals gave a satisfactory description of the data. Moreover, the superposition of four uniaxial anisotropies coalescing into one ESR line by exchange narrowing helped to understand the linewidth anomalies. From a theoretical point of view, the trigonal distortion of the $CrS_6$ octahedra (along the four <111> directions) partially lifts the degeneracy of the spin quartet into two doublets and thus gives rise to a locally uniaxial anisotropy.

B. Spin-phonon coupling and the spin Jahn-Teller effect

As outlined above, the chromium spinels are expected to be Mott insulator with a half-filled $t_{2g}$ shell, with $S = 3/2$, an almost spherical charge distribution and negligible spin-orbit coupling (SOC). In fact, one can expect that in these compounds orbital effects are quenched and they behave like spin-only systems with weak coupling to the lattice only. However, as documented in the preceding chapter, rather unexpectedly the magnetic transitions in all compounds are strongly coupled to the lattice and the complex spin-ground states can only be explained taking considerable spin-lattice coupling into account. This unusual spin-lattice coupling in chromium spinels has early been recognized by Kino and Lüthi [194] analysing ultrasonic experiments in $ZnCr_2O_4$. These authors observed a strong softening of the shear elastic constant ($c_{11}-c_{12}$), which at that time remained unexplained. Later on this strong spin-phonon coupling in chromium spinels has been treated on the basis of a spin-driven JT effect [36,37,195], which subsequently has been called magnetic JT or spin JT effect. These theories proposed an elegant mechanism for lifting the frustration of spins on a pyrochlore lattice through a coupling between spin and lattice degrees of freedom. The high symmetry of the pyrochlore lattice and the spin degeneracy drive a distortion of the corner-sharing tetrahedra via a magnetic-type JT effect. The resulting low-temperature state exhibits a transition from cubic to tetragonal symmetry and the development of bond order in the spin system with unequal spin correlations <$S_i · S_j$> on different bonds of the tetrahedra. In the ordered phase, there are 4 (2) strong and 2 (4) weak bonds per tetrahedron. It was proposed that this modelling can explain the unusual magneto-structural distortions in a number of AFM chromite and vanadium spinels.

Yamashita and Ueda [36] proposed that this spin-JT effect is responsible for the phase transitions observed in the vanadium spinels $ZnV_2O_4$ and $MgV_2O_4$. Both compounds are normal spinels with the vanadium ions with spin $S = 1$ localized at the B sites, at the pyrochlore lattice. At low temperatures, both compounds reveal strongly coupled structural and magnetic transitions [196,197,198]. To prove these theoretical predictions, Reehuis et al. [199] reported neutron-powder diffraction experiments to determine the crystallographic and magnetic structure of $ZnV_2O_4$. The main results of this study are reproduced in Fig. 20. The temperature dependence of FC and ZFC magnetic susceptibilities is documented in the left panel. Two anomalies at $T_s$ = 51 K and $T_N$ = 40 K indicate a sequence of structural and magnetic phase transitions. Splitting of FC and ZFC susceptibilities starts already at slightly higher temperatures and probably results from magnetic fluctuations driving the spin-JT effect [199]. The absence of any low-temperature Curie tails indicates a high purity of the polycrystalline sample. The crystallographic and magnetic structures were determined by Rietveld refinement of the diffraction profiles at 60 and 1.8 K. At 60 K the cubic spinel structure is characterized by space group $Fd\bar{3}m$. The low-temperature tetragonal phase is described by space group $I4_1/amd$. At 1.8 K the tetragonal unit cell is just one half of the cubic one with cell dimensions $a/\sqrt{2} \times a/\sqrt{2} \times a$. The magnetic structure can be described by AFM spin order along *b* accompanied by a slight buckling of the surrounding oxygen octahedral (see middle frame of Fig. 20). The spin structure within a single tetrahedron is shown in the left frame.

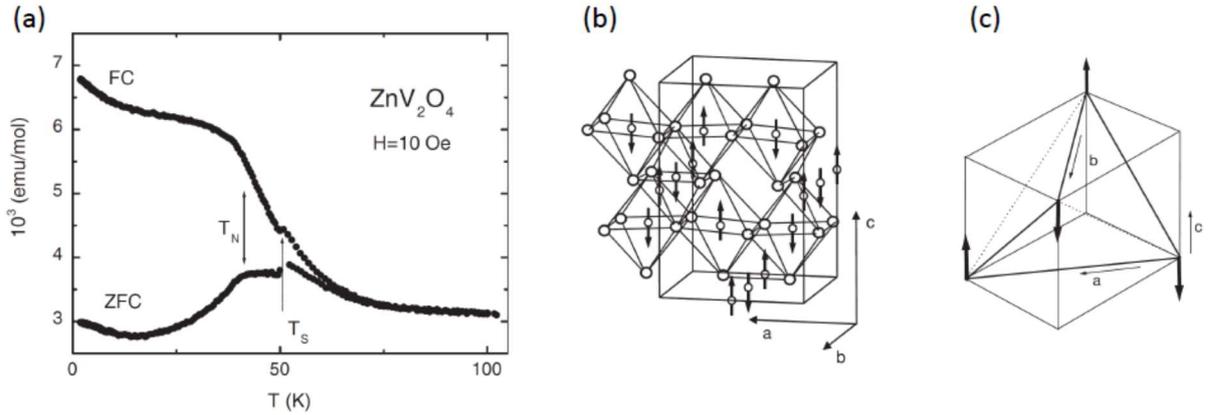

Fig. 20. Magnetic susceptibility, crystallographic and spin structures of $ZnV_2O_4$.
(a) Field cooled (FC) and zero-field cooled (ZFC) magnetic susceptibilities as measured in an external magnetic field of 10 Oe. The structural and magnetic phase transitions at $T_S$ = 51 K and $T_N$ = 40 K, respectively are indicated by arrows. (b) Magnetic structure as determined by Rietveld refinement. The magnetic moments of the vanadium ions exhibit antiferromagnetic order along *b*. The figure also evidences the buckling of the $VO_6$ octahedra. (c) Spin structure within a single tetrahedron of the corner-sharing tetrahedral network. The labels of the axes refer to the tetragonally distorted low-temperature structure. The full and dotted lines of the tetrahedron refer to strong and weak bonds of neighbouring spins, respectively. Reprinted figures with permission from Reehuis et al. [199]. With kind permission of The European Physical Journal (EPJ).

Later on, a controversy arose about the role of the vanadium orbitals in these vanadium-oxide spinels. Tsunetsugo and Motome [200] proposed that the structural transition in $ZnV_2O_4$ observed at 50 K should be attributed to the onset of OO introducing spatial modulations of spin-exchange couplings, which partially releases frustration and leads to subsequent spin order at 40 K. The scenario was corroborated by inelastic neutron-scattering experiments showing that the spin-excitation spectra adopt a one-dimensional characteristics [201]. This ongoing controversy was discussed in detail by Radaelli [202] as well as by Maitra and Valenti [203].

In a series of subsequent optical experiments, it was proven that the most direct evidence concerning spin-driven JT transitions can be obtained by analysing the temperature dependence and possible splittings of optical phonons. On a number of chromium spinels it was demonstrated that lattice vibrations can provide qualitative and even quantitative information about spin correlations by measuring the eigenfrequencies of IR-active phonons [38,161,204,205,206]. When crossing the magnetoelastic phase transitions as function of temperature or external magnetic field, a large splitting of phonon modes involving magnetic ions has been observed. From the magnitude of the splitting it was even possible to infer the absolute value of the nearest-neighbour spin correlations [204] and a universal linear dependence of the exchange-driven phonon splitting on the magnetic exchange was reported for a number of AFM transition-metal monoxides and for the strongly frustrated AFM oxide spinels $CdCr_2O_4$, $MgCr_2O_4$, and $ZnCr_2O_4$ [207].

Some archetypal examples of the splitting of phonon modes at the magneto-structural transitions in a number of spinel compounds are documented in Fig. 21 [38]. This phonon splitting is driven by the spin JT effect, which is active for antiferromagnetically coupled spins located at the vortices of a pyrochlore lattice. At room temperature, the four IR-active modes of the cubic spinel lattice are observed. Please note that the eigenfrequencies strongly depend on the anion: The mode frequencies are stiff for the oxide and rather soft for the selenide. The continuous increase on cooling results from conventional anharmonic interactions of the phonons [38]. For $CdCr_2O_4$, $ZnCr_2S_4$ and $ZnCr_2Se_4$ a clear splitting of some of the phonons can be observed below the magneto-structural phase transitions, which certainly depends on symmetry and on the degree how much the chromium ion carrying the localized spin is involved in the specific displacement pattern of a given excitation. In most cases, the phonon splitting when observed is of order 10 cm$^{-1}$ corresponding to energies of ~ 1 meV. Sushkov et al. [204] argued that the significant phonon splitting of some modes in $ZnCr_2O_4$ is caused through modulation of the direct Cr-Cr exchange.

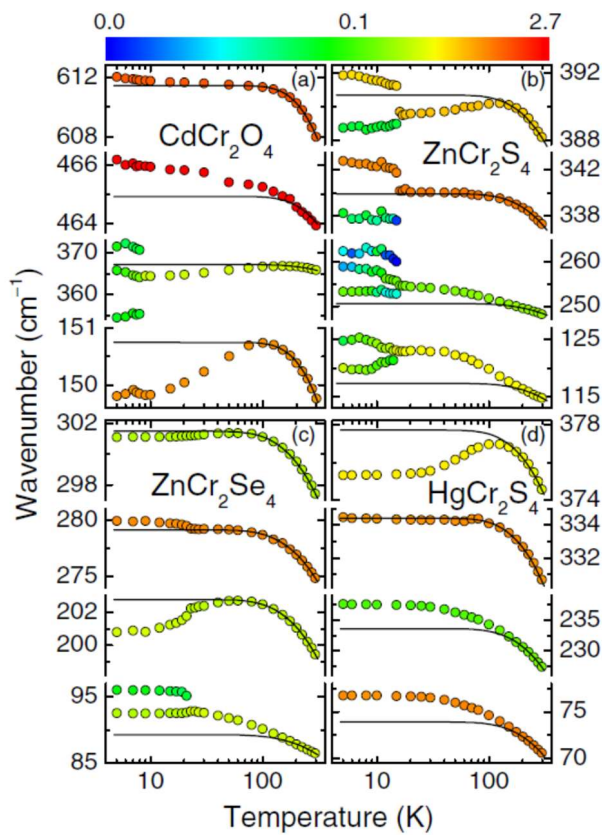

Fig. 21. Phonons in spinel compounds. Temperature dependence of the phonon eigenfrequencies on a semi-logarithmic scale for (a) $CdCr_2O_4$, (b) $ZnCr_2S_4$, (c) $ZnCr_2Se_4$, and (d) $HgCr_2S_4$. For (a), (b) and (c) some of the phonon modes clearly split when crossing the magneto-structural transition. No splitting is observed for $HgCr_2S_4$, which is dominated by FM fluctuations (see text). The mode strength is indicated by a colour code extending from 0.001 (dark blue) to 2.7 (bright red) as indicated in a colour bar on top of the figure. The solid lines schematically illustrate purely anharmonic behaviour of the phonon eigenfrequencies.
Reprinted figure with permission from Rudolf et al. [38]. Copyright (2007) by the Institute of Physics (IOP).

Theoretically, Fennie and Rabe [208], studied the influence of spin correlations on the phonon frequencies of $ZnCr_2O_4$ using a LSDA + U method. They provided evidence that the anisotropy induced by AFM ordering of spins can account for the large phonon anisotropy measured in IR experiments. To calculate optical phonon shifts $\Delta\omega = \omega_0 - \omega = \lambda \langle S_i \cdot S_j \rangle$ from first principles they developed an approach using experimentally accessible quantities and applied this method to calculate the spin-phonon coupling parameter $\lambda$ for $ZnCr_2O_4$. In contrast to what is seen experimentally, theoretically the phonon frequencies of all four modes are split. However, the splitting of the two high-frequency modes is significantly smaller than the splitting of the two low-frequency modes. It seems reasonable to assume that the splitting of the high-frequency modes is too small to be experimentally observed. Returning to Fig. 21, it seems important to note that despite the fact that strong anomalies were identified in the temperature dependence of the phonon frequencies of most compounds with an AFM ground state, no splitting was observable in $HgCr_2S_4$, an antiferromagnet with a helical spin-spiral ground state. The absence of any experimentally detectable phonon splitting probably results from the fact that already at high temperatures, this compound is dominated by strong FM fluctuations, but finally evolves into a spin-spiral ground state close to 21 K. The anomalies in the temperature dependence of the phonon frequencies appear just below 100 K [see Fig. 21(d)] and probably signal the onset of FM fluctuations. That indeed, no phonon splitting can be observed in FM chromium spinels has been documented in detail by Rudolf et al. [209] for $CdCr_2S_4$. In FM compounds, the strong spin-phonon coupling becomes obvious via significant anomalies in the temperature dependence of the phonon eigenfrequencies. In $CdCr_2S_4$ these anomalies appear at the onset of long-range FM order.

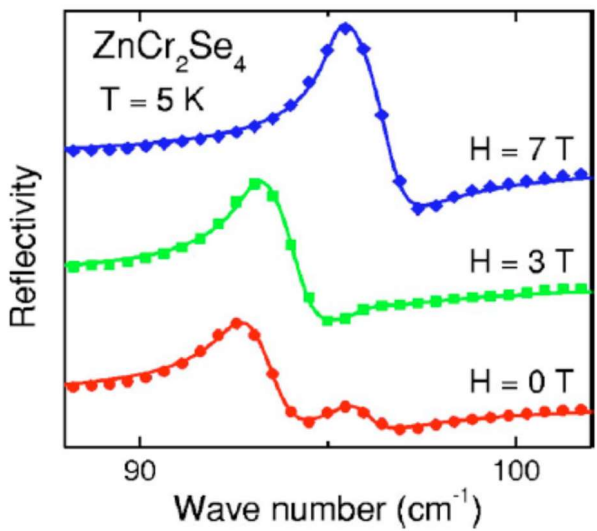

Fig. 22. Magnetic-field dependence of optical reflectivity in $ZnCr_2Se_4$. Reflectivity *vs* wave number as measured around the low-frequency phonon mode at 5 K and in magnetic fields of 0 T (red circles), 3 T (green squares), and 7 T (blue diamonds). With increasing external magnetic fields, the splitting of the phonon mode becomes fully suppressed. The solid lines represent Lorentzian fits.
Reprinted figure with permission from Rudolf et al. [205]. Copyright (2007) by the American Physical Society.

That spin-lattice coupling can also be suppressed by external magnetic fields has been impressively documented by Rudolf et al. [205] for $ZnCr_2Se_4$. In this compound AFM ordering can be suppressed in an external magnetic field of 7 T and the spin system enters into a field-polarized FM phase [179]. This transition from the AFM to the spin-polarized state can be mirrored by optical spectroscopy and can be identified via the splitting of the lowest-frequency phonon mode. A representative result is documented in Fig. 22, where the splitting of the mode close to 95 cm$^{-1}$ is suppressed in an external magnetic field of 7 T [205]. Later on, utilizing magnetic susceptibility, heat capacity, and thermal expansion experiments, it was documented that the spin spiral state is separated from the field-polarized phase by a QCP [187] close to 6 T. At this QCP the magnon modes completely soften and the cubic symmetry of crystalline lattice is restored [188], fully compatible with the suppression of the phonon splitting as documented in Fig. 22.

C. Fractionalized magnetization plateaus in chromite spinels

Stimulated by high-field magnetization experiments from Ueda et al. on $CdCr_2O_4$ [210] and $HgCr_2O_4$ [173], which were later summarized by Shannon et al. [211], Penc et al. [212,213] proposed a theoretical model documenting that in pyrochlore lattices a half-magnetization plateau arises quite naturally, once spin-lattice coupling is taken into account. As documented above, in frustrated magnets the coupling to the lattice provides a very efficient mechanism for lifting the ground-state degeneracy. Of course, external magnetic fields also will remove magnetic frustration and in case the external field exceeds the strength of the magnetic exchange, the spin system will develop into a field-induced ferromagnet. The corner-sharing network of tetrahedra is thought as a three-dimensional generalization of the spin-Peierls problem and the spin system always can gain energy by distorting the tetrahedral bonds and ordering the spins. Under these assumptions, a Heisenberg Hamiltonian in externals magnetic fields $h$ with a biquadratic-bilinear term has been proposed [212]:

$$\mathcal{H} = \sum_{i,j} J[S_i S_j - b(S_i S_j)^2] - h \sum_i S_i$$

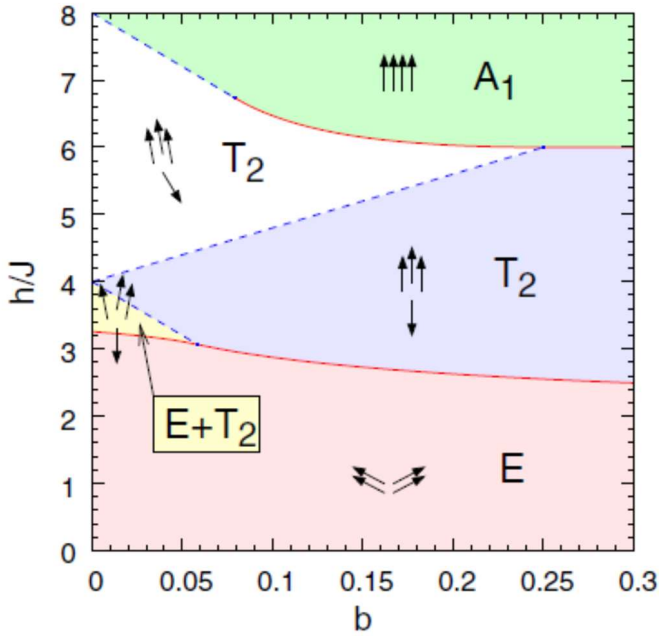

Fig. 23. Magnetic field vs. spin-lattice coupling phase diagram of the pyrochlore lattice. Spin structures as a function of magnetic field $h$ (in units of magnetic exchange $J$) and the dimensionless spin-lattice coupling $b$. Solid lines denote first-order and dashed lines, second-order transitions. Spin configurations and irreducible representations of the order parameter in each phase are shown. The spin configurations indicate the orientation of the four spins on each corner of a single tetrahedron.
Reprinted figure with permission from Penc et al. [212]. Copyright (2005) by the American Physical Society.

Here $S_i$ and $S_j$ are neighbouring spins on a single tetrahedron, $J$ is the magnetic exchange and $b$ measures the strength of the spin-lattice coupling. The sum $<ij>$ runs over all nearest-neighbour bonds of the pyrochlore lattice. The resulting generalized phase diagram is shown in Fig. 23. For small fields $h$ the lattice has overall tetragonal symmetry with distorted tetrahedra, with two long FM bonds and four short canted AFM bonds. In this regime, the magnetization remains linear, but with reduced slope [212]. For $h \sim 3 J$ and $b \geq 0.05$, the system undergoes a first-order transition into a collinear phase with three spins up and one spin down irrespective of the external magnetic field $h$. This 3:1 phase exhibits an extended and stable half-magnetization plateau over a broad field range. In this half-magnetization plateau phase, each tetrahedron has three long FM and three short AF bonds, giving rise to an overall trigonal lattice distortion. The width of the half-magnetization plateau shrinks linearly with decreasing coupling strength and at low values of $b$ two further phases, a coplanar 2:1:1 canted phase, and a coplanar 3:1 canted phase are stabilized. Finally, at large fields there is a transition into a cubic fully saturated FM phase. Later on the ordered and disordered phases of this bilinear and biquadratic Heisenberg model on the pyrochlore lattice were studied at finite temperatures and fields [214]. In these calculations, the authors found a rich collection of unconventional spin states, like nematic and vector-multipole phases with distinct symmetries, separated by a half-magnetization plateau. All of these states were deduced from a proper understanding of the geometry of the pyrochlore lattice and the properties of a single tetrahedron and along these lines the zero-temperature phase diagram can be explained from an assembly of ordered tetrahedral into complex states of higher symmetry.

This extended half-magnetization plateau was clearly observed in high-field magnetization experiments up to 50 T in the chromium oxide spinels $CdCr_2O_4$ [210] and $HgCr_2O_4$ [173,211] as documented in Fig. 24. In the former compound, the half-magnetization plateau evolves beyond 30 T. In the mercury spinel the plateau extends roughly between 10 and 30 T. It is interesting that no magnetization plateaus up to 50 T were observed in $ZnCr_2O_4$ and $MgCr_2O_4$. However, in these compounds half magnetization is still not reached even at the highest field values of these experiments and it can be expected that similar half-magnetization plateaus will be reached at significantly higher fields. In addition, magnetization plateaus were not observed in $ZnCr_2S_4$ and $ZnCr_2Se_4$. These compounds are characterized by a continuous increase of the magnetization until the fully field-polarized FM state is reached. The missing evolution of fractionalized magnetization values in these latter compounds already signals a significantly different frustration mechanism. Indeed, while in the oxides the spins on the pyrochlore lattice are strongly antiferromagnetically coupled and frustration is released via a spin-JT effect, in $ZnCr_2S_4$ and $ZnCr_2Se_4$ FM and AFM exchange interactions compete (see Fig. 19) and in increasing external fields the FM exchange will be continuously strengthened.

The theoretically predicted structural phase transition at the evolution of the half-magnetization plateau in $CdCr_2O_4$ was measured utilizing a high-field x-ray diffraction technique by Inami et al. [215]. Theoretically, in the 3:1 phase with three long FM and three short AFM bonds a rhombohedral or cubic lattice symmetry is expected, depending on the sign of the next-nearest neighbour interaction. If the plateau phase was rhombohedral, the 440 reflection would split into two peaks, which however, experimentally was not observed. Hence, it was concluded that the half-magnetization plateau exhibits cubic symmetry, or a splitting, which is too small to be experimentally observed [215]. Spin order and lattice symmetry of the half-magnetization plateau, which evolves between 10 and 30 T in $HgCr_2O_4$ has also been investigated in detail by high-field neutron and synchrotron x-ray scattering techniques [174]. These authors found that the magnetic structure has the same symmetry as the crystal structure: When entering the field-induced half-magnetization plateau the symmetry of the system changes from the orthorhombic Néel state with total zero spin by a first-order transition into a cubic collinear AFM 3:1 state with half the magnetization of the total spin system.

In the following years, the chromium oxide spinels were investigated in ultra-high external magnetic fields up to 600 T utilizing magneto-optical techniques, like Faraday rotation or magneto-optical absorption techniques [216,217,218,219,220,221,222]. These experiments were mainly devoted to establish the full ($H,T$)-phase diagrams of compounds governed by strong magnetic couplings between nearest neighbouring chromium ions, e.g., like $ZnCr_2O_4$, MgCr$_2$O$_4$ and of CdCr$_2$O$_4$, where full saturation of the field-polarized phase could not be reached with external magnetic fields up to 50 T (see Fig. 24). All these compounds exhibit stable half-magnetization plateaus and in all compounds, finally the fully polarized field could be reached, which is established at 90 T in the cadmium compound [216], at ~ 130 T in the magnesium compound [221] and at fields of ~ 400 T in the zinc compound [218]. In addition to the observation of the sequence of known spin patterns like the 3:1 or the canted 3:1 phase, theoretically predicted phases, like a canted 2:1:1 phase, were identified in these experiments [217,218,221]. Some of these magnetic phases were discussed as symmetry analogues to the quantum phases of liquid helium, like super-solid or super-liquid phases [217,219]. These quantum phases will be discussed in the multiferroics chapter in more detail.

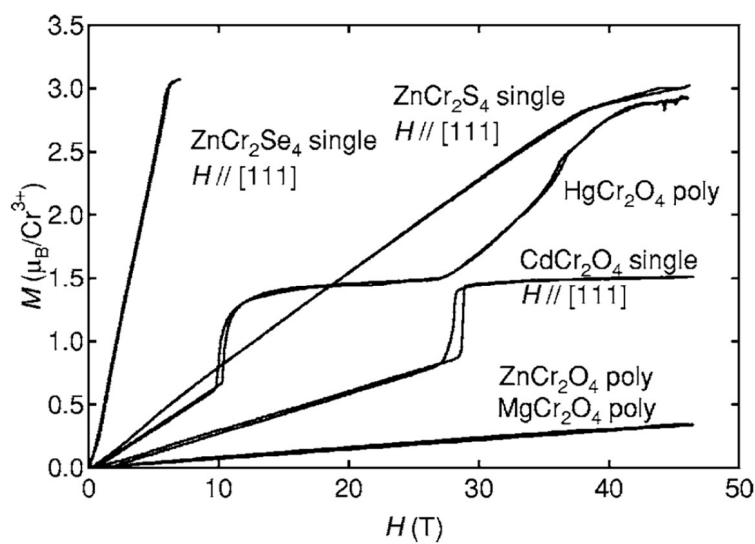

Fig. 24. Magnetization curves of various chromium spinels at high magnetic fields. $HgCr_2O_4$ and $CdCr_2O_4$ are measured at 1.8 K, all other compounds at 4.2 K. Only $HgCr_2O_4$ and $CdCr_2O_4$ exhibit half-magnetization plateaus. For $ZnCr_2O_4$ and $MgCr_2O_4$ these will be reached at higher fields. The sulphide and selenide compounds exhibit continuously increasing magnetization up to full spin polarization. Reprinted figure with permission from Ueda et al. [173]. Copyright (2006) by the American Physical Society.

There were a number of high-field magnetization and ultrasound studies in spinel chromium sulphides and selenides with AFM ground states [223,224]. As is documented in Fig. 24, neither in ZnCr$_2$S$_4$ nor in ZnCr$_2$Se$_4$, up to full field-polarization fractional magnetization plateaus were observed, but in both compounds, the magnetization rather increases continuously. In the selenide, the field-polarized phase is already established in external fields well below 10 T, while in the sulphide the field-polarized phase is reached beyond 40 T [173]. Compared to the chromium oxides, these spinel compounds are characterized by a strongly enhanced indirect FM exchange. In the zinc sulphide AFM and FM exchange are of equal strength, resulting in an almost zero CW temperature. In zinc selenide the FM exchange dominates and the CW temperature is of order 100 K (see Table 1). In the latter, the strong FM exchange is mirrored by the low saturation field of 10 T. It is obvious that the spin-JT mechanism, which releases frustration in the chromium oxides, is not active in the AFM sulphide and selenide. In these compounds, frustration must be released by another mechanism depending on the strength of bond frustration. However, like in the chromium oxide spinels, in both compounds the transition into the AFM ground state is also accompanied by significant structural distortions.

Temperature and field-dependent sound-velocity measurements are highly sensitive to unravel possible structural deformations occurring at magnetic transition either as function of temperature or magnetic field. From magnetic susceptibility, thermal-expansion and heat-capacity experiments, it is well known that in ZnCr$_2$S$_4$ two subsequent magnetic transitions appear close to 15 and 8 K [176]. The temperature dependence of the longitudinal sound velocity as observed in experiments in a single crystal along the [001] direction, which in cubic crystals is determined by the elastic constant $c_{11}$, is documented in Fig. 25. In zero external magnetic field and on decreasing temperatures, the sound velocity undergoes strong softening when approaching the AFM phase transition and reveals a narrow dip-like minimum at $T_{N1}$, accompanied by a spike-like maximum in the sound attenuation. On further decreasing temperatures, the sound velocity passes through a broad minimum close to 10 K and increases again on further cooling. The second AFM transition at $T_{N2}$ is hardly visible in the temperature dependence of the sound velocity, but is indicated by a broad maximum in the temperature dependent damping. On increasing fields, both AFM transitions are slightly shifted to lower temperatures and the minimal sound velocity between the two phase transitions becomes less significant, but the damping strongly increases. We would like to recall that at $T_{N1}$ the sample transforms from the cubic and PM state into the tetragonal helimagnetic phase, while at $T_{N2}$ ZnCr$_2$S$_4$ transforms into an orthorhombic structure. Also from the sound-velocity measurements, it becomes clear that both magnetic transitions are accompanied by significant structural changes.

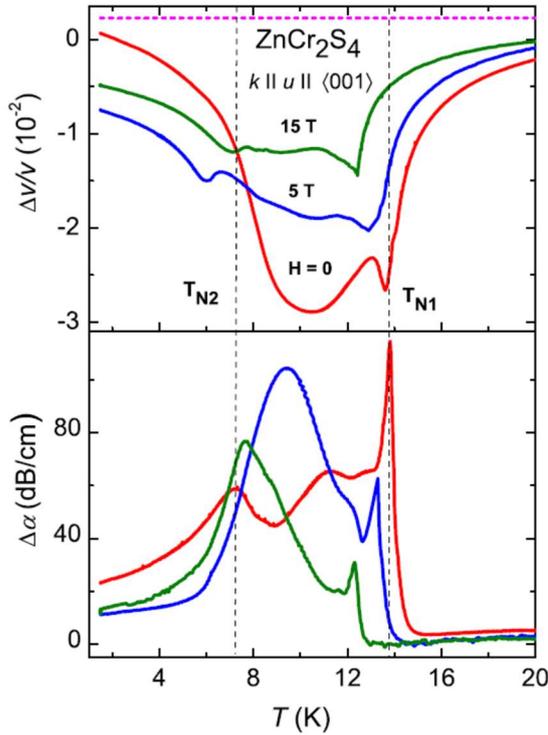

Fig. 25. Sound velocity and damping in $ZnCr_2S_4$. Temperature dependence of the relative change of sound velocity $\Delta v/v$ (upper panel) and sound attenuation $\Delta\alpha$ (lower panel) measured in different static magnetic fields up to 15 T. Vertical dashed lines mark the magnetic phase transitions $T_{N1}$ and $T_{N2}$ as observed in zero field. The horizontal dashed line in the upper panel shows the extrapolated undisturbed sound velocity estimated from a fit to the data above 60 K.
Reprinted figure with permission from Tsurkan et al. [223]. Copyright (2011) by the American Physical Society.

Figure 26 documents the field-dependent experiments in $ZnCr_2S_4$ up to 50 T for a series of temperatures around the onset of magnetic order [223]. Here, sound-velocity changes [Fig. 26(a) and (b)] and damping [Fig. 26(d) and (e)] as observed as function of magnetic field for temperatures between 1.5 and 20 K are shown together with the field-dependent magnetization [Fig. 26(c)] and its derivative dM/dH [Fig. 26(f)]. Despite the fact that the magnetization continuously increases up to its saturation, which is reached close to 50 T, a series of anomalies, indicative of subsequent structural phase transitions are observed in the field-dependence of sound velocity and damping. In Fig. 26 these anomalies are indicated by arrows and labelled with numbers 1 to 4. Figs 26 (a) and (b) document rather significant changes of the sound velocity with field, proving strong magnetoelastic coupling in $ZnCr_2S_4$. At 1.5 K, the sound velocity shows a non-monotonic behaviour with the field: starting from zero field, $\Delta v/v$ first decreases, but beyond 30 T increases again characteristic for an even stiffer material beyond 50 T. Both sound velocity and damping exhibit four prominent anomalies approximately at 7, 27, 38, and 44 T, suggesting changes in the spin configuration with accompanying structural phase transitions.

Comparing up and down field sweeps of the velocity in magnetic fields, at 1.5 K and at magnetic fields above 27 T, the anomalies 1 - 3 do not reveal hysteretic behaviour, whereas the anomaly 4 displays a marked hysteresis, which indicates a first-order magneto-structural transitions induced by the magnetic field. With increasing temperature, all sound-velocity anomalies shift to lower fields and at 20 K in the PM phase, all anomalies have vanished yielding an almost constant sound velocity and damping. Remarkably, at 10 K plateaus develop in the field dependence of the sound velocity indicating crystallographic structures with constant stiffness, while the continuous increase of magnetization [Fig. 26(c)] signals that these structural changes are not accompanied by significant changes in the spin structure. It seems important to summarize that in $ZnCr_2S_4$, while lacking any fractionalized magnetization plateau, plateaus are observed in the sound velocity, revealing a constant field-independent structure. Based on these field-dependent ultrasound and magnetization experiments, the authors of Ref. [223] constructed a (H,T)-phase diagram with a series of subsequent magnetic phases before the field-polarized phase is reached.

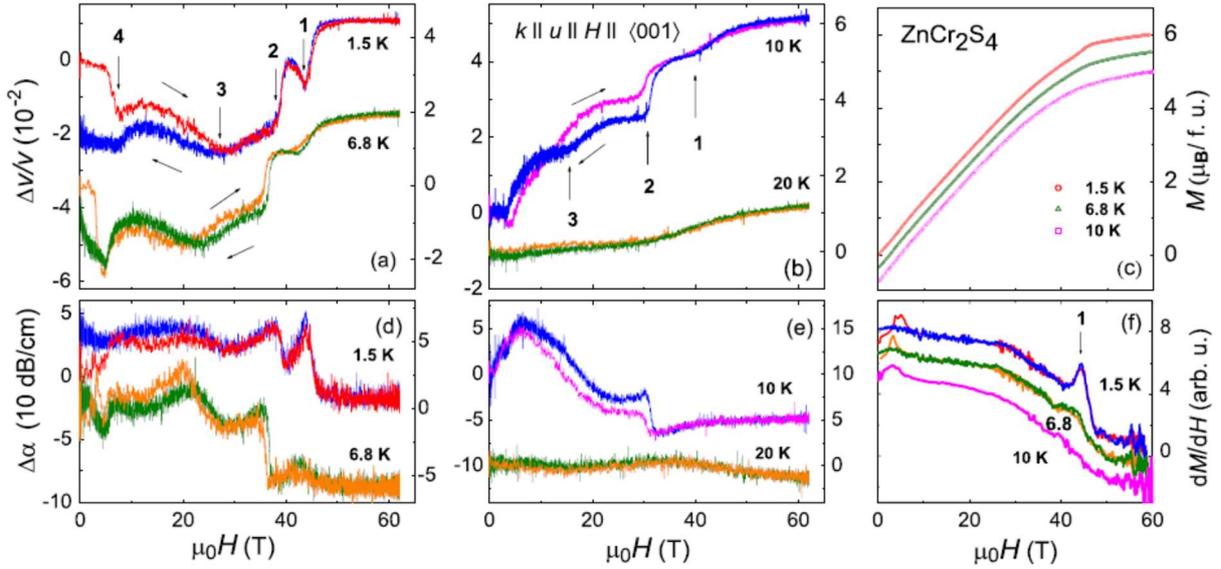

Fig. 26. Sound velocity, damping, and magnetization of $ZnCr_2S_4$ in high magnetic fields. Relative change of the longitudinal sound velocity $\Delta v/v$ [(a) and (b)] and attenuation $\Delta \alpha$ [(d) and (e)] as measured along the crystallographic <001> direction vs. magnetic field at 1.5 and 10 K (left scale) and at 6.8 and 20 K (right scale). The vertical arrows mark the magneto-structural anomalies labelled with numbers from 1 to 4.
(c) Magnetization curves with the magnetic field aligned along <001> and (f) derivatives of the magnetization dM/dH as obtained for 1.5, 6.8, and 10 K. For clarity, the curves are shifted along the vertical axis. Data for magnetic field sweeps up and down are shown. Reprinted figure with permission from Tsurkan et al. [223]. Copyright (2011) by the American Physical Society.

Field-dependent ultrasound experiments were also performed in $ZnCr_2Se_4$ [224]. In this compound, with much stronger FM exchange interactions, the field-polarized phase is already reached at 10 T. On the basis of these experiments a (*H*,*T*)-phase diagram was constructed [224], where the helical phase is followed by a conical helix with the spins canting in field direction. It was speculated that a nematic phase separates the conical phase from the fully spin-polarized state.

It is worth mentioning that similar temperature and field-dependent ultrasound experiments were conducted in $CdCr_2O_4$, a system with a well-developed half-magnetization plateau [225]. From a detailed comparison of the experimentally observed propagating of the longitudinal acoustic mode along the <111> direction with a theory based on exchange-striction, it was possible to estimate the strength of the magnetoelastic interaction. The derived spin-phonon coupling constant was in good agreement with previous determinations based on infrared experiments. Interestingly, an extended plateau in the longitudinal sound velocity measured along the <111> direction extends from 30 to 60 T, exactly correlating with the half-magnetization plateau in the magnetization experiments. In further sound velocity experiments on $CdCr_2O_4$ as function of magnetic field and temperature [226], a further magneto-structural transition close to 3 K and evolving under external magnetic fields at least up to 16 T was observed. It was speculated that the experimentally observed significant anomalies in sound velocity and damping, indicate the transition into a further low-temperature spin spiral phase.

### 3.3.2. Frustration in *A*-site spinels: Magnetic ions at the diamond lattice

In *A*-site spinels only the *A* site of the spinel lattice is occupied by magnetic ions, while non-magnetic ions are located at the *B* site. Over the years, there were not too many investigations on *A*-site spinels due to the fact that *A*-*A* exchange interactions were expected to be rather weak due to an indirect exchange path through three intermediate ions resulting in low magnetic ordering temperatures. An illustrative example of *A*-*A* magnetic exchange paths in $AB_2O_4$ spinels is given in Fig. 27 [227]. First interest in this class of materials arose because an unexpectedly high AFM ordering temperature of $T_N$ = 40 K was found in $Co_3O_4$ [227]. In this normal spinel compound $Co^{2+}$ ions ($d^7$) are located at the tetrahedral *A* site and carry a spin S = 3/2. The $Co^{3+}$ ions ($d^6$) occupy the octahedral *B* sites and have zero magnetic moment as a consequence of the large splitting of the 3*d* levels by the octahedral crystal field and a resulting low-spin configuration. Later on, Roth [228] investigated a series of transition-metal aluminates with $AAl_2O_4$ (*A* = Mn, Fe, Co) to check if high magnetic ordering temperatures are also observed in these compounds or if this anomalous strong *A*-*A* interaction is an unique property of $Co_3O_4$. Indeed, the *A*-*A* interactions in these compounds were an order of magnitude weaker and it was concluded that the anomalous strong interaction in $Co_3O_4$ results from indirect exchange coupling through the octahedral $Co^{3+}$ complex.

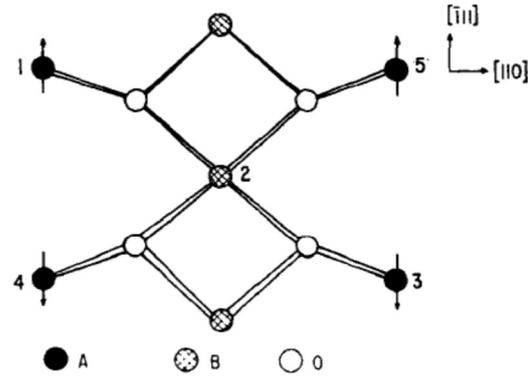

Fig. 27. Magnetic exchange in spinels. *A*-*A* magnetic superexchange paths in the spinel structure where only *A*-site ions carry magnetic moments. Reprinted from Roth [227]. Copyright (1964), with permission from Elsevier.

Later on detailed magnetic susceptibility and heat capacity experiments [62], as well as neutron scattering experiments [229] were performed on the aluminium-oxide spinels. In these studies, the problem of possible inversion was avoided, by synthesizing the samples at the lowest possible temperatures, which still guaranteed reasonable diffusion rates and in addition, all samples were ultra-slowly cooled to room temperature [62]. The magnetic susceptibility measurements as reported by Tristan et al. [62] are shown in Fig. 28. These measurements were taken between 1.8 and 400 K in external magnetic fields of 1 T. The high-temperature susceptibilities were fitted by a CW behaviour and the resulting fit parameters, namely CW temperatures $\Theta$, PM moments $p_{eff}$ and resulting *g* values are given in Table 2. The *g* value for the manganese compound is close to the value expected for a spin-only compound with a half-filled shell. The *g* values of the iron and cobalt compound are significantly larger than 2. This enhancement can be understood by a residual SOC, as expected for magnetic ions with a more than half-filled shell. From the susceptibilities alone, it would be rather uncertain to determine the magnetic ordering temperatures. Hence Tristan et al. [62] performed additional magnetization, FC and ZFC experiments, as well as ESR experiments to determine the exact ordering temperatures and the nature of the magnetic phase transitions. From this analysis, it was concluded that the manganese compound undergoes AFM order close to 40 K, while iron and cobalt compounds reveal spin-glass freezing close to 5 and 12 K, respectively.

Spin-glass freezing in the iron and cobalt compound was further substantiated by heat-capacity experiments [62], which are reproduced in Fig. 29. Analysing these heat capacity experiments, it can be concluded that only $MnAl_2O_4$ reveals a well-defined λ-type anomaly indicating a transition into a long-range ordered spin state. The heat capacities of $FeAl_2O_4$ and $CoAl_2O_4$ show broadened cusp-like maxima, which appear slightly above the freezing temperatures as determined in the FC and ZFC experiments [62] indicative for spin-glass freezing. However, in canonical spin glasses the heat capacity follows a linear $T$ dependence, while the two compounds shown in Fig. 29 clearly reveal a $T^2$ dependence. A squared temperature dependence of the heat capacity was also found in geometrically frustrated two-dimensional spin glasses with spinel and distorted Kagomé-based structures [230, 231]. As documented in Tab. 2, the frustration parameters $f$ of this class of aluminium oxide spinels are rather different, ranging from 3.6 up to 22. Following Table 2, only the cobalt and iron compounds can be assigned to be strongly frustrated magnets.

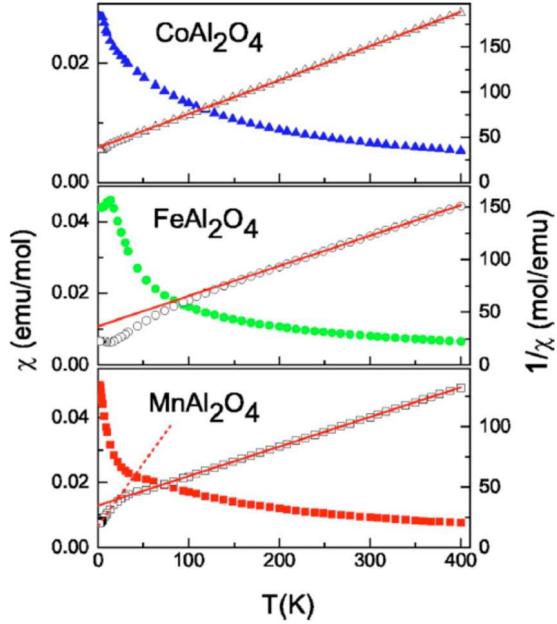

Fig. 28. Magnetic susceptibility in aluminium oxide spinels. Temperature dependence of magnetic susceptibility (closed symbols, left scale) and inverse susceptibility (open symbols, right scale) of $CoAl_2O_4$ (upper frame), $FeAl_2O_4$ (middle frame), and $MnAl_2O_4$ (lower frame). The solid lines represent high-temperature CW fits. In the lower frame, the dashed line indicates a second CW-like law below $T_N$. Reprinted figure with permission from Tristan et al. [62]. Copyright (2005) by the American Physical Society.

From a detailed analysis of the temperature-dependent heat capacity of these aluminium-oxide spinels including entropy considerations, it became clear that in the iron compound orbital degrees of freedom play an important role. The spin-glass behaviour of the iron and cobalt compound was further elucidated in detailed quasi-elastic and inelastic neutron scattering investigations, documenting a liquid-like structure factor for these two compounds [229]. Overall, these investigations documented that $A$-site spinels reveal significant magnetic frustration, a behaviour that was later-on studied experimentally and theoretically in a group of scandium $A$-site spinels.

| Compound | $CoAl_2O_4$ | $FeAl_2O_4$ | $MnAl_2O_4$ |
|---|---|---|---|
| Spin, $S$ | 3/2 | 2 | 5/2 |
| Curie-Weiss temperature, $\Theta_{CW}$ (K) | -104(2) | -130(1) | -143(5) |
| Ordering temperature, $T_m$, $T_N$ (K) | 4.8(2) | 12(0.5) | 40(0.5) |
| Frustration parameter, $f$ | 22 | 11 | 3.6 |
| Paramagnetic moment, $p_{eff}$ ($\mu_B$) | 4.65(9) | 5.32(3) | 5.75(10) |
| Effective g factor, $g_{eff}$ | 2.40(4) | 2.17(1) | 1.94(6) |

Table 2: Spin values $S$, characteristic magnetic ordering temperatures $T_m$ and $T_N$, PM moments $p_{eff}$, CW temperatures $\Theta_{CW}$ and $g$ factors as determined from high-temperature CW fits to the magnetic susceptibilities of $AAl_2O_4$ spinels ($A$ = Co, Fe, Mn). In addition, the frustration parameter $f$ is indicated. Taken from Tristan et al. [62].

After the prediction of a spin-spiral liquid state in *A*-site spinels by Bergman et al. [22], which will be discussed in more detail in the next section, the ground states of pure and doped $CoAl_2O_4$ gained considerable attention and were re-investigated by various experimental methods including, magnetic, thermodynamic, ESR, x-ray diffraction, as well as by detailed quasi-elastic and inelastic neutron-scattering techniques [232,233,234,235,236,237]. These experiments on the pure cobalt-oxide spinel were supplemented by systematic experiments on the mixed spinel compounds $Co(Co_{1-x}Al_x)_2O_4$ [233,234], as well as by Monte Carlo simulations [234]. While all the different experimental investigations generally agreed on some type of disordered ground state, the true microscopic nature still is under debate. The latest proposal is that at the freezing temperature, AFM correlations become arrested and the low-temperature scattering contributions can be explained assuming the freezing of domain-wall motions preventing the formation of true long-range AFM order [237].

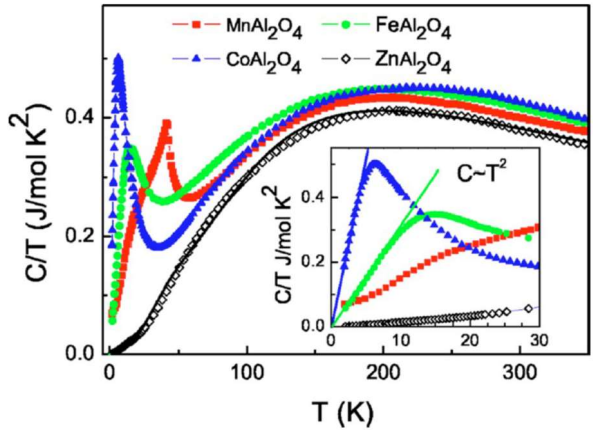

Fig. 29. Heat capacities of aluminium oxide spinels. $C/T$ vs. $T$ for $AAl_2O_4$ for $A$ = Mn (red squares), Fe (green circles), and Co (blue triangles). In addition, $C/T$ of the non-magnetic reference zinc compound (empty rhombs) is shown. The inset shows the low-temperature heat capacity on an enlarged scale, where the quadratic regime with $C \sim T^2$ is indicated by straight lines. Reprinted figure with permission from Tristan et al. [62]. Copyright (2005) by the American Physical Society.

The research on *A*-site spinels was heavily intensified by the report of Fritsch et al. [40] documenting strong spin frustration effects in $MnSc_2S_4$ and $FeSc_2S_4$. In a series of subsequent magnetization and heat capacity measurements [238], as well as NMR [239] and neutron scattering experiments [240] it was documented that both compounds reveal strong frustration parameters: The manganese compound has a CW temperature close to 20 K and reveals complex magnetic order below 2 K. The iron compound is characterized by a CW temperature of 50 K, revealing no long-range magnetic order down to the mK regime. In addition, $Fe^{2+}$ with a $d^6$ electronic configuration is tetrahedrally coordinated by sulphur $S^{2-}$ ions and reveals a twofold orbital degeneracy (see Fig. 6). Consequently, a JT transition leading to long-range OO is expected at low temperatures, which however, is not observed. In FeSc$_2$S$_4$, in addition to the spin degrees of freedom, also the orbital degrees of freedom seem to be frustrated and this iron-spinel compound belongs to the rare class of SOLs down to the lowest temperatures.

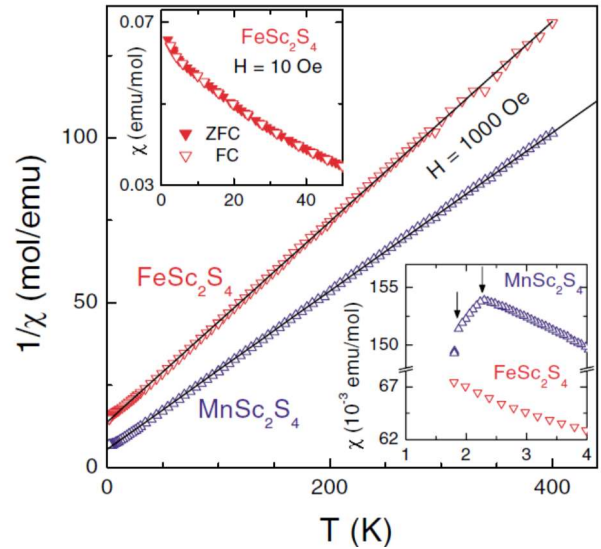

Fig. 30. Magnetic susceptibility in *A*-site spinels. Temperature dependence of the inverse magnetic susceptibilities $1/\chi$ of $MnSc_2S_4$ (triangles up) and $FeSc_2S_4$ (triangles down). The solid lines are fits with a CW law. Upper inset: FC and ZFC susceptibility in $FeSc_2S_4$ measured at 10 Oe. Lower inset: Susceptibility $\chi$ vs $T$ at low temperatures. Reprinted figure with permission from Fritsch et al. [40]. Copyright (2004) by the American Physical Society.

Fig. 30 shows the inverse magnetic susceptibilities for the manganese and the iron compounds. Both susceptibilities reveal almost ideal CW behaviour at elevated temperatures. At the lowest temperatures, slight deviations appear in MnSc$_2$S$_4$ with a clear fingerprint of subsequent AFM ordering transitions at 2.3 and 1.8 K (see lower inset of Fig. 30). No indication of a magnetic or structural phase transition can be identified in FeSc$_2$S$_4$ and the close agreement of ZFC and FC measurements document the absence of any spin-glass like freezing process (upper inset of Fig. 30).

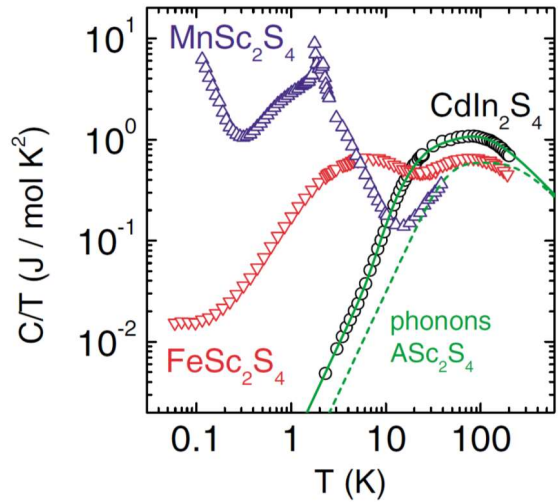

Fig. 31. Specific heat in $A$-site spinels. Specific heat plotted as $C/T$ vs. the logarithm of $T$ for MnSc$_2$S$_4$ (triangles up), FeSc$_2$S$_4$ (triangles down), and for non-magnetic CdIn$_2$S$_4$ (circles). The solid line represents the calculated specific heat of the non-magnetic reference compound CdIn$_2$S$_4$. The dashed line provides an estimate of the phonon contribution for $A$Sc$_2$S$_4$ ($A$ = Mn, Fe) scaled with the adequate mass factor. Reprinted figure with permission from Fritsch et al. [40]. Copyright (2004) by the American Physical Society.

The results of the corresponding heat-capacity experiments on both compounds are shown in double-logarithmic representation in Fig. 31, together with measurements of the non-magnetic reference compound CdIn$_2$S$_4$. In the two magnetic compounds, phonon contributions govern the heat capacity above 10 K, while below this temperature spin and orbital contributions dominate. In the manganese compound, a double-peak structure close to 2 K indicates two subsequent magnetic transitions. A broad hump in the iron compound probably results from the orbital degrees of freedom [40]. There is no evidence for any type of phase transition down to 50 mK. In this respect these two scandium-based $A$-site spinels are prototypical examples of strongly frustrated magnets with frustration parameters of $f > 10$ for the manganese compound and at least larger than 1000 for the iron compound.

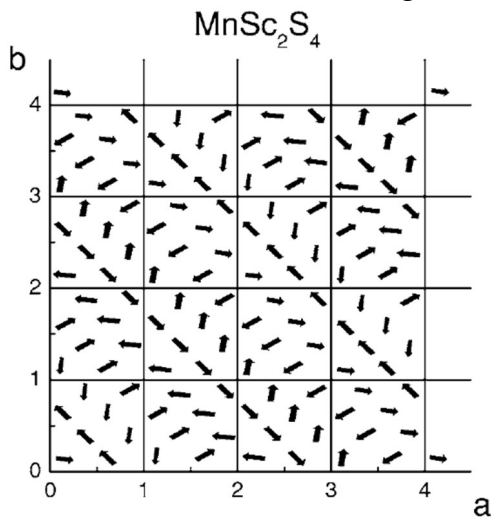

Fig. 32. The magnetic structure of MnSc$_2$S$_4$. Spins within the $ab$ plane showing a cycloidal arrangement. Reprinted figure with permission from Krimmel et al. [241]. Copyright (2006) by the American Physical Society.

Subsequently, the magnetic structure and the excitation spectrum of the AFM state of MnSc$_2$S$_4$ have been determined by elastic neutron-scattering experiments by Krimmel et al. [241]. The spiral spin structure formed at low temperatures is shown in Fig. 32. The magnetic structure found at $T = 1.5$ K is a spiral within the $ab$ plane characterized by a propagation vector of $\mathbf{q} = (3/4, 3/4, 0)$. The temperature dependence of the magnetization roughly follows mean-field behaviour with a saturation value close to 5 $\mu_B$ as expected for a half-filled $d$ shell. A slight anomaly in the temperature dependence of the magnetization shows up close to 1.8 K indicating a second magnetic phase transition in good agreement with the heat-capacity experiments [241].

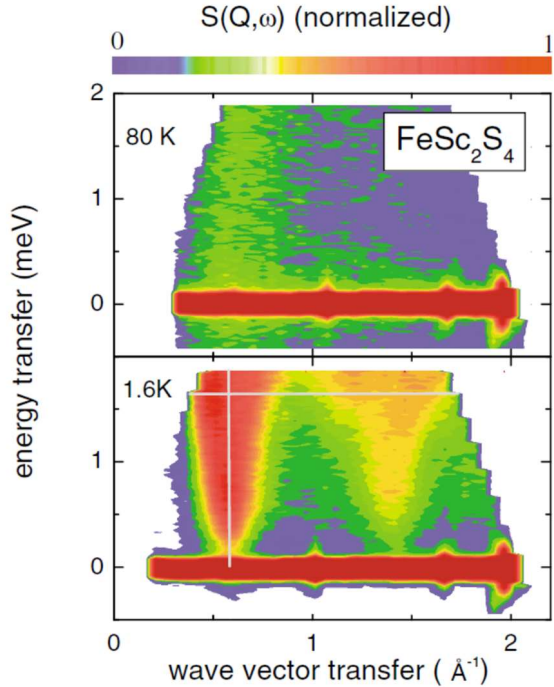

Fig. 33. Neutron-scattering results in FeSc$_2$S$_4$. Contour plot of the dynamic structure factor $S(Q, \omega)$ in the low-$Q$ and low-energy region at $T = 80$ K (upper panel) and $T = 1.6$ K (lower panel). The scattered intensity is colour coded as indicated in the colour bar above the figure.
Reprinted figure with permission from Krimmel et al. [240]. Copyright (2005)

Quasielastic and inelastic neutron scattering experiments were also performed in polycrystalline FeSc$_2$S$_4$ [240]. The most fascinating results showing the evolution of the dynamic structure factor as function of temperature are documented in Fig. 33 and show strong dispersion effects of the magnetic excitation spectrum with strong softening on cooling. At the lowest temperature of this investigation, the magnon response has almost completely softened at the zone boundary with a very small excitation gap of order 0.2 meV at the lowest temperatures. In the color-coded contour plot of Fig. 33 this softening is revealed by the extended intensity minima appearing close to the hypothetical AFM wave vectors (lower frame of Fig. 33). These intensity minima indeed correspond to strongly dispersing magnon-like modes around the AFM Bragg points, as has been documented by constant-energy scans in [240]. At 80 K, again at the zone boundary of the high-temperature PM phase, corresponding to the AFM ordering wave vector, considerable spin fluctuations show up, characteristic for an antiferromagnet just beyond the onset of magnetic order (upper frame of Fig. 33). At the lowest temperatures, well below the modulus of the CW temperature, soft excitations dominate the spectrum, which clearly indicate that FeSc$_2$S$_4$ is very close to magnetic and, in addition close to OO. Indeed, subsequent detailed theoretical work of Balents and coworkers [242,243] documented that the iron compound is close to quantum criticality and that a critical spin-orbital spectrum, like that documented in Figure 33 evolves close to the hypothetical AFM ordering wave vector.

Interestingly it has been suggested that the SOL state in FeSc$_2$S$_4$ does not result from frustration, but is the consequence of a strong competition between the local and on-site SOC, and intersite spin-orbital exchange [242,243]. SOC is responsible for splitting of the electronic states, while spin-orbital exchange is the reason for significant dispersion effects. While spin and orbital exchange, both favour long-range ordered ground states, strong SOC can result in a SOL. A schematic $(x,T)$ phase diagram as taken from Mittelstädt et al. [244] is plotted in Fig. 34. Here the control parameter $x$ represents the ratio of magnetic exchange $J$ to the effective spin-orbit interaction $\lambda$ [242,243]. In this phase diagram, a quantum-critical region separates the disordered ground state, the SOL or spin-orbital singlet (SOS) from the ground state with long-range AFM and concomitant OO. In Fig. 34 the dashed vertical line indicates the suggested location of FeSc$_2$S$_4$ in this phase diagram. Here it is proposed that this iron-scandium-sulphide spinel exhibits a SOL ground state, with neither magnetic nor orbital long-range order.

In order to understand the excitation spectrum of the strongly coupled spin and orbital degrees of freedom it is useful to consider the CEF splitting including spin-orbit effects. This has been done in detail in Ref. [245]. As documented in Fig. 6, the electronic ground state of Fe$^{2+}$ in a tetrahedral crystal field is an orbital doublet followed by an excited triplet. Considering the $d^6$ electron configuration of this ion, Hund's-rule coupling and assuming a high-spin configuration leads to a spin $S = 2$ system with a hole in the ground-state doublet. This CEF configuration is JT active and OO is expected at low temperatures, which however, experimentally is not observed in FeSc$_2$S$_4$. Consequently, the orbital doublet is split by second-order spin-orbit interaction into five equally spaced levels separated by $\lambda = 6\,\lambda_0^2/\Delta$, where $\Delta$ is the crystal-field splitting and $\lambda_o$ is the atomic SOC constant. Using reasonable numbers of crystal field and SOC, the splitting $\lambda$ was calculated to be of order 15 – 20 cm$^{-1}$ [244].

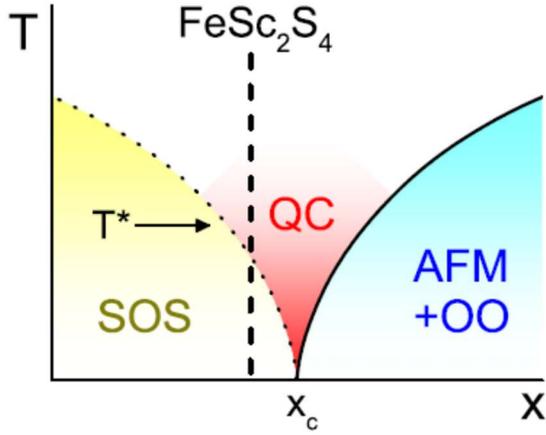

Fig. 34. ($x$,$T$) phase diagram of a spin system with strong spin-orbit interactions. $x$ is a control parameter determined by the ratio of magnetic exchange to effective spin-orbit coupling. SOS, QC, AFM, and OO denote spin-orbital singlet, quantum-critical, antiferromagnetic, and orbitally ordered states, respectively. The vertical dashed line gives the tentative position of FeSc$_2$S$_4$ in this phase diagram.
Reprinted figure with permission from Mittelstädt et al. [244]. Copyright (2015) by the American Physical Society.

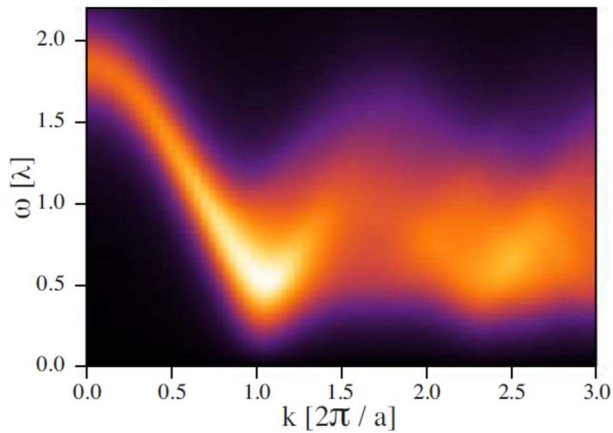

Fig. 35. Imaginary part of the angular-averaged dynamic spin susceptibility of FeSc$_2$S$_4$ in zero magnetic field in the frequency vs. momentum space. The calculated frequency and wave-vector dependent spin susceptibility can directly be compared with the experimental results shown in Fig. 33.
Reprinted figure with permission from Chen et al. [243]. Copyright (2009) by the American Physical Society.

The experimental excitations of FeSc$_2$S$_4$ documented in Fig. 33 reveal strong dispersion effects of the entangled spin and orbital degrees of freedom. For comparison of the neutron-scattering results with model calculations of these spin-orbital entangled excitations, which were measured on polycrystalline samples, the imaginary part of the angular-averaged spin susceptibility was calculated in zero external magnetic field [243]. The results are shown in Fig. 35. The excitation minima near wave vectors $k = 2\pi/a$ and $5\pi/a$ agree well with the inelastic neutron-scattering data [240] and can directly be compared with the contour plots derived from these neutron-scattering experiments that are documented in Figure 33. From these findings, it was concluded that FeSc$_2$S$_4$ is close to the quantum critical point, but still in the SOS phase as shown in Fig. 34. It is interesting to note that theoretically no magnetic field-induced transition to an ordered state was found or predicted [243]. Indeed, NMR experiments in magnetic fields up to 8.5 T showed no signs of magnetic ordering [239], but still indicated a PM and orbitally disordered low-temperature phase.

To investigate the microscopic nature of the spin-orbital coupled ground state in FeSc$_2$S$_4$, detailed THz spectroscopic studies were performed [244,245]. As documented in Figs. 33 and 35 and as outlined in the crystal-field scenario above, a strongly wave-vector dependent singlet-triplet excitation is expected. This propagating mode is almost zero at the AFM zone boundary and strongly enhanced at the zone centre. A zone-centre value of 1.9 $\lambda$ can be deduced from Fig. 35, was calculated in Ref. [243], and should be observable by THz spectroscopy. Indeed a well-defined excitation has been identified in THz experiments [244,245] close to 35 cm$^{-1}$ corresponding to ~ 4.5 meV. Its triplet character has been demonstrated by field-dependent experiments documenting a clear splitting of this mode [245]. Its temperature dependence has been studied in detail by Mittelstädt et al. [244] and is shown in Fig. 36. Here the temperature dependent dielectric loss is shown as determined in THz transmission experiments on polycrystalline FeSc$_2$S$_4$ samples. A well-defined excitation close to 35 cm$^{-1}$ evolves from a continuously increasing background at low temperatures. In Ref. [244] this excitation of strongly entangled spin and orbital degrees of freedom was called spin-orbiton. The strongly increasing background as function of frequency was ascribed to a spin-orbital continuum, a characteristic signature of quantum criticality [244].

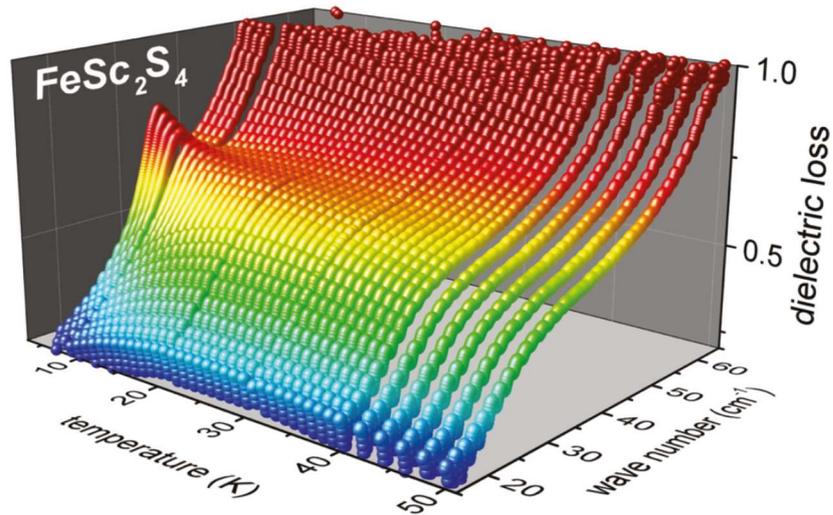

Fig. 36. Three-dimensional plot of the dielectric loss of FeSc$_2$S$_4$ vs wave number and temperature. The dielectric loss is colour coded to indicate equal-loss contours above the temperature and wave-number plane. The strong increase of the background was ascribed to a spin-orbital excitation continuum, a characteristic signature of quantum criticality.
Reprinted figure with permission from Mittelstädt et al. [244]. Copyright (2015) by the American Physical Society.

   The A-site spinels gained further attention by the theoretical prediction of a possible spin-spiral liquid state in A-site spinels [22,246]. It was argued that the formation of a spiral spin liquid strongly depends on the ratio of the strengths of the intra $J_1$ and inter-site interactions $J_2$ of the bipartite diamond lattice. Exchange ratios $J_2/J_1 > 1/8$ result in a spin-liquid state, in which the ground state is an enormously degenerate set of coplanar spin spirals with propagation wave vectors lying on a two-dimensional surface in momentum space [22]. From this modelling, it seemed that MnSc$_2$S$_4$ should be the prime candidate to observe this exotic spin-liquid state. The authors of Ref. [22] predicted that the spiral propagation vectors form a unique spiral surface of strong spin fluctuations in momentum space, which they proposed could be observed by neutron-scattering experiments. However, at the time of this theoretical prediction, single crystals of appropriate size and quality to perform inelastic and quasielastic neutron-scattering experiments were not available. After 10 years of hard work it was possible to grow single crystals, large enough to perform quasi-elastic neutron scattering [247] and neutron-scattering experiments have been performed at the Heinz Maier-Leibnitz Zentrum in Munich, as well as at the spallation neutron source SINQ of the Paul Scherrer Institute in Villigen (CH) [28]. One main result of this work is shown in Fig. 37: Here the experimental neutron scattering results are compared to the theoretical predictions. Fig. 37(a) shows the two interpenetrating fcc lattices of the diamond lattice with the intra-exchange $J_1$ and the inter-exchange $J_2$, within one and in between the two sublattices. $J_3$ indicates magnetic exchange with the third-nearest neighbour. Frame (b) shows the plane in reciprocal space carrying the spin-spiral surface where the experiments were conducted. The frames (c) and (d) compare experimentally observed diffuse scattered intensities with theoretical predictions. Frame (d) documents Monte Carlo simulations, which were performed for spin correlations in the ($HK$0) plane using the model with the exchange ratio $|J_2/J_1|=0.85$ and for temperatures $T/|J_1|=0.55$. Finally, Fig. 37(f) shows cuts through the diffusive spots as function of temperature. As documented specifically in frames (c) and (d) by the almost perfect agreement between theory and experiments, the experimentally observed diffuse scattering intensities directly prove the existence of the spiral surface and the spiral spin-liquid state in MnSc$_2$S$_4$. It is really fascinating to see that 10 years after the theoretical prediction of the existence of spin-spiral liquids, their existence has been proven experimentally by quasielastic neutron-scattering experiments.

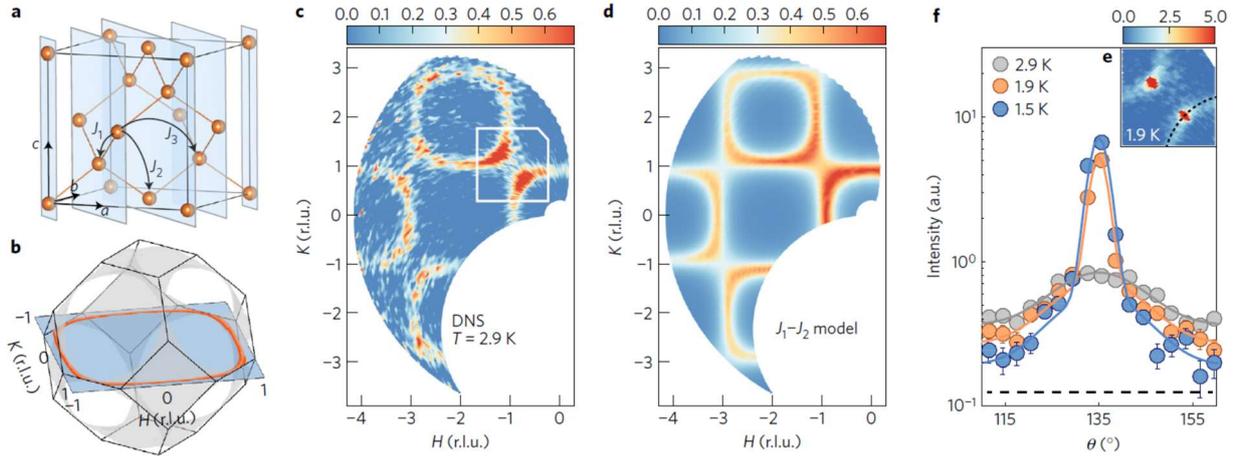

Fig. 37 Magnetic exchange paths and neutron scattering results in MnSc$_2$S$_4$.
a) Diamond lattice with two interpenetrating fcc lattices and magnetic exchange paths $J_1$, $J_2$ and $J_3$. The (110) planes are shaded blue. b) Spiral surfaces (grey) predicted by theory with the ratio $|J_2/J_1|=0.85$. The red ring emphasizes a cut within the ($HK$0) plane (blue).
c) Diffuse neutron scattering intensities in the ($HK$0) plane measured at 2.9 K. The indicated white square outlines the area shown in frame e). d) Monte Carlo simulations for spin correlations in the ($HK$0) plane using the model with the ratios $|J_2/J_1|= 0.85$ and $T/|J_1|=0.55$.
e) Diffuse scattering intensities around (110) (contour outlined by the white line in c) measured at 1.9 K, showing the coexistence of Bragg peaks and diffuse signal. The dashed arc describes the path for the cut presented in f). f) Comparison of the 1D cut at $T$ = 2.9, 1.9 and 1.5 K, showing the diffuse signal.
Material from: S. Gao et al. [28], Copyright © 2016, Springer Nature.

Very recently, by combining neutron-scattering measurements with extensive Monte Carlo simulations, it was demonstrated that at low temperatures (< 1 K) and at moderate external magnetic fields ranging from approximately 4 – 6 T, a fractional antiferromagnetic skyrmion lattice is stabilized in MnSc$_2$S$_4$. This very rare and incipient meron (half skyrmion) structure is stabilized through anisotropic couplings and is composed of three antiferromagnetically coupled sublattices, with each sublattice showing a triangular skyrmion lattice that is fractionalized into two parts with an incipient meron character [248].

## 3.4. Multiferroic spinels

### 3.4.1. Ferroelectricity in spinels

The use of FM metals even in laminated form as high-permeability core materials is restricted to rather low frequencies because of eddy-current losses. The ideal core material is an insulating ferromagnet (or ferrimagnet) with high permeability and such low conductivity that eddy-current loss can be completely ignored. After Snoek's proposal [249,250] to use semiconducting spinel ferrites, which have been developed in the Philips Research Laboratories over the years (see e.g., Verwey and Heilman [5], Polder [251]), as high-permeability core materials, there has been a tremendous increase on research activities on their magnetic properties as outlined in the preceding chapters of this review. The resistivity of the majority of this class of materials was sufficiently high to neglect eddy-current losses in inductance applications and hence, resistance or dielectric impedance experiments were seemingly unnecessary and hence, rare. However, the increasing interest in microwave applications of ferrites stimulated careful dielectric characterization yielding rather astonishing and unexpected results with the observation of unusually large dielectric constants at low frequencies, finally promoting speculations about the existence of local dipolar moments in spinels.

More than 80 years ago, Blechschmidt [252] found an anomalous enhancement of the dielectric constants towards low frequencies in certain manganese ferrites. Comparing these results with similar measurements in inhomogeneous graphite-paraffin or copper-paraffin mixtures, he concluded that the manganese ferrites under investigation are inhomogeneous compounds consisting of metallic and insulating regions. More than a decade later, in the early fifties, large dielectric constants were measured in manganese-zinc ferrites by Brockman et al. [253], in nickel-zinc ferrites by Koops [254], in cooper-zinc ferrites by Möltgen [255], and in nickel and cobalt ferrites by Kamiyoshi [256]. These authors proposed that the strong frequency dependence of the dielectric constants reaching large or even colossal values at low frequencies can be explained assuming an inhomogeneous dielectric structure as discussed by Maxwell [257] and Wagner [258]. A simple model of grains of conducting material separated by poorly conducting sheets was able to explain some of the results [254] and the assumption of an electronic circuit consisting of a resistance and a capacitance in parallel, provided a qualitative understanding of the reported dispersion effects. Later on the dielectric properties of a variety of ferrites were summarized and analysed in detail by van Uitert [259] and by Fairweather et al. [260]. Finally, Peters and Standley [261] performed broadband dielectric experiments on magnesium-manganese ferrites ranging from 100 Hz to 100 MHz and again found significant dispersion effects and dielectric constants strongly increasing towards decreasing frequencies. These authors questioned the previous analysis given in terms of an inhomogeneous material with conducting and insulating particles and, for the first time, speculated about the existence of intrinsic relaxation phenomena of permanent dipole moments. A critical discussion of dielectric measurements in semiconducting materials and relaxation spectra as observed in transition-metal compounds can be found by Miles et al. [262]. The experimentally observed behaviour was described as being typical for conducting materials through which the free flow of current is impeded, for example, by grain boundaries and barrier layers. They showed that the dispersion caused by a Maxwell-Wagner type double layer is identical with that of a dipolar relaxation. In the late fifties, this debate about the existence and the origin of large dielectric constants in semiconducting spinels faded away without a final answer and apparently became forgotten. This discussion about large dielectric constants and possible intrinsic dipolar relaxations in spinel compounds was revived in the early seventies when speculations about possible off-centre positions of chromium ions in some oxide and sulphide spinels came up.

As a side remark, it is worth mentioning that in the first decade of the 21$^{stt}$ century this debate on colossal dielectric constants (CDCs) suddenly became revived by the observation of CDCs in perovskite-derived transition metal oxides [263]. Large dielectric constants of order $10^6$ were observed in $CaCu_3Ti_4O_{12}$ in a broad range of frequencies and temperatures, with Arrhenius-type relaxation dynamics on decreasing temperatures. The temperature and frequency dependencies of the complex dielectric response were interpreted in terms of a relaxor-like dynamical slowing down of dipolar fluctuations of nanosize domains [263]. In subsequent work, structural and electronic properties of $CaCu_3Ti_4O_{12}$ were calculated using density-functional theory within the local spin-density approximation. No direct evidence was found for intrinsic lattice or electronic mechanisms rationalizing this polar behaviour, a result rather pointing towards an extrinsic effect of the observation of CDCs [264]. Indeed, in further detailed dielectric experiments on this compound [265,266] it was proven that the CDCs and the concomitant relaxation phenomena can be explained by Maxwell-Wagner type phenomena due to internal or external barrier-layer capacitors, i.e., due to grain boundaries or surface-layer effects at the electrodes. In addition, it was documented that this finding of CDCs is quite common in semiconducting transition-metal oxides [267]. Nevertheless, this early work on CDCs stimulated enormous research activities, which were reviewed, e.g. by Lunkenheimer et al. [268].

The observation of CDCs and their possible origin in local dipolar moments in spinel compounds in principle can be viewed as precursor phenomena of possible polar order in spinel compounds. The true story about ferroelectricity in spinels started with early ESR [269] and optical spectroscopy investigations [270] of chromium-doped $MgAl_2O_4$. In this material, the chromium ions ($Cr^{3+}$, $3d^3$) substitute aluminium at the $B$ site of the normal spinel structure and are embedded in a regular octahedron of oxygen ions. Hence the three electrons of the chromium $d$ shell occupy the three $t_{2g}$ levels of the electronic ground state (see Fig. 6) and constitute a spin-only value of $S = 3/2$ with negligible SOC. The octahedral site symmetry for an ideally cubic spinel is $D_{3d}$, which indeed was identified in the low impurity limit. However, for chromium concentrations > 2 %, both sets of data, ESR as well as optical results, were consistently explained by significant trigonal distortions of the local crystal field along the crystallographic [111] direction [269,270]. Later on, performing detailed structural and optical investigations in the mixed molecular compounds $MgCr_xAl_{2-x}O_4$ Grimes and coworkers [271,272,273] substantiated these findings of a locally distorted $C_{3v}$ symmetry at the chromium site. It was concluded that the chromium ions in this series of alloys behave like JT ions, despite the fact that chromium has a half-filled shell, characterized by zero or small SOC and no JT activity. Grimes and Collett [273] documented that the infrared spectra exhibit increasing complexity with increasing chromium concentration and that the IR spectra in accordance with the structural data can be satisfactorily explained assuming local distortions with $C_{3v}$ symmetry at the chromium site. It was concluded that the over-all cubic symmetry of the spinel structure and its optical isotropy are the consequence of a statistical distribution of local trigonal distortions along all symmetry-equivalent <111> directions. This observation together with an increasing number of experimental evidence led Grimes [274] finally to the conclusion that the spinel compound $MgCr_2O_4$ is not satisfactorily described by the space group $Fd\bar{3}m$, but most likely exhibits the space group $F\bar{4}3m$. It was further documented that the displacements from conventional positions of the cubic spinel structure cannot be larger than 0.10 Å and it was stated that the microscopic origin of this trigonal distortion remains to be settled. Finally, the claim that a large number of spinel compounds contrary to conventional description cannot be referred to as space group $Fd\bar{3}m$, was further substantiated by a detailed analysis of the Debye-Waller factors of a large number of spinel compounds [275], including a number of ferrites, cobaltates and chromates, with $ZnCr_2O_4$ and $MgCr_2O_4$ being the most prominent examples. These found significant deviations from cubic symmetry, the observation of a possible non-centrosymmetric space group together with anomalous dielectric results discussed above, led

Grimes finally to conclude about a polar, specifically antiferroelectric ground state in some spinel compounds [276]. This idea of a space group $F\bar{4}3m$ with no inversion center in some spinel compounds was further elaborated by Schmid and Ascher [277]. The occurrence of off-centre chromium was rationalized by electrostatic potential considerations in semiconducting spinel compounds. Hence, it seemed important to check the symmetries in metallic spinel compounds, where metallic electrical conductivity might help to retain the high cubic symmetry. Williams and Grimes [278] published a detailed x-ray diffraction study of the metallic thiospinels $CuCr_2S_4$ and $CuCo_2S_4$ and indeed found that the space group $Fd\bar{3}m$ describes the structure of these metallic compounds satisfactorily.

This final proposal about a polar ground state of some insulating or semiconducting spinel oxides started a long-standing dispute with a large number of pros and cons, which, to our knowledge, up to now, has not been settled satisfactorily and will be shortly summarized below. We would like to recall, that as outlined in chapter 3.4.1 B, various chromium oxide spinels are strongly frustrated magnets and, according to recent wisdom, undergo a spin-driven JT effect, with a structural distortion concomitantly with magnetic ordering. However, these effects arise at low temperatures in the magnetically ordered state. As outlined above, many of the chromite spinels instead reveal this so-called JT activity already at high temperatures, far above the onset of magnetic order, a fact that has been widely overlooked in the recent literature and has neither been discussed nor was taken into consideration in all theoretical models dealing with the spin-JT effect. In what follows, we shortly summarize the controversy on the assignment of the correct space group in spinel compounds: The controversy mainly referred to the observation of "forbidden" reflections in diffraction experiments and the question whether their occurrence is simply due to multiple-scattering events or is a hallmark for lower symmetry without inversion centre. Utilizing neutron-diffraction experiments on $Fe_3O_4$ and on $MgAl_2O_4$, Samuelson [279], Samuelson and Steinsvoll [280,281] could not observe intrinsic {200} reflections indicative for space group $F\bar{4}3m$ and they concluded that the loss of inversion symmetry must be based on other evidence than the presence of these forbidden reflections. This conclusion was confirmed by Smith [282], performing electron-diffraction experiments on a natural spinel of composition $Mg_{0.85}Fe_{0.15}Al_2O_4$ and by Tokonami and Horiuchi [283] as well as De Cooman and Carter [284] again with electron diffraction on $MgAl_2O_4$. These conclusions concerning the space group $Fd\bar{3}m$ derived from diffraction experiments were further confirmed by electron-nuclear double-resonance experiments on $Cr^{3+}$ in natural $MgAl_2O_4$ [285]. The observation of intrinsic "forbidden" reflections, which were not explained by double or multiple diffraction processes were reported utilizing electron diffraction [286,287,288,289] as well as x-ray diffraction techniques [290]. Results from lattice-energy calculations [291] also supported the loss of inversion symmetry and $F\bar{4}3m$ being the appropriate symmetry for $MgAl_2O_4$. From our point of view, this dispute about the true symmetry of some of the spinel compounds was never finally settled. Over the years there were a number of experimental observations, which will be detailed and mentioned below, indicative for some structural effects in a variety of spinels, like off-centre chromium ions, which appear at elevated temperatures in the ~150 K range and could be interpreted in terms of a loss of inversion symmetry.

3.4.2. Concomitant polar and magnetic order in spinels

The possibility of the existence of a linear coupling of electric and magnetic fields for substances with certain types of magnetic crystal symmetry was known since Landau and Lifshitz. Antiferromagnetic $Cr_2O_3$ was the first material, which was predicted [292] and has been proven to show this linear ME effect [293]. In a ME material with linear coupling, the magnetization depends on the external electric field and the polarization on the external magnetic field. The cross coupling between the magnetic and electric order parameters provides an additional degree of freedom, which seems promising to design future electronic devices. ME materials are limited by symmetry requirements, where spatial inversion as well as time reversal symmetry need to be broken simultaneously. A linear ME effect can exist in materials without spontaneous electric polarization as observed in AFM $Cr_2O_3$ and were routinely reported in MF materials with concomitant magnetic and polar order. In recent years a number of $A$-site spinels, including $Co_3O_4$, $MnAl_2O_4$, $MnGa_2O_4$ [294] and $CoAl_2O_4$ [295] were recorded. In the latter compound, the authors documented that the linear ME effect depends on the $A$-site disorder and becomes suppressed with increasing disorder. This fact supports the conclusion that $CoAl_2O_4$ with low site disorder exhibits a long-range ordered AFM state at low temperatures [295]. In this review, we mainly will focus on MF spinels exhibiting concomitant spin and polar order.

The attempts to synthesize a material exhibiting both, magnetic and polar order (as homogeneous bulk samples or as heterogeneous ceramics) started in the 1960s, mainly by the groups of Smolenskii [296] and Venevtsev [297,298] in the former Soviet Union. Materials combining these different ferroic properties were later on called multiferroics. It was argued that in these MF materials the cross coupling of magnetization to electric fields and of polarization to magnetic fields would be strongly enhanced allowing incorporation into practical devices providing novel device paradigms for future technological applications. After these early attempts and being quiet silent for many years, in the first decade of the 21$^{st}$ century the research on multiferroics was highly accelerated by a revived discussion on novel materials. It started with the observation of a relatively strict exclusion principle of magnetic and ferroelectric (FE) order, mainly focusing on the broad class of perovskites: In this class of compounds the canonical ferroelectrics, e.g., $BaTiO_3$, are characterized by a $d^0$-ness, while magnets, like e.g. $LaMnO_3$, are found in systems with incompletely filled $d$ shells. These facts were summarized and elaborated in an illuminating article by Hill [299]. In addition and shortly after, spin-driven ferroelectricity was found in $TbMnO_3$ [300] where the onset of spin-spiral order destroys inversion symmetry and allows for FE polarization in the magnetically ordered phase. Ferroelectricity at the onset of spin-spiral order theoretically was explained in terms of a spin-current mechanism [301] or by an inverse Dzyaloshinskii–Moriya (iDM) interaction [302,303]. In these models, spin-driven ferroelectricity with macroscopic polarization $\boldsymbol{P}$ is induced at the spin-spiral ordering transition between adjacent magnetic moments $\boldsymbol{S}_i$ and $\boldsymbol{S}_j$ via:

$$\boldsymbol{P} \sim \boldsymbol{e}_{ij} \times (\boldsymbol{S}_i \times \boldsymbol{S}_j)$$

Here $e_{ij}$ is the unit vector connecting the two sites $i$ and $j$ occupied by localized magnetic moments. This vector product immediately makes clear that helical spin structures, with the plane of rotation of the spins perpendicular to the propagation vector (i.e., $\boldsymbol{S}_i \times \boldsymbol{S}_j \parallel \boldsymbol{e}_{ij}$) are chiral but not polar. On the other hand, cycloidal spin structures with the propagation vector within the basal plane induce macroscopic FE polarization. Later on it was shown that there exist many more possibilities to induce spin-driven ferroelectricity, for instance, via an exchange-striction mechanisms [304] or via charge ordering [305]. These paths to spin-driven polar order with a list of the most prominent examples are summarized and discussed in detail in [304]. There exists a tremendous amount of work on this class of spin-driven multiferroics and any detailed survey would largely overburden this review. In addition, there is a number of excellent reviews focusing on the basic principles and possible future applications of multiferroics in bulk materials and as solid thin films [304,306,307,308,309,310]. Here we will concentrate on ME effects or on multiferroicity as observed in spinel compounds only. Very recently a review on ME and MF properties of spinel compounds, including the large class of lacunar spinels has been published by Athinarayanan and Ter-Oganessian [311].

## A) Spin driven ferroelectricity via an inverse Dzyaloshinskii–Moriya mechanism

$ZnCr_2Se_4$: The first reports of the observation of a significant ME effect in spinel compounds concerned $ZnCr_2Se_4$. As outlined above, this chromium selenide undergoes spiral spin ordering below 21 K (see Table 1). The IC spin helix propagates along one of the cube axes with the plane of rotation of the spins perpendicular to the propagation vector and hence, in zero external magnetic field will be non-polar. The screw spin structure with a unique propagation vector has still two domains with clockwise or counter-clockwise rotation of the spins. Utilizing neutron-diffraction techniques, Siratori et al. [312] were able to control the evolution of screw domains by ME cooling, i.e., by cooling the sample through the magnetic-ordering temperature in the presence of combined external magnetic and electric fields. Later on Siratori and Kita [313] measured the temperature and the magnetic-field dependence of the ME coupling. An external magnetic field will induce a conical spin structure and, in addition, can tilt the cone axis away from the propagation vector. In this way, a macroscopic FE polarization can be induced, if the external magnetic field is not parallel to the propagation vector. Indeed, Siratori and Kita [313] measured the ME coupling coefficient as function of the cone axis with respect to the external magnetic field. Almost 30 years later a MF characterization of $ZnCr_2Se_4$ was performed in full detail by Murakawa et al. [314]. These authors measured the FE polarization with respect to the angular dependence of the crystallographic orientation of the crystal with respect to the external magnetic field. According to the standard rules of the iDM interaction, in a helical spin phase, with the spins rotating in a plane perpendicular to the propagation vector, no FE polarization is induced. This is also true for conical spin structures, derived by external magnetic fields parallel to the propagation vector. However, finite polarization is induced, when the magnetic field is tilted with respect to the propagation vector and induces a tilting of the cone axis with respect to the propagation vector. This model behaviour of the spin-driven ferroelectricity has been impressively documented in Ref. [314].

We have already discussed earlier that the chromium-oxide spinels $MnCr_2O_4$ and $CoCr_2O_4$ crystallize in the normal spinel structure with values of the effective exchange parameter $u$, which according to the LKDM theory are in the locally unstable region (see Fig. 7). These compounds undergo a FiM-type of spin-spiral ordering, where the magnetic moments of the ions at the $A$ and the $B$ site form FM conical spirals, with their longitudinal FM moments pointing opposite to each other [102]. The polar ground state and multiferroicity was investigated in detail by Yamasaki et al. [315] in $CoCr_2O_4$. This compound reveals a FiM transition at 93 K and on further cooling undergoes a transition into cycloidal spin states at 26 K, where both spirals evolve into longitudinal conical structures with opposite spin directions as described above. This means that the ground state exhibits a uniform spontaneous magnetization (like in conventional ferrimagnets) plus a transverse spin-spiral component. The IC propagation of the spirals is along the crystallographic [110] direction, while the basal plane of the spin rotation is in one of the cube planes, perpendicular to [001]. Hence, according to the vector rules of the iDM interaction there should appear spin-driven FE polarization along [$\bar{1}10$]. At the transition to the conical spin-spiral state at 26 K this polarization indeed was observed [315]. In addition, due to the existence of a spontaneous FM magnetization due to FiM order of the longitudinal spin components it was possible to reverse the polarization by the reversal of rather small external magnetic fields of order 0.1 T.

$LiCuVO_4$ is a very well-documented example of a spin-driven FE spinel. It crystallizes in an orthorhombically distorted inverse spinel structure. The nonmagnetic $V^{5+}$ ions occupy the tetrahedrally coordinated $A$ sites, while nonmagnetic $Li^+$ and $Cu^{2+}$ ($3d^9$ configuration with $S = 1/2$) occupy the $B$ positions within the oxygen octahedra of the spinel structure in a completely ordered fashion. The orthorhombic distortion results from a cooperative JT effect of the $Cu^{2+}$ ions at the octahedral sites. In this inverse spinel structure, the CuO$_6$ octahedra form independent and infinite chains along the crystallographic $b$ direction and the title compound behaves like a one-dimensional $S = 1/2$ Heisenberg antiferromagnet. The magnetism of LiCuVO$_4$ is well characterized by magnetic susceptibility [316,317], NMR [318,319,320], ESR [317, 318] and neutron scattering techniques [321]. It reveals the characteristic magnetic susceptibility of a low-dimensional spin system and, due to weak inter-chain couplings, below ~ 2.5 K exhibits three-dimensional long-range magnetic order. At this phase transition, a cycloidal spin order is established, with a propagation vector of the spiral along $b$ and with the spin-rotation plane within the $a,b$ plane, with the normal vector $e$ parallel to the crystallographic $c$ direction. Hence, according to the symmetry rules of spiral magnets, polar order is expected with FE polarization $P$ aligned along the $a$ direction.

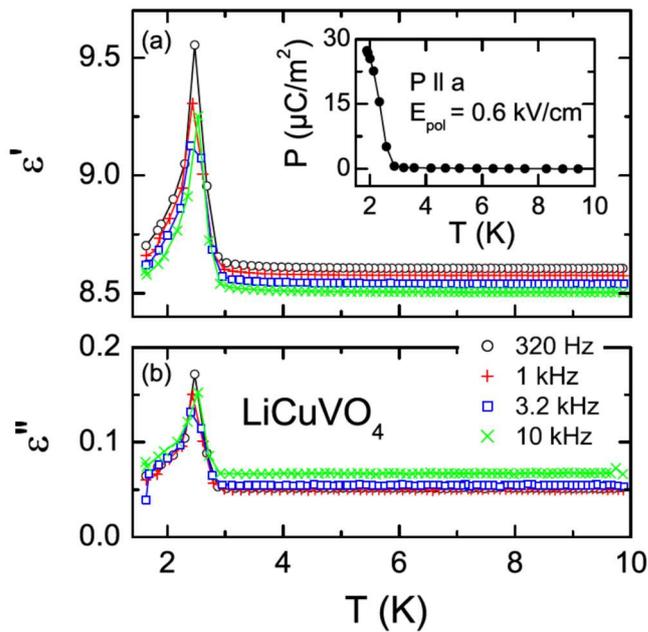

Fig. 38. Complex dielectric constants in LiCuVO$_4$.
Temperature dependence of (a) dielectric constant and (b) dielectric loss for $T \leq 10$ K, for a series of frequencies with the electric field $E$ along the crystallographic $a$ direction. The inset shows the temperature dependent polarization $P$ along $a$, measured via the pyrocurrent after polarizing the sample during cooling with an electric field of 0.6 kV/cm.
Reprinted figure with permission from Schrettle et al. [324]. Copyright (2008) by the American Physical Society.

In subsequent dielectric experiments this polar order indeed was observed [322,323,324]. Fig. 38 shows the temperature dependence of the real [Fig. 38(a)] and the imaginary part [Fig. 38(b)] of the dielectric constant at a series of frequencies between 320 Hz and 10 kHz below 10 K. At about 2.5 K, the real part of the dielectric constant shows a $\lambda$-like anomaly and a concomitant loss peak with very little frequency dependence. The absolute values of dielectric constant and of the loss close to the peak maxima are relatively low and do not indicate any kind of dipolar criticality, as would be observed in ferroelectrics driven by soft phonon modes or in order-disorder FE phase transitions. These small and frequency-independent anomalies, without any precursor phenomena like dipolar fluctuations in the paraelectric (PE) phase, are characteristic for an incipient FE, where polar order is not the primary order parameter, but driven and induced by another ordering phenomenon, here the onset of spin order. The inset of Fig. 38(a) shows the temperature dependence of the FE polarization, which evolves along the crystallographic $a$ direction as expected by applying the symmetry rules of the iDM interaction. To probe FE order, the pyro-current was integrated to determine the temperature dependence of the spontaneous polarization [324].

In a series of detailed dielectric experiments as function of temperature and external magnetic field, as well as in a series of pyro-current and magneto-current experiments [324] a complex ($H,T$)-phase diagram of LiCuVO$_4$ was established, which is shown in Fig. 39. At low magnetic fields (H < 2 T), as long as the anisotropy of the system dominates the spin structure, the basal plane of spin helix always lies within the $ab$ plane ($e \parallel c$). As soon as the external magnetic field overcomes the magnetic anisotropy of the system, in fields approximately above 2 T, the basal plane of the spin helix follows the external magnetic field. In Ref. [324] the authors were able to document that the polar order can be switched by switching the external magnetic field with a concomitant switching of the FE polarization. Hence, when the spin helix follows the external magnetic field, also the polarization becomes switched following the vector rules of the iDM interaction. Overall, the findings in this system provide a nice confirmation of the theoretically predicted symmetry relations for spin-driven FE in spiral magnets. Finally, above 8 T the spin system undergoes a modulated collinear ordering with no induced spin-driven polarization, and the title compound remains PE down to the lowest temperatures.

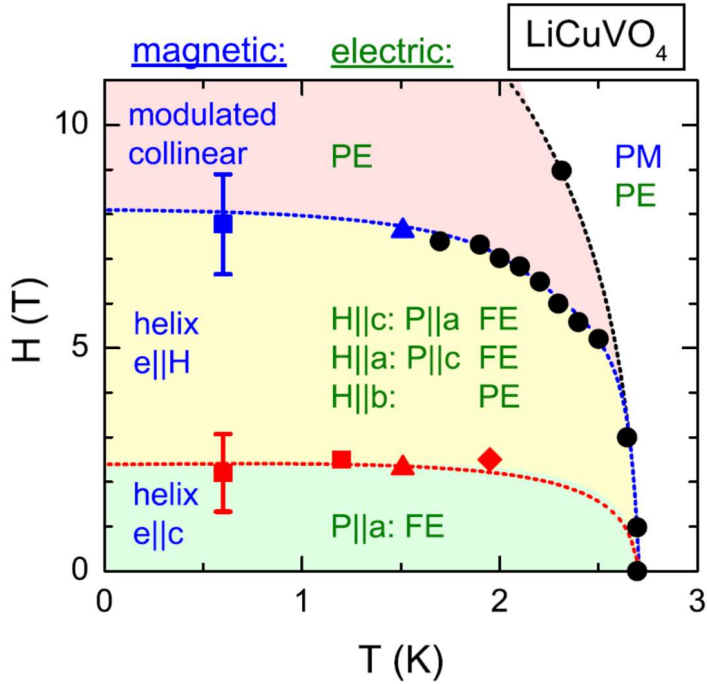

Fig. 39. ($H,T$)-phase diagram of LiCuVO$_4$.
Phase diagram determined from a series of field and temperature-dependent dielectric measurements at temperatures below 3 K and external magnetic fields below 10 T. Polar and magnetic phases are indicated with respect to the different directions of electric and magnetic fields.
Reprinted figure with permission from Schrettle et al. [324]. Copyright (2008) by the American Physical Society.

B. Multiferroic spinels with unconventional ferroelectricity

In this chapter, we will discuss MF behaviour of spinel compounds, which cannot be straightforward explained by a canonical spin-driven mechanism, i.e. by an iDM effect. Instead, probably different routes to polar order have to be taken into account, like charge- and/or orbital-ordering phenomena. However, other spin-driven mechanisms, like exchange-striction, also could play a role and in many cases the true driving force for polar order is not finally established. In this respect, the dipolar relaxation dynamics due to off-centring of chromium ions, have largely been overlooked and may play a more important role as a precursor phenomenon of a polar ground state, which becomes enhanced and activated via the onset of magnetic interactions or via charge ordering phenomena.

Fe$_3$O$_4$, the most popular spinel ferrimagnet, undergoes FiM ordering close to 780 K and the Verwey transition [30], a complex type of MIT close to 125 K [325]. Early on there were reports on ferroelectricity and large ME effects in magnetite: First observations of linear and bilinear ME effects observed in magnetically biased Fe$_3$O$_4$ were published by Rado and Ferrari [326]. Subsequently, Kita et al. [327,328] reported measurements of FE hysteresis loops in magnetite at low temperatures, well below the Verwey transition. Finally the temperature dependence of the ME effect under applied strain was measured in [329,330] and in [331] it was shown that the ME polarizations can be successfully reversed by reversing the dc electric field, documenting that magnetite at low temperatures is a true and switchable FE. A detailed dielectric spectroscopy study below the charge-ordering Verwey transition, in a wide range of probing frequencies covering more than 10 decades of frequency, was reported by Schrettle et al. [332]. Their main result of the temperature dependence of the real part of the dielectric constant below the Verwey transition, for a series of measuring frequencies between mHz and GHz, is documented in Fig. 40. On cooling, for the higher frequencies ε'(T) shows a single steplike decrease from the static dielectric constant $ε_s$ of the order of several thousands to its high-frequency limit $ε_∞ ≈ 60$. This feature shifts to lower temperatures with decreasing frequency. From the fact, that this step-like decrease strongly depends on the electrical contacts used for these experiments, it was concluded that the large or even CDCs of $Fe_3O_4$ at elevated temperatures, result from non-intrinsic Maxwell-Wagner like effects [332], as have been observed in a variety of semiconducting transition-metal oxides [267].

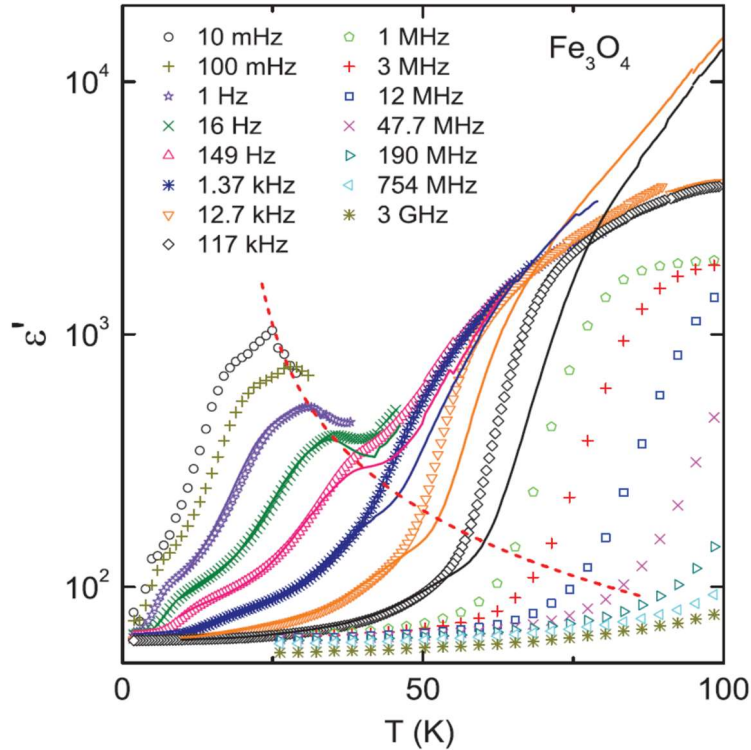

Fig. 40. Dielectric constant in $Fe_3O_4$. ε'(T) of magnetite for various frequencies between mHz and GHz as indicated in the figure obtained with silver-paint (symbols) and sputtered gold contacts (solid lines). The dashed line was calculated assuming a CW law and illustrates the estimated temperature dependence of the static dielectric constant of the intrinsic relaxation. Reprinted figure with permission from Schrettle et al. [332]. Copyright (2011) by the American Physical Society.

For frequencies below about 100 kHz, a second relaxation-like phenomenon appears in the temperature dependence of ε'(T) below 50 K, developing into a broad cusp-like maximum for the lowest frequencies. With decreasing frequency, the height of this peak increases, while shifting to lower temperatures. This is the typical behaviour of relaxor ferroelectrics, which are characterized by a diffusive polar phase transition and the freezing-in of short-range cluster-like FE order [333]. In these relaxor-like materials, the dipolar degrees of freedom are believed to freeze-in with local but not long-range order, establishing a nano- or micro-domain polar phase at low temperatures. They usually show sizable polarization and switchable ferroelectricity. It was concluded that FiM magnetite has a polar ground state, however with short-range FE order only. The temperature dependence of the peak maxima as documented in Fig. 40 follows a CW law with a critical ordering temperature, which is located at rather low temperatures. However, from Fig. 40 it is clear that the freezing process of the dipolar degrees of freedom is very slow and below 20 K the dipolar dynamics is characterized by time constants $τ$ well beyond 100 s [because $τ = 1/(2πν)$]. Hence, for any realistic experimental time scale the structure can be viewed as quasi-static and exhibiting short-range polar order. The coexistence of switchable and macroscopic FE polarization with macroscopic magnetic hysteresis has been documented by Alexe et al. [334] in $Fe_3O_4$ thin films. A representative result is shown in Fig. 41. Sizable FE polarization appears at temperatures below 38 K and reaches saturation values close to 11 μC/cm². The squared temperature dependence of the polarization [inset in Fig. 41(a)] has been interpreted as fingerprint of a second order FE phase transition with a critical temperature of 38 K. Interestingly, this critical temperature is very close to a possible divergence of the dipolar relaxation strength as outlined in [332] and as documented in Fig. 40. From this relaxor-like relaxation it is clear that polar hysteresis effects should be observable on macroscopic time scales below ~ 30 K. Figs. 40 and 41 document that $Fe_3O_4$ indeed is a MF with switchable magnetization and switchable polarization. However, sizable MF order sets in deep in the magnetically ordered phase and well below the Verwey transition. It seems clear that the onset of polar order is driven by charge-ordering phenomena. However, it still is unclear why macroscopic polarization appears far below the onset of charge order (CO), which definitely sets in at the Verwey transition.

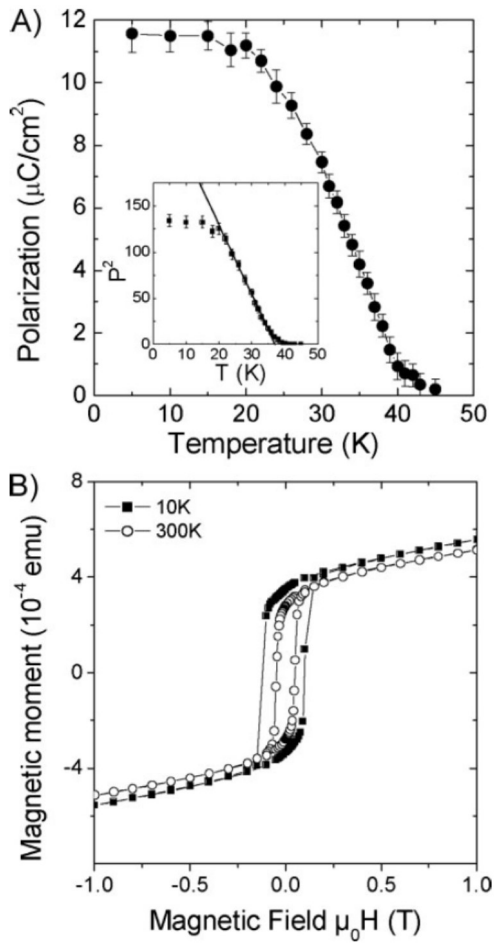

Fig. 41. Multiferroic properties of a $Fe_3O_4$ thin film.
A) Ferroelectric polarization $P$ versus temperature. The inset documents a plot of $P^2$ vs. temperature and a linear fit suggesting a second-order phase transition and a critical phase-transition temperature close to 38 K.
B) Magnetization hysteresis curves (magnetization vs. magnetic field) recorded at 10 and 300 K. On cooling, the magnetic hysteresis becomes slightly narrower, however, with similar values of the saturated magnetic moment.
Reprinted figure with permission from Alexe et al. [334]. Copyright (2009) John Wiley and Sons.

These results certainly may help to resolve the long-standing problem about the low-temperature crystal symmetry of magnetite: For cluster-like ferroelectricity, the inversion symmetry can be lost on a local scale, while on the average monoclinic symmetry can be retained [335]. These findings also could have important implications for the CO in magnetite: It's polar and charge degrees of freedom are intimately related and, thus, the CO should also be of short-range type only. This fact seems to be even more plausible, having in mind that charge ordering happens on a strongly frustrated pyrochlore lattice. Anyway, charge ordering seems to be the driving force of ferroelectricity established in FiM $Fe_3O_4$ and the model developed by Efremov et al. [336] to explain ferroelectricity in certain charge-ordered perovskites may well be appropriate for magnetite below its Verwey transition [306]. By means of first-principles simulations, it was unambiguously shown that improper ferroelectricity in magnetite is driven by charge-ordering phenomena and that the polarization arises because of charge disproportionation of electronic charges between octahedral Fe sites, leading to a non-centrosymmetric $Fe^{2+}/Fe^{3+}$ charge-ordered pattern [337]. Finally, the complexity of this charge ordering transition was further elucidated by Attfield and coworkers in a series of fundamental contributions [325,338,339], which will be discussed later in this review in the chapter dealing with the MIT in spinel compounds.

$CdCr_2S_4$: An astonishing and rather unexpected experimental result was the observation of polar order and FE polarization in FM $CdCr_2S_4$ [32]. The most important findings elucidating its MF behaviour are documented in Fig. 42. $CdCr_2S_4$ is dominated by FM exchange interactions, documented by a positive CW temperature of 155 K and exhibits long-range FM order below ~ 85 K [see Fig. 42(a) and Tab. 1], with an ordered moment of 6 $\mu_B$ per formula unit [see inset in Fig. 42(a)] as expected for two $Cr^{3+}$ ions per unit cell, that are octahedrally coordinated and have half-filled $t_{2g}$ levels (see Fig. 6). FE polarization evolves in the magnetically ordered phase and practically follows the evolution of spontaneous FM magnetization, however, slightly shifted to lower temperatures [Fig. 42(b), left scale]. The onset of FE polarization is accompanied by a marked anomaly in the temperature dependence of the dielectric constant [Fig. 42(b), right scale]. The dielectric constant passes through a broad maximum well below the onset of FE ordering, a temperature evolution significantly different to what is expected in second-order displacive polar phase transitions, where a maximum dielectric constant is expected just at the ordering temperature. At 25 K, well below the MF ordering transition a FE polarization hysteresis was measured [Fig. 42(c)], supporting an interpretation in terms of macroscopic polar order.

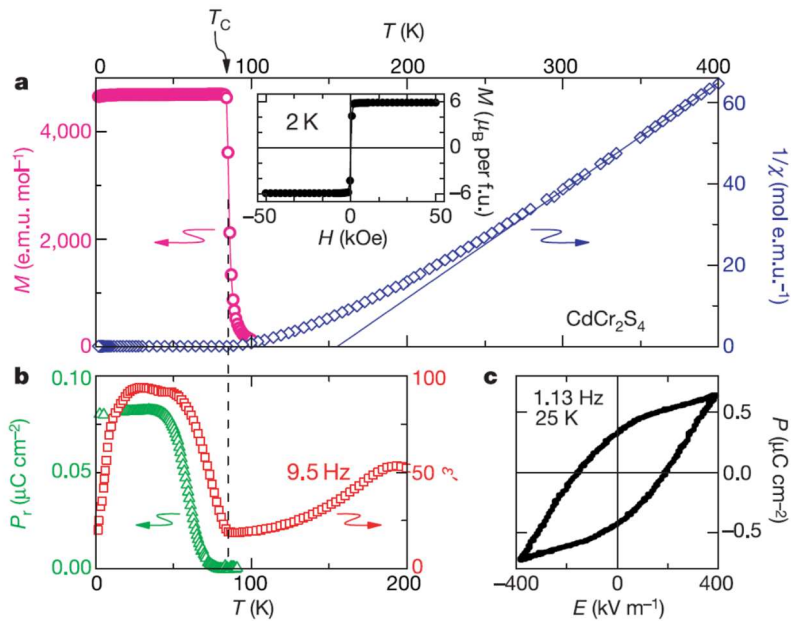

Fig. 42. Magnetic and dielectric characterization of $CdCr_2S_4$. (a) Right scale: inverse magnetic susceptibility versus temperature measured at 100 Oe. The solid line indicates a fit for $T > 300$ K) using a Curie–Weiss law, with a CW temperature of 155 K. Left scale: magnetization versus temperature measured at 100 Oe. Inset: ferromagnetic hysteresis at 2 K, indicating a saturated moment of 6 $\mu_B$ per formula unit (f.u.). (b) Right scale: temperature dependence of the real part of the dielectric constant measured at a frequency of 9.5 Hz. Left scale: remnant ferroelectric polarization versus temperature measured after cooling in an electric field of 50 kV/m. (c) Polarization versus electric field, showing a ferroelectric hysteresis deep within the multiferroic phase. Material from: Hemberger et al. [32]. Copyright © 2005, Springer Nature.

Despite the fact that the polar order detected in $CdCr_2S_4$ obviously appears below the onset of FM order and in this sense certainly has the characteristics of a spin-driven ferroelectricity, from the very beginning it was clear that a conventional spin-driven mechanism via an iDM interaction as outlined above can be excluded and one has rather to think of polar order driven by an exchange-striction mechanism. However, $CdCr_2S_4$ belongs to the broad class of spinel compounds, which seem to be prone to polar order already at much higher temperatures, most probably driven by an off-centre motion of the Cr ions in double-well potentials (see Chapter 3.4.1). For some of these sulphide and selenide spinels, it was argued that the relatively large radii of the *A*-site ions destabilize the cubic structure and support the tendency towards ionic off-centre displacements. Indeed, utilizing detailed x-ray diffraction experiments, for $CdCr_2S_4$ and $CdCr_2Se_4$ negative thermal expansion and an anomalous broadening of Bragg reflections was detected well above the magnetic ordering temperature, indicating the softening of the spinel structure of these compounds towards lower temperatures [340].

Having these claims of possible polar order in spinel compounds in mind, magneto-capacitive effects in $CdCr_2S_4$ were studied in detail already well above the onset of FM order [32,341]. The temperature and field-dependent dielectric constants are shown in Fig. 43 in a broad range of temperatures, magnetic fields and probing frequencies. Indeed, very unusual dipolar relaxations as function of temperature and magnetic fields were found, indicative for the existence of reorienting dipolar moments already at elevated temperatures. Starting below room temperature, a significant dipolar relaxation was observed, which appears as a broad hump-like peak in the real part [Fig. 43(a)] and as a symmetrically shaped maximum in the imaginary part of the dielectric constant [inset of Fig. 43(a)]. On cooling, these features reveal a strong increase and are shifted to lower temperatures. This observation was explained by a relaxor FE-type dielectric function with a characteristic temperature of 135 K, where the static dielectric constant would diverge in case thermodynamic equilibrium could be reached. Interestingly, this critical temperature is not too far from the onset of long-range magnetic order and it is clear that the dipolar relaxations become very slow and the sample should fall out of thermodynamic equilibrium at somewhat higher temperatures [the mean relaxation time is of the order of 1s already at 151 K, see inset of Fig. 43(a)]. From these experiments [32,341] it was concluded that in $CdCr_2S_4$ only short-range polar order in nano- or mico-domains can be established. However, it is evident that this dipolar relaxational behaviour is another fingerprint of a possible FE behaviour of spinel compounds as discussed earlier in this chapter and it seems obvious that short-range ordered dipolar moments are present already well above the onset of long-range magnetic order, an observation which certainly makes $CdCr_2S_4$ an unconventional MF. As becomes obvious from Fig. 43(b), this spinel compound reveals a very large ("colossal") magnetocapacitive coupling. While the relaxor behaviour at $T > T_c$ is almost unaffected by an external magnetic field, around $T_c$ strong magnetocapacitive effects show up. The corresponding relative magnetocapacitance, shown in the inset of Fig. 43(b), reaches up to 450%.

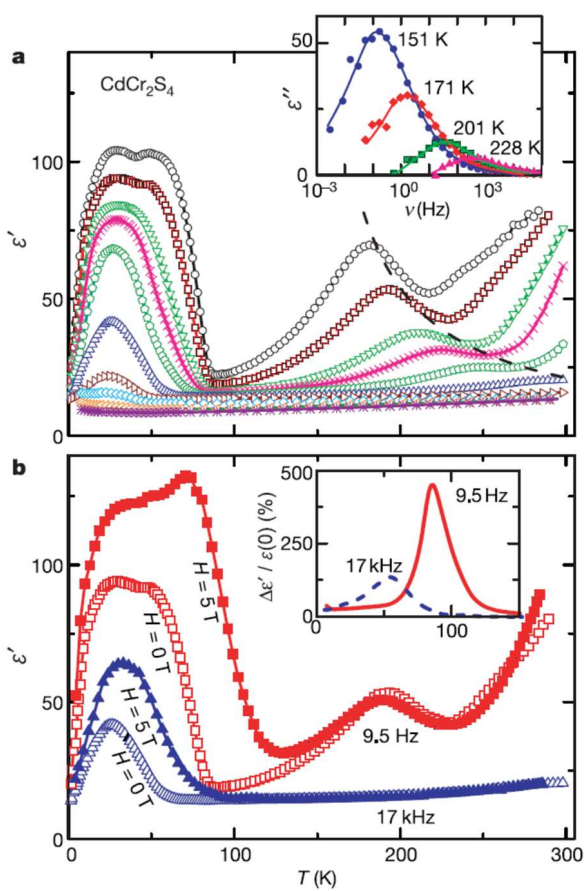

Fig. 43. Magnetocapacitive behaviour of $CdCr_2S_4$. (a) Temperature dependence of $\varepsilon'$ at various frequencies (from top to bottom: 3 Hz, 9.5 Hz, 53 Hz, 170 Hz, 950 Hz, 17 kHz, 950 kHz, 12 MHz, 83 MHz, 3 GHz). The dashed line indicates the static dielectric susceptibility following a Curie–Weiss-like law with a characteristic temperature of 135 K. The inset shows $\varepsilon''(\nu)$ for various temperatures (lines are drawn to guide the eye). (b) Dielectric constant versus temperature at 9.5 Hz and 17 kHz, measured at zero field and in an external magnetic field of 5 T, directed perpendicular to the electric field. The inset provides a measure of the magnetocapacitive effects, with $\Delta\varepsilon'/\Delta\varepsilon'(0T)$, for the two frequencies shown in the main panel.
Material from: Hemberger et al. [32]. Copyright © 2005, Springer Nature.

The relaxation dynamics of the dipolar entities in $CdCr_2S_4$ was further detailed using broadband dielectric spectroscopy [341]. In this work, the authors provided strong evidence that the variation of the dielectric constant at the magnetic transition and the concomitant colossal magnetocapacitive effect are caused by a drastic speeding-up of the polar relaxation dynamics under the formation of FM spin order. Hence, the dipolar relaxations freeze-in already well above the onset of magnetic order, forming a relaxor-like polar state. However, with the onset of FM order the relaxation dynamic becomes significantly faster again allowing for a macroscopic growth of the micro-domains documented by the FE hysteresis experiments shown in Fig. 42(c). Stimulated by these results, in subsequent experimental work the magnetic field-dependent dielectric and magnetic properties in $CdCr_2S_4$ were investigated by different groups [342,343,344]. Sun et al. [342] reported on the observation of a glassy polar state at the onset of FM order, followed by subsequent FE order close to 56 K, which is further enhanced by external electric fields. The authors explained these field dependent MF states by strong spin-lattice coupling and exchange striction effects. In subsequent work, this group reported the observation of colossal magnetoresistance and colossal electroresistance effects, which are induced by external electric fields in $CdCr_2S_4$ [343]. Utilizing local probe techniques, Oliveira et al. [344] established that the dynamic off-centring of the $Cr^{3+}$ ions play a key role in establishing a MF ground state in this material.

Similar observations of dipolar relaxations and large magnetocapacitive effects were also reported for FM $CdCr_2Se_4$ [345], for $HgCr_2S_4$ with helical spin order [151,191] and for FM indium doped $CdCr_2S_4$ [151]. We would like to recall that $CdCr_2S_4$, indium doped $CdCr_2S_4$ and $CdCr_2Se_4$ exhibit long-range ordered FM ground states. Only $HgCr_2S_4$ exhibits a helical spin structure, however with the basal plane perpendicular to the propagation vector. In addition, very small fields switch the mercury compound into an induced FM state. We conclude that in these compounds the conventional iDM mechanism cannot induce polar order. It always was suspected that these spinel compounds are close to ferroelectricity due to relaxing off-centre chromium moments and that finally the onset FM spin order induces polar ground states. This obviously happens via a drastic acceleration of the relaxation dynamics with the onset of FM order, which appeared to be almost frozen-in at temperatures well above the magnetic phase transition.

In this review, we do not want to dissimulate concerns about a possible extrinsic nature of the reported polar effects. This concerns were also triggered by reports that in polycrystalline samples these significant magneto-capacitive effects are very small or even missing [151]. From the very beginning, after observing these effects in $CdCr_2S_4$, the authors of Ref. [32,341] tried to prove the intrinsic nature of the reported effects and to provide experiments to exclude Maxwell-Wagner-like phenomena or that the reported anomalies stem from inhomogeneous samples. It is important to recall that these authors belonged to the first groups being aware of the importance of Maxwell-Wagner phenomena in transition-metal oxides [266,267]. In due course, they provided a variety of experimental results on differently synthesized samples, differently doped samples, and measurements with different electrical contacts and different contact geometries [57,151,345] to exclude extrinsic effects and finally concluded about the intrinsic character of the reported results in $CdCr_2S_4$. The first criticism was raised in a brief communication arising by Catalan and Scott [346] to the work of Hemberger et al. [32] and was answered and commented by Hemberger et al. [347]. This controversy finally stimulated a dispute about ferroelectrics going bananas, where Scott [348] stated that most of the work reported on FE hysteresis reported in literature concerning transition-metal oxides could equally well be measured in bananas. Responding to this note, Loidl et al. [349] provided measurements on a real banana skin documenting that when care is taken in the appropriate dielectric measurements the correct behaviour can be deduced.

Subsequently the phonon properties of $CdCr_2S_4$ were studied in full detail as function of temperature by IR [209] and Raman-scattering techniques [350], as well as by ab-initio calculations within frozen-phonon and Berry-phase techniques within the framework of LSDA+U [351]. The first-principle phonon calculations of the IR active phonons within the cubic spinel structure found that all phonons should be rather well behaved and cannot account for the large dielectric responses observed at lower frequencies, hence suggesting that the origin for the observed relaxor behaviour in $CdCr_2S_4$ cannot result from a displacive polar soft mode. However, the experimental Raman and IR work identified the observation of new modes and significant anomalies in the temperature-dependent phonon properties. Anomalies in the temperature dependence of IR active phonon eigenfrequencies and damping, i.e. striking deviations from the conventional anharmonic behaviour, were reported by Rudolf et al. [209]. These anomalies appeared at temperatures ~ 130 K, well above the onset of FM order, and appeared in a temperature regime where anomalous thermal expansion and a broadening of Bragg reflections were reported earlier [340]. It was speculated that these phonon anomalies may also result from chromium off-centre relaxations, which locally break the cubic symmetry. Possible relaxations of these dipolar modes at elevated temperatures can be identified in the dielectric spectra shown in Fig. 40. The appearance of new modes at low temperatures in Raman experiments [350] was interpreted in terms of a loss of inversion symmetry and as an indication of Cr off-centring, supporting the MF ground-state picture in $CdCr_2S_4$.

$FeCr_2S_4$, with its FiM transition close to 165 K and the JT transition at 9 K has been discussed in some detail in chapter 3.2. Further interest in this compound arose due to the fact that after reports of a structural phase transition in the 50-60 K range from a cubic high-temperature to a triclinic low-temperature phase [140], a concomitant transition from a collinear ferrimagnet to a spin-spiral state were observed in Mössbauer [134,136] and muon-spin rotation experiments [135]. These structural and magnetic phase transitions, which possibly are triggered by the onset of orbital fluctuations, are also clearly visible in magnetic susceptibility measurements (see Fig. 14). Finally, from detailed neutron diffraction, thermal expansion, magnetostriction, dielectric, and specific-heat measurements on polycrystalline $FeCr_2S_4$ as function of temperature and external magnetic fields it was concluded that it undergoes FE ordering when passing into the orbitally ordered phase [34]. The multiferroicity of $FeCr_2S_4$ for $T < T_{OO}$ was concluded from a series of dielectric experiments with the electric field parallel or perpendicular to the magnetic field and finally resulted in a ($H$,$T$)-phase diagram which is reproduced in Fig. 44. This phase diagram reveals the sequence of PM, collinear FiM, non-collinear FiM in an OL, which is characterized by strong orbital fluctuations, and finally an orbitally ordered phase, which is non-collinear FiM and FE. This phase diagram implies that FE order exists up to higher magnetic fields (> 5 T) where collinear FiM order is re-established. This observation excludes a conventional spin-driven iDM mechanism. In this case one also would expect that macroscopic FE polarization evolves at the magnetic phase transition close to 60 K, where the spins become non-collinear. It rather seems that in the case of $FeCr_2S_4$, FE polarization is coupled to the onset of OO.

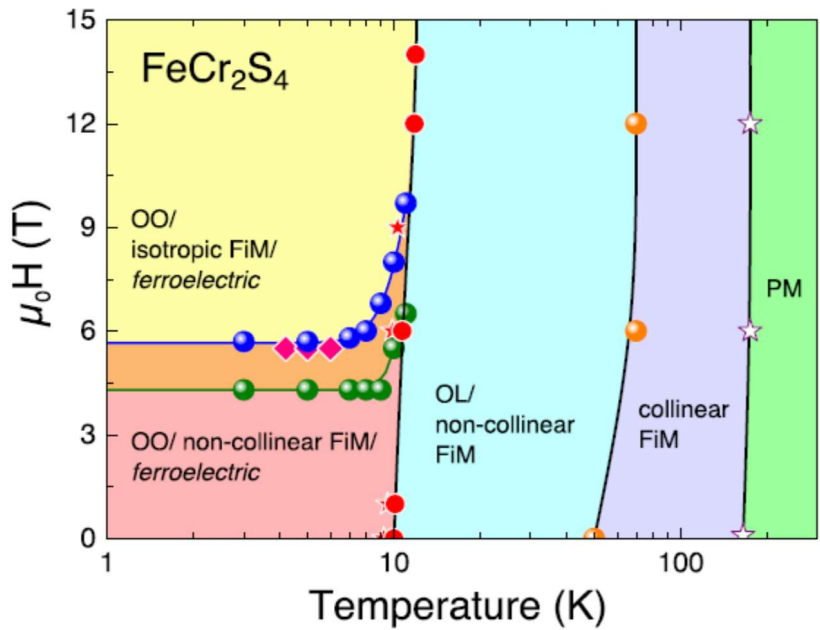

Fig. 44. (*H*,*T*)-phase diagram of FeCr$_2$S$_4$ on semilogarithmic scales. OO: orbital order, FiM: ferrimagnet, OL: orbital liquid, PM: paramagnet. Values were taken from the respective temperature and field dependencies of thermal expansion, magnetostriction (spheres), and specific heat (stars). Lines are drawn to guide the eyes.
Material from: Bertinshaw et al. [34]. Copyright © 2014, Springer Nature.

Parallel to the work of Bertinshaw et al. [34], Lin et al. [352] published a detailed dielectric, magnetic and thermodynamic study of FeCr$_2$S$_4$ arriving at very similar conclusions concerning a MF ground state in the orbitally ordered phase. These authors found that the FE polarization has two components. One component arises from the non-collinear conical spin order and in combination with SOC depends on the external magnetic field. The other component most likely can be attributed to the JT distortion, which induces symmetry breaking in the orbitally ordered phase.

MnCr$_2$S$_4$ has been discussed in detail in Chapter 3.1. It belongs to the rare examples of magnets, that exhibit a triangular YK phase at low temperatures [25,113]. FiM ordering sets in below ~ 65 K and is followed by a low-temperature magnetic phase transition at 5 K where a triangular spin structure is formed: The manganese moments, which are aligned antiparallel to the chromium spins, undergo canting and the chromium moment together with two manganese moments form a spin triangle, the so-called YK phase. It is though that in the FiM phase for temperatures above the YK phase, the chromium moments are aligned parallel to the external field and concerning the manganese moments, only the longitudinal components are ferrimagnetically ordered, while the transverse components still are dynamically disordered (see Fig. 10 for details).

Recently, the (*H*,*T*)-phase diagram of MnCr$_2$S$_4$ was investigated utilizing field-dependent sound velocity and pulsed magnetization experiments in external magnetic fields up to 60 T [26]. Some representative results are shown in Fig. 45. In Fig. 45(a) the field-dependence of the magnetization *M* as measured at 1.5 K is shown as blue solid line. After a strong increase of *M* due to domain-orientation effects below 1 T, *M*(*H*) reveals a continuous increase up to ~ 11 T, where obviously spin-reorientation occurs, visible via a significant change of slope (labelled Nr. 1). This field regime characterizes the YK phase, which obviously extends up to 11 T. On increasing magnetic fields, the angle between the manganese moments opens and hence, the magnetization increases. Beyond 11 T, *M*(*H*) further increases up to ~ 25 K followed by an extremely wide magnetization plateau extending from 25 to 50 T. The change of slope again signals a change in the spin configuration (labelled by "2"). The plateau regime exhibits a fixed and constant magnetization of 6 $\mu_B$, a value signalling FM alignment of the chromium moments in the external field with perfect AFM alignment of the two manganese sublattices. Beyond 50 T the magnetization increases further (labelled "3"). The field dependence of the magnetization was investigated at temperatures up to 28 K, where a continuous increase of the magnetization with increasing fields was detected [Fig. 45(b)]. The field dependence of the longitudinal sound velocity is documented in Fig. 45(a) (left scale) and reveals the same characteristic temperatures indicative for magnetic phase transitions coupled to the lattice. Specifically, the anomaly 2, just when reaching the magnetization plateau, indicates strong softening of the lattice followed by a constant sound velocity, a fingerprint of a stable structural phase in the plateau regime. Like in the chromium oxides, where the half-magnetization plateau is accompanied by a significant structural phase transition [215], also in $MnCr_2S_4$ the transition into the magnetization plateau comes together with a significant structural phase transition.

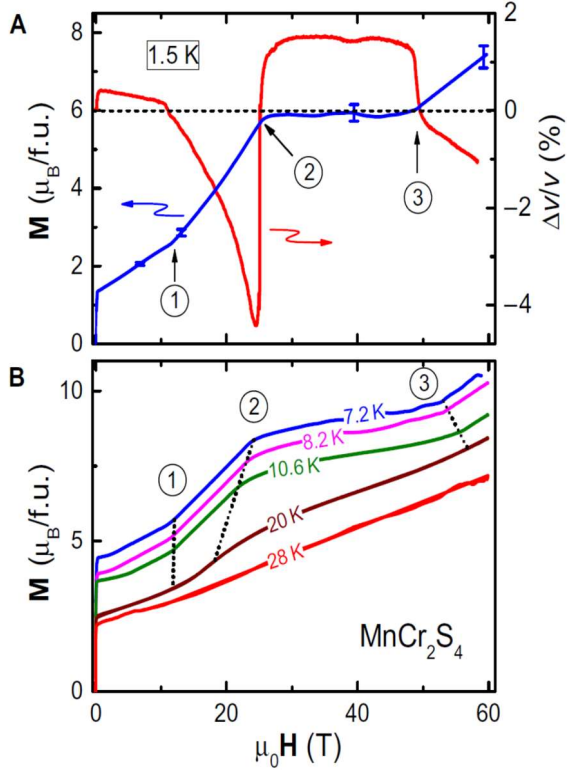

Fig. 45. Field dependence of magnetization and ultrasound velocity in $MnCr_2S_4$ at 1.5 K. (A) Magnetization (left scale), in Bohr magneton per f.u. and relative change of the sound velocity $\Delta v/v$ of the longitudinal acoustic mode along <111> (right scale), parallel to the external magnetic field. (B) Magnetization for different temperatures between 7.2 and 28 K. For clarity, the curves at $T < 28$ K are vertically shifted. Anomalies at 11, 25, and 50 T are consecutively numbered. Dotted lines indicate their temperature-dependent shifts.
Figure taken from Tsurkan et al. [26]. Copyright © (2017) The Authors.

In subsequent work, it was attempted to interpret [26,353] and to model [354] the spin structures of $MnCr_2S_4$ as function of external magnetic fields. A schematic sketch of the spin structures on increasing fields, taken from Ref. [353], is reproduced in Fig. 46. In Fig. 46(a) the spin structure of the YK phase is plotted in a specific crystallographic plane containing the chromium as well as both manganese ions, which are situated on the two sublattices of the diamond lattice: The two canted manganese moments have an overall FiM alignment with respect to the chromium spins. When compared to the structure of pure *A*-site spinels as shown in Fig. 27, we see the influence of the strong AFM exchange of chromium spins acting on the manganese moments. In principle the chromium exchange tends to align both manganese moments antiparallel, but is competing with the weaker AFM Mn-Mn exchange. Fig. 46(b) shows the spin alignment of all magnetic moments within the three-dimensional unit cell. Finally, Fig. 46(c) shows the spin orientations with increasing external fields. It was assumed that the chromium moments are always aligned parallel to the external field. The triangular YK structure is followed by an intermediate phase (IM1), where the triangle, defined by the manganese spins, is rotated with respect to the chromium moments. The AFM phase characterizes the stable magnetization plateau [Fig. 45(a)] and here the manganese moments exhibit ideal AFM order and do not contribute to the magnetization. In the plateau region, the external field compensates the chromium exchange and the manganese moments do not feel any additional field except their own exchange field, resulting in pure AFM order. Finally, the spin structure passes into a further intermediate phase (IM2) and subsequently forms an inverse YK (IYK) structure before reaching the completely spin-polarized FM phase. The complete ($H,T$)-phase diagram up to 100 T will be discussed later in full detail.

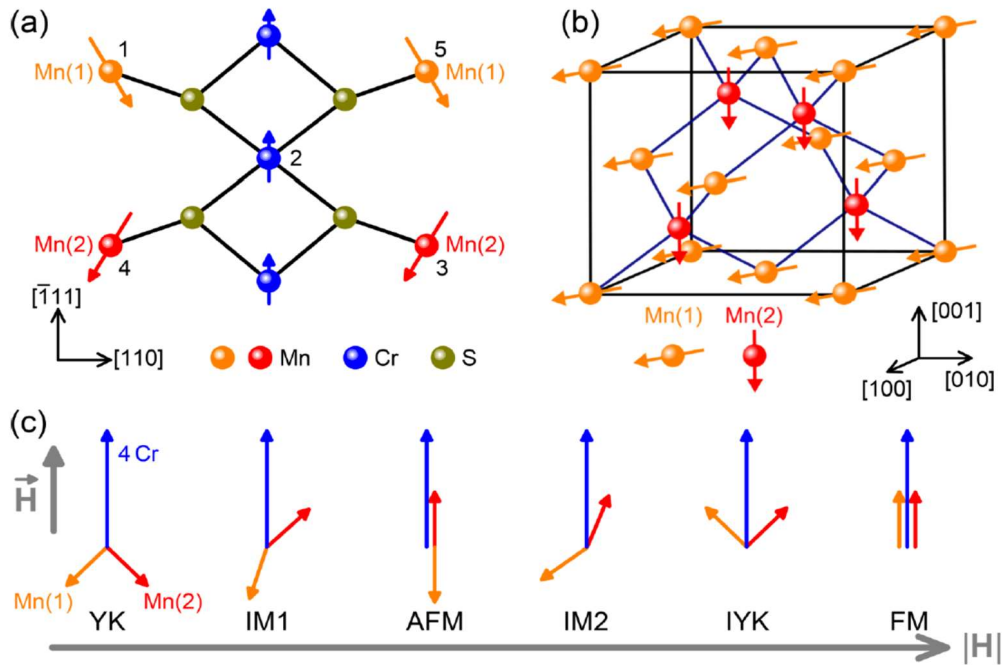

Fig. 46. (a) Exchange paths between the two manganese sublattices, Mn(1) and Mn(2) of the bipartite diamond lattice. Chromium, manganese, and sulphur ions, including the localized spins at the manganese and chromium sites in the Yafet-Kittel (YK) phase, are indicated. Chromium spins are aligned in the external magnetic field along [111]. Via exchange paths of the predominantly 180° bonding angles (1-2-3 or 4-2-5), the two manganese sites are antiferromagnetically coupled resulting in a YK structure in low external magnetic fields $H$ and at low temperatures. (b) The two manganese sites of the diamond lattice within the unit cell of the normal spinel structure. The chromium spins (not shown) are aligned along the crystallographic [111] direction (body diagonal of the fcc lattice). The shown spin canting of the manganese spins corresponds to YK order at low magnetic fields. (c) Schematic spin order as function of an external field $H$: On increasing fields, Yafet-Kittel (YK) order is followed by an intermediate phase 1 (IM1), by antiferromagnetic (AFM) order, by an intermediate phase 2 (IM2), by an inverse YK (IYK) order, and finally, by the field-polarized ferromagnetic order (FM). Reprinted figure with permission from Ruff et al. [353]. Copyright (2019) by the American Physical Society.

Here it should be noted that some of the spin structures as depicted in Fig. 46(c) can be interpreted as magnetic analogues of superliquids and supersolids, so-called spin superfluids and spin supersolids. They are characterized by complex spin structures, with independent order of the longitudinal and transverse spin components and were identified earlier in some chromium oxo-spinels [219,220,221]. The identification of supersolid spin states can be based on the analogy of bosonic and spin systems, as outlined in the pioneering work of Matsuda and Tsuneto (355) and Liu and Fisher (356). In spin systems, the longitudinal AFM order is an

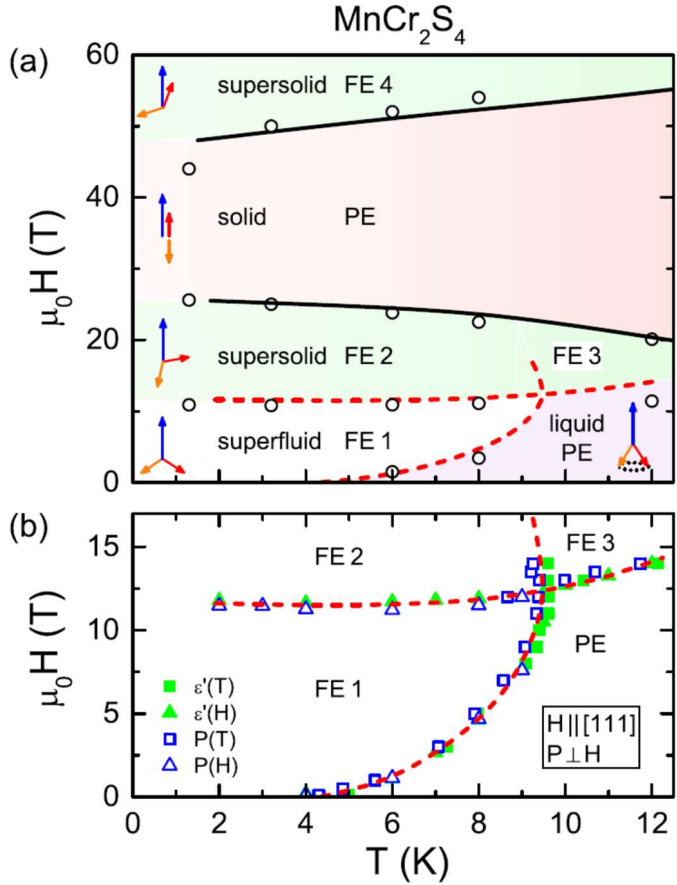

Fig. 47 (a) Schematic ($H,T$)-phase diagram of MnCr$_2$S$_4$. Phase boundaries are indicated by solid lines. Anomalies deduced from high-field pyrocurrent experiments are shown by empty circles. Spin configurations [arrows: chromium (blue), manganese (red and orange)], and the polar order (PE, FE) are indicated. Four ferroelectric phases (FE1, FE2, FE3, and FE4) with different polarizations were identified. Dashed lines are derived from the dielectric and pyrocurrent experiments in static fields. (b) Low-field phases as determined from the temperature and field dependence of the dielectric constant ε and polarization $P$ up to 14 T. Dashed lines are drawn to guide the eye. Paraelectric (PE) and ferroelectric phases FE1, FE2, and FE3 are indicated.
Reprinted figure with permission from Ruff et al. [353]. Copyright (2019) by the American Physical Society.

analogue of the crystalline order, which, in the quantum lattice-gas model, corresponds to diagonal long-range order, whereas off-diagonal long-range order, which is characteristic of superfluidity, is the counterpart of transverse (XY) AFM order without longitudinal AFM components. Naturally, this analogy also applies for the PM and the supersolid phases. The former, with both orders destroyed, corresponds to a liquid, whereas the latter, hosting both orders, is the spin analogue to the supersolid. Note that in this analogy, the FM phase is equivalent to the paramagnet.

In this paragraph, we are interested in possible MF phases in the $H,T$ phase diagram of MnCr$_2$S$_4$. It seems natural to check whether the non-collinear spin states as plotted in Fig. 46(c) possibly are also FE. Ruff et al. [353] recently performed a careful dielectric investigation of MnCr$_2$S$_4$ in a broad temperature range and in high external magnetic fields. This study was made possible by the development of experimental techniques, which allow to measure magneto currents in pulsed magnetic fields up to 60 T [357]. The phase diagram as derived in Ref. [353] is shown in Fig. 47. The lower frame of this figure [Fig. 47(b)] shows the phase diagram as determined in static magnetic fields up to 14 T. The upper frame [Fig. 47(a)] documents the phase diagram as deduced in pulsed high-magnetic fields experiments up to 60 T. Fig. 47(b) was derived from a series of experiments measuring the dielectric constant as function of temperature and field as well as measuring the pyrocurrent at a series of magnetic fields or the magnetocurrent at different temperatures. In all these experiments, the external magnetic field was aligned parallel to the crystallographic [111] direction and the electric field was aligned perpendicular to the magnetic field [353]. In the low-temperature and low-field diagram [Fig. 47(b)], three FE phases (FE1, FE2 and FE3) were identified, in addition to the PE phase, which is stable beyond the YK phase boundary. The YK phase exhibits ferroelectricity and hence is MF. It also shows that the two intermediate phases, IM1 and IM2, which can be assigned as spin-supersolid phases, are also FE. The broad magnetization plateau [Fig. 45(a)] with ideal collinear AFM order of the manganese spins is paraelectric. It seems clear that all MF phases displayed in Fig. 47 are spin driven. However, as outlined by the authors of Ref. [353] the title compound does not show conventional spin-spiral order, but canted magnetism of spin pairs located on a bipartite diamond lattice and it was speculated that this type of multiferroicity corresponds to a new type of spin-driven ferroelectricity. At high magnetic fields, beyond the extended plateau regime the inverted YK phase again seems to be a polar phase [353] and in the phase diagram is indicated as FE4 [Fig. 47(a)]. The origin and nature of the MF phases of $MnCr_2S_4$ are still under discussion: As documented in Fig. 47, FE and hence, MF behaviour appears in all non-collinear phases, while the strictly antiparallel phase is paraelectric. In addition, in model calculations to explain the spin structure as function of an external magnetic field [354] it was shown that only incorporating spin-lattice coupling can explain the complex spin structures including the ultra-stable magnetization plateau. However, it was proposed that this term also induces deviations of the chromium moments from the magnetic field direction, at least in the intermediate (supersolid) phases. In these complex triangular spin structures, iDM interactions still could be responsible for spin-driven ferroelectricity. Very recently, in addition to FE polarization below the onset of YK order, macroscopic polarization has been reported in the FiM phase below ~ 65 K in polycrystalline $MnCr_2S_4$ [358]. It was argued that FM order of the chromium moments probably induces FE polarization via an exchange striction mechanism. This polarization, which was also measured in external magnetic fields up to 9 T, seems to be much stronger than the polarization induced by the triangular YK spin structure. This recent experimental report documents that much more work is needed to identify possible polar phases in spinel compounds.

### 3.4.3. Vector-chiral phases

As detailed above, $LiCuVO_4$ is an inverse spinel and exhibits spin $S = ½$ chains along the crystallographic *b* direction. Due to residual inter-chain coupling this compound undergoes 3D ordering at 2.5 K, a temperature which also marks the onset of spin-driven ferroelectricity with polarization *P* along *a*. The MF ground state can be explained in terms of a conventional iDM mechanism and the polarization is switchable by external magnetic fields. Very recently, Ruff et al. [359] detected the onset of macroscopic FE polarization along the crystallographic *a* direction for external magnetic fields > 3 T along the *c* direction, at temperatures well above the onset of the MF ground state. Some representative results are shown in Fig. 48. This figure clearly documents the onset of ferroelectric polarization as function of temperature or as function of an external magnetic field: Fig. 48(a) shows the evolution of macroscopic polarization for magnetic fields > 2 T and temperatures below 11 K. Even in the context of spin-driven improper ferroelectricity the polarization is very small and the strong polarization increase below 2.5 K is due to the onset of conventional spin-driven FE. As was documented in Fig. 38, the polarization in the MF ground state reaches values of about 30 $\mu C/m^2$. The field dependent measurements [Fig. 48(b)] document the onset of polar order at 3 T. The inset in Fig. 48(b) shows a marked anomaly of the field-dependent dielectric constant when passing into this field-induced polar phase. This dielectric anomaly provides even more significance for a possible phase boundary in low magnetic fields and at temperatures well above the MF ground state. It was argued [359] that this could be experimental evidence for a long-sought vector-chiral (VC) phase, where ferroelectricity is induced by the chirality alone and not by coherent 3D spin ordering.

Vector chirality, defined by the vector product of spins at adjacent lattice sites, $S_i \times S_j$, is one of the key concepts in quantum magnetism and an important symmetry element of spin spirals. Chirality or the handedness of objects is of prime importance in life science, biology, chemistry, and physics. VC phases, with the twist (either clock- or counter clock-wise) between neighbouring spins being ordered, but with disorder with respect to the angles between adjacent spins, have been predicted almost five decades ago. In 1D magnets, both, chirality and spin order exhibit long-range order. However, either finite temperatures or quantum fluctuations can completely suppress the long-range spin order of a spiral, while leaving the chiral twist less affected, leading to a VC phase. Long ago, for 1D helimagnets Villain [360] predicted such a phase appearing between the high-temperature paramagnet and the conventional long-range ordered 3D helical spin solid. Villain argued that this chiral order is similar to the ordering of 2D XY helimagnets and that the order parameter is not the real 2D spin, but the fictive Ising spin, which can be related to the chirality and shows up in the four-spin correlation function.

A possible proof of Villain's conjecture of a VC phase was reported, supported by the observation of a two-step magnetic process in the helimagnet Gd(hfac)$_3$NITEt [361]. However, in light of the mentioned FE polarization induced by chiral spin twists the most promising route seems to search for VC phases by identifying FE polarization at temperatures slightly above well-known 3D-ordered spin-driven MF phases [362], with LiCuVO$_4$ being a possible candidate. However, some critical remarks have to be made: Screw-type helical spin structures, with the spin plane perpendicular to the screw axis, are chiral but not polar. On the other hand, cycloidal spin structures break inversion symmetry and are polar but not chiral. Hence, cycloidal spin structure can show clock-wise or counter clock-wise rotation, but strictly should not be termed chiral.

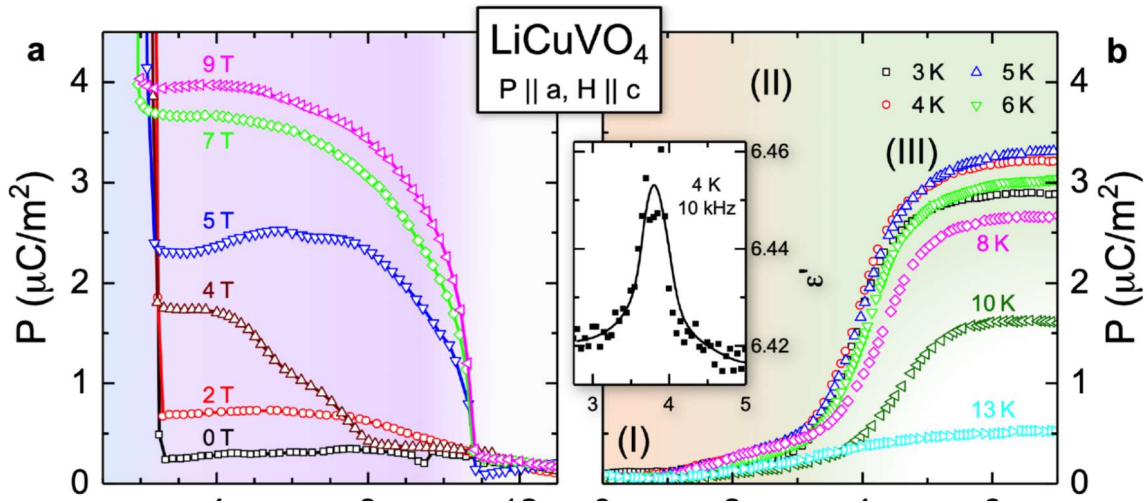

Fig. 48.| Polarization of LiCuVO$_4$ in the region of the VC phase. a) Temperature dependence of the electric polarization along the crystallographic ***a*** direction, ***P*** || ***a***, for magnetic fields between 0 and 9 T with ***H*** || ***c***. b) Field dependence of the polarization ***P*** || ***a*** for temperatures between 3 and 13 K. The inset shows the magnetic-field dependence of the dielectric constant measured at 4 K, well above $T_N$. Regime (I) denotes the region where neither long-range magnetic nor polar order is detectable and is denoted as vector-chiral liquid (VCL), (II) a transition region indicated by a metamagnetic anomaly with marginal polarization only and (III) the PM VC phase with distinct FE polarization.
Figure taken from Ruff et al. [359]. Copyright © (2019) The Authors.

The experimentally determined (*H*,*T*)-phase diagram, with a focus on temperatures below 12 K and magnetic fields below 9 T, as taken from Ref. [359] is documented in Fig. 49. The phase diagram has been taken with the external magnetic field along the crystallographic ***c*** direction and with the electrical field parallel to ***a***. This corresponds exactly to the field configurations, where below 2.5 K, with the onset of 3D magnetic order, long-range FE order was identified. The phase boundaries were identified by pyrocurrent as well as magnetocurrent experiments in addition to dielectric, heat capacity, and magnetization experiments. It was assumed that at zero magnetic fields, cycloidals are formed, however, in equal amounts of clockwise and counter clockwise spin rotation and, hence, with zero macroscopic polarization. At higher fields, above 4 T small macroscopic FE polarization is induced, documenting an imbalance of clockwise and counter clockwise cycloidals. It is thought that a finite polarization is induced similar to the bulk polarization induced by magneto-electric cooling in $ZnCr_2Se_4$ [312]. However, in $LiCuVO_4$ the polarization results from correlations of the handedness of the spin system only, and is not induced by long-range spin order.

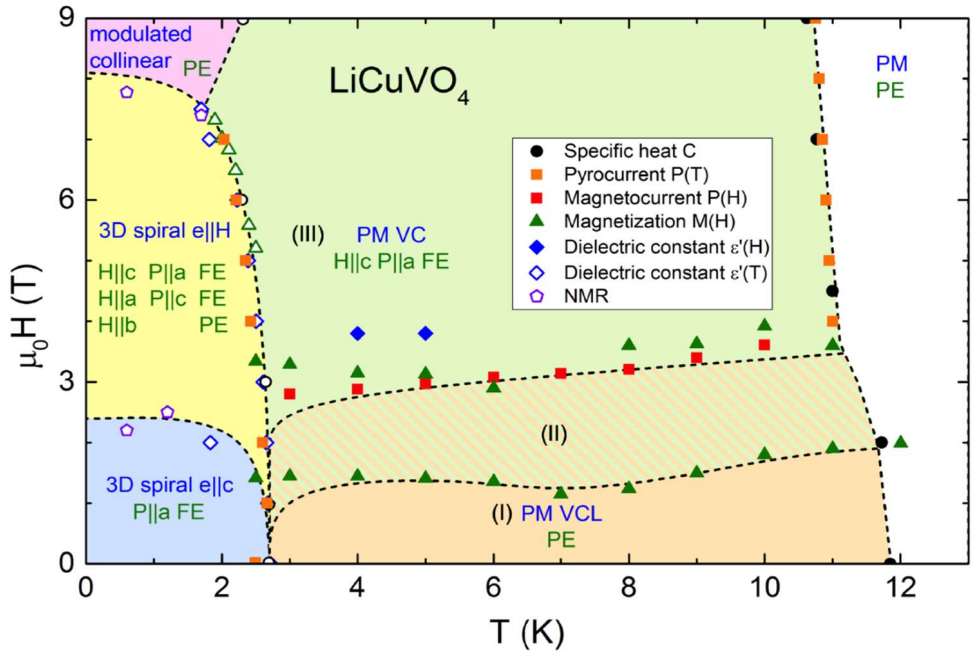

Fig. 49. (*H*,*T*)-phase diagram of LiCuVO4. Anomalies observed by different measuring techniques are characterized by different symbols (see figure legend). Magnetic (blue lettering) and electric phases (green lettering) are indicated in the different regimes. The VC phase is indicated by the light-green area (III). This phase evolves at 3 T and extends at least up to 9 T. At low external fields up to 1.5 T a VCL phase (I) with fluctuating chirality is established. The intermediate phase (II) (dashed area) could not be uniquely identified. Figure taken from Ruff et al. [359]. Copyright © (2019) The Authors.

## 3.5. Spinel derived thin films and heterostructures

As documented in the previous chapters, spinel compounds constitute a class of materials with significant application potential ranging from technologies such as transformer cores, high-frequency microwave devices like circulators or phase shifters, millimetre-wave integrated circuitry, and magnetic recording. While some of these applications can be realized utilizing bulk technologies, most of them rely on the synthesis of thin films and heterostructures. For these applications it seems important that the electronic properties of the bulk compounds equally hold for thin films or heterostructures. These relevant bulk properties include strong magnetic exchange with concomitant high transition temperatures and high permeabilities, high resistivity, and low-loss characteristics even at high frequencies. Having this outstanding future potential and these possible applications in mind, during the last decades considerable efforts have been undertaken, especially in the synthesis and characterization of ferrite thin films optimized for specific applications. Quite generally, ferrites are ceramic magnetic oxides with iron as an important constituent. In the early1950s, ferrites referred mainly to spinel-structure crystals, but later-on the wording ferrites was used to classify the entire family of iron oxides including spinels, garnets, hexaferrites, and orthoferrites. In this review we will only discuss spinel thin films. Their properties, including basic information on synthesis and characterization, are documented in an early review given by Suzuki [363].

From a more fundamental and basic-research dominated perspective, epitaxial thin films and heterostructures also can represent model systems, which exhibit novel phenomena, such as modified super-exchange interactions or perpendicular exchange couplings. In addition, since the early days of high-temperature superconductivity, epitaxial heterostructures composed of complex transition-metal oxides have fascinated theoreticians and experimenters over decades and triggered efforts to realize emergent many-body phenomena not attainable in the bulk. In this respect and during the last decade, the physics of frustrated magnets was an important research topic and a number of groups investigated the role of dimensionality when moving from truly 3D bulk materials, reducing dimensionality and finally approaching the 2D limit. Apart from stabilizing artificial structures along the conventional <001> direction, tuning the growth direction along unconventional crystallographic axes was found to be a promising route to realize novel quantum many-body phases and represents a rapidly developing field of geometrical lattice engineering. This geometrical lattice-engineering approach based on the synthesis of spinel films along <111> crystallographic directions became increasingly popular, e.g., because the $B$ sublattice of spinel compounds along the <111> direction consists of alternating kagome and triangular atomic planes. Further interest in the growth of heterostructures on the basis of spinel-derived transition-metal oxides came along with the rapidly developing field of multiferroics, specifically combining polar and magnetically ordered systems. To enhance ME coupling multilayering techniques or vertically architectured nanocomposite systems were synthesized, which will be discussed later in more detail.

3.5.1. Spinel thin films for possible application in electronic devices

As outlined earlier, due to their unique properties including high resistivity, low loss at high frequency, high corrosion resistance, and good mechanical stability, magnetic spinel compounds have received great attention in view of their potential applications in a variety of microwave devices, perpendicular recording, and magneto-optical media. As bulk ferrite components are not compatible with planar circuit design, for integration in electronic devices, it seemed important to grow these materials in form of thin films and to characterize their structural and electronic properties as function of film thickness, substrate, as well as substrate orientation and temperature. For these purposes and using thin-film preparation techniques, it seemed important that the unique magnetic properties of the bulk materials remain relatively unchanged even in the form of thin films. As documented in the previous chapters, many of the spinel oxides constitute an important class of insulating magnetic oxides, widely used as soft magnets at high frequencies. Their general formula is $A$Fe$_2$O$_4$ where the iron is ferric Fe$^{3+}$ and the $A$ site of the spinel structure is occupied by a divalent transition-metal cation, such as Mg$^{2+}$, Mn$^{2+}$, Fe$^{2+}$, Co$^{2+}$, Ni$^{2+}$, or Zn$^{2+}$. As documented in detail in the previous chapters, when the $A$ and $B$ sites of the spinel structure are occupied by magnetic ions, all the spinel ferrites are FiM with high Curie temperatures. At room temperature, all these compounds with the exception of magnetite, which is a bad metal, are insulating. Only the cobalt compounds, due to their orbital degrees of freedom exhibit sizeable magnetic anisotropy. In transition-metal oxides in general and specifically in oxide spinels, due to the strong interactions of the internal degrees of freedom, like spin, charge, orbital moment, and lattice degrees of freedom, the electronic properties are extremely sensitive to structural distortions and strain. Because of the inevitable lattice mismatch between spinel films and the substrates used, epitaxial thin films are an ideal platform to study the effects of strain on a given structure. Misfit strains induced by substrates are often employed to tune the structure and concomitant electronic properties of the thin films under investigation. Ferrite films were grown in form in nano- or polycrystalline form as well as epitaxial single crystalline thin films. To grow polycrystalline as well as epitaxial thin films, a number of techniques, including electroplating, magnetron sputtering, pulsed laser deposition, evaporation, and molecular beam epitaxy have been developed over the years and were used extensively. An overview and details of film-preparation techniques can be found in [363].

So far, mainly (Mn:Zn)Fe$_2$O$_4$ ferrites [364,365,366,367], CoFe$_2$O$_4$ (365,366,368,369,370,371,372], ZnFe$_2$O$_4$ [373], NiFe$_2$O$_4$ [371,374], Fe$_3$O$_4$ and Co doped magnetite [375] were grown on a variety of substrates, like SrTiO$_3$, MgAl$_2$O$_4$, or MgO by a variety of techniques. Most of these thin films were structurally characterized by x-ray diffraction, and mainly by various magnetization experiments to determine magnetic permeabilities, saturation magnetization, coercivity, and anisotropy. In many of the studies this characterization was performed as function of substrate, substrate orientation and temperature, film thickness, and microstructure. To grow high-quality single-crystalline spinel thin films, in exceptional cases spinel-structure buffer layers have been used [365]. In recent years, new spinel-based ferrite films have been synthesized with astonishing low-loss magnetization dynamics and strong magnetoelastic coupling, which possibly will enable the development of power-efficient ME and acoustic spintronic devices. In this respect, thin films with nominal composition of Ni$_{0.65}$Zn$_{0.35}$Al$_{0.8}$Fe$_{1.2}$O$_4$ [376] and MgAl$_{0.5}$Fe$_{1.5}$O$_4$ [377] were synthesized by pulsed laser epitaxy on single crystal MgAl$_2$O$_4$ substrates.

### 3.5.2. Novel phenomena of spinel thin films and heterostructures

In a further approach, utilizing thin-film growth techniques, it was attempted to synthesize and identify new model systems with emerging physical phenomena or many-body quantum states non-existent in the bulk. In this respect, promising modern venues are so-called interface engineering or geometrical-lattice engineering of thin films and heterostructures. Here thin films were grown on unconventional substrate planes resulting in a significant mismatch of lattice parameters between the layers, creating heavy strain and inducing unusual interactions. Furthermore, it seems interesting to investigate ultra-thin films and heterostructures along specific orientations or the stacking of very specific numbers of atomic layers. In the upcoming field of multiferroics, nanostructures and multilayers were grown on specific substrates, always attempting to enhance the ME coupling via coupling of the polar and magnetic order parameters. Consequently, there has been growing interest in the coupling of different functionalities by interfacing materials with appropriate functions, either via multilayering of different compounds or via synthesis of vertical nanocomposites.

In this latter field, the synthesis of a nanostructured $BaTiO_3$-$CoFe_2O_4$ ferroelectromagnet was reported combining a FE and AFM material [378], with a potential coupling between FE and magnetic order parameters. These nanostructures were deposited on single-crystalline $SrTiO_3$ by pulsed laser deposition. As a result, the films showed hexagonal arrays of $CoFe_2O_4$ nanopillars embedded in a $BaTiO_3$ matrix, with the nanopillars showing uniform size and average spacing of 20 to 30 nanometers. Since then, a number of isostructural analogues of the $CoFe_2O_4$ and $BaTiO_3$ heterostructures have been synthesized [379]. The growth and characterization of the MF properties of magnetite thin films has been reported in chapter 3.4.2. and is documented in Fig. 41 of this review [334]. A detailed review on the progress in growth, characterization, and understanding of thin-film multiferroics has been given by Ramesh and Spaldin [309].

One of the first attempts in geometrical lattice engineering has been reported by Lüders et al. [380] reporting the epitaxial growth of FiM $CoCr_2O_4$ films on $MgAl_2O_4$ resulting in self-organized {111}-faceted pyramids with almost micron-sized base dimensions. Similar work concerning the epitaxial growth of (111)-oriented $MgCr_2O_4$ thin films has recently been reported by Wen et al. [381]. These authors systematically characterized the structural and electronic properties of thin films grown along the <001> and <111> directions. A short review on geometrical lattice engineering of complex oxide heterostructures has been published by Liu et al. [382]. These authors document the rapidly developing field of geometrical lattice engineering with the emphasis on the design of complex oxide heterostructures along the <111> orientation to design new quantum phases. Another modern branch of thin-film heterostructures refers to the search for emergent magnetic states of ultra-thin films sandwiched between mostly insulating and non-magnetic spacer layers. Here it is worth mentioning a recent work by Liu et al. [383] reporting emergent magnetic states in $CoCr_2O_4$ thin films sandwiched between $Al_2O_3$ spacer layers. In this work, polarized neutron reflectometry revealed a significant enhancement of magnetization compared to the bulk value. Further synchrotron x-ray magnetic circular-dichroism measurements prove the appearance of FM order of both, Co and Cr spins. The observed phenomena were interpreted as being due to magnetic frustration invoked by quantum confinement effects, which are unattainable in the bulk.

### 3.6. Metallic and superconducting spinels

As is clearly documented in this review, the majority of spinel compounds are insulators or semiconductors with a broad class of semiconducting magnets, interesting for both, applications and basic research. Usually the transition-metal ions exhibit well-localized $d$ shells, forming localized magnetic moments interacting via direct overlap or via SE interactions given rise to an astonishingly rich variety of magnetic ground state. The complexity of the magnetism arises due to frustration effects and due to competing interactions. Many cupreous oxides are magnetic semiconductors. However, all non-oxidic copper spinels Cu$M_2X_4$ are metals with significantly different properties: within this class, the spinels with $M$ = Cr are metallic ferromagnets with magnetic-ordering temperatures well above room temperature. All the other known compounds reveal a temperature-independent Pauli paramagnetism and do not undergo magnetic order down to the lowest temperatures. In addition, CuRh$_2$S$_4$ and CuRh$_2$Se$_4$ exhibit superconductivity at low temperatures. We would like to recall that for the explanation of itinerant electron behaviour, specifically of CuCr$_2$S$_4$, two competing explanations have been proposed: Lotgering [104], and Lotgering and van Stapele [107] assumed monovalent copper at the $A$ site and mixed valent Cr$^{3+}$ and Cr$^{4+}$ at the $B$ site, while Goodenough proposed divalent copper at the $A$ site and conventional trivalent chromium at the $B$ site [108]. In this latter interpretation, the copper ions have the formal valence 2+ but their $3d$ electrons are delocalized and form itinerant electron bands. In this review, we will focus only on some relevant results and will not discuss the physics of these metallic spinels in great detail. A detailed (but early) review on thiospinels including the metallic samples has been given by van Stapele [110]. Later on, using hydrothermal synthesis methods for the growth of single crystals, metallic conductivity has been observed in LiV$_2$O$_4$ [384], which later turned out to be the first heavy-fermion (HF) transition-metal oxide and will be discussed later in full detail. In the early seventies, with LiTi$_2$O$_4$ the only oxide-spinel superconductor has been identified by Johnston and coworkers [385,386] with a relatively high superconducting-transition temperature of the order of 10 K.

All copper-chromium-chalcogenide spinels, CuCr$_2$S$_4$, CuCr$_2$Se$_4$, and CuCr$_2$Te$_4$, are FM metals with magnetic-ordering temperatures far above room temperature. These spinel-type metals were first synthesized by Hahn et al. [387], who determined the crystallographic structure and identified them to be normal spinel compounds. Subsequently, the magnetic properties of the sulphide and selenide compounds were determined by Lotgering [104]. These compounds exhibit spontaneous magnetization with Curie temperatures $T_c$ = 375, 430, and 365 K, for the sulphide, the selenide and the telluride, respectively. At liquid helium temperatures, saturated moments of 4.58, 4.94, and 4.93 $\mu_B$ were determined [110]. All compounds reveal a very well-developed CW behaviour of the susceptibility below 1000 K with CW temperatures very close to the FM ordering temperatures [110]. From the very beginning, it was clear that the sulphide is relatively hard to synthesize in pure form and often contains AFM CuCrS$_2$ as an impurity phase.

The contradictory interpretation of the itinerant electron behaviour of these compounds is mentioned above. In the light of some more recent soft x-ray absorption spectroscopy experiments on CuCr$_2$S$_4$ and CuCr$_2$Se$_4$ [388], it was concluded that the chromium $2p$ core excited spectral features of the metallic ferromagnets are similar to those of semiconducting chromium spinels, partially supporting Goodenough's conjecture of Cr$^{3+}$ at the octahedral sites. However, these studies showed that the tetrahedral sites are not occupied by divalent copper, but rather by monovalent ions with a small number of holes inducing a finite magnetic moment. It was proposed that the extra holes go to the chalcogenide $p$ band, classifying these compounds as charge-transfer-type materials [388].

To shed further light on the valencies of copper and chromium in these compounds, a series of solid solutions were prepared and investigated, documented in detail in Ref. [110].

Among these investigations, solid solutions between FiM and insulating $FeCr_2S_4$ and FM and metallic $CuCr_2S_4$ played an outstanding role. These experiments recently invoked renewed interest, as giant magnetoresistance was reported for $Fe_{1-x}Cu_xCr_2S_4$ [31], which will be discussed later. The first magnetic experiments on these solid solutions were performed by Haacke and Beegle [389]. The main result of these investigations was the fact that only for small and large copper concentrations $x$, CW behaviour of the susceptibility was observed in reasonably wide temperature ranges, while at intermediate concentrations the inverse susceptibilities display no linear regions at all. In addition, throughout the concentration regime, the saturated moments evolve in a completely non-linear fashion. In this work, it was speculated that the chromium ions may be in a low-spin state at some concentrations [389].

A recent proposal of the phase diagram, connecting orbitally-ordered $FeCr_2S_4$, with a magnetic spin-spiral configuration at low temperatures, and FM and metallic $CuCr_2S_4$, is constructed and documented in Fig. 50. The critical temperatures for this $(x,T)$-phase diagram were taken from magnetization [389,390] and magneto-resistance measurements [454], as well as from Mössbauer and µSR experiments [134,135,136,391]. This phase diagram documents the intrinsic complexity of spinel compounds, which is by far not completely understood. In the pure iron compound ($x = 0$), we find the well-known sequence of PM and FiM phases, followed by a spin-spiral configuration coexisting with an OL, which can be thought to correspond to dynamic short-range OO ($\lesssim 60$ K) and finally by a spin-spiral ground state in combination with coherent long-range OO ($\lesssim 10$ K). Here it should be clearly stated that the spiral-like spin configuration below 60 K was proposed by µSR experiments [135] and still awaits a microscopic proof, e.g. by neutron-scattering techniques. At least there exists experimental evidence that below this characteristic transition temperature, orbital fluctuations induce some type of non-collinear spin arrangement. Below the FiM ordering temperature, with the onset of significant spontaneous polarization, the electrical resistance decreases on decreasing temperature, exhibiting a metallic-type behaviour, which will be described below in Chapter 3.8 by magneto-resistance effects. Below 100 K the resistance increases again, probably due to localization processes of the charge carriers.

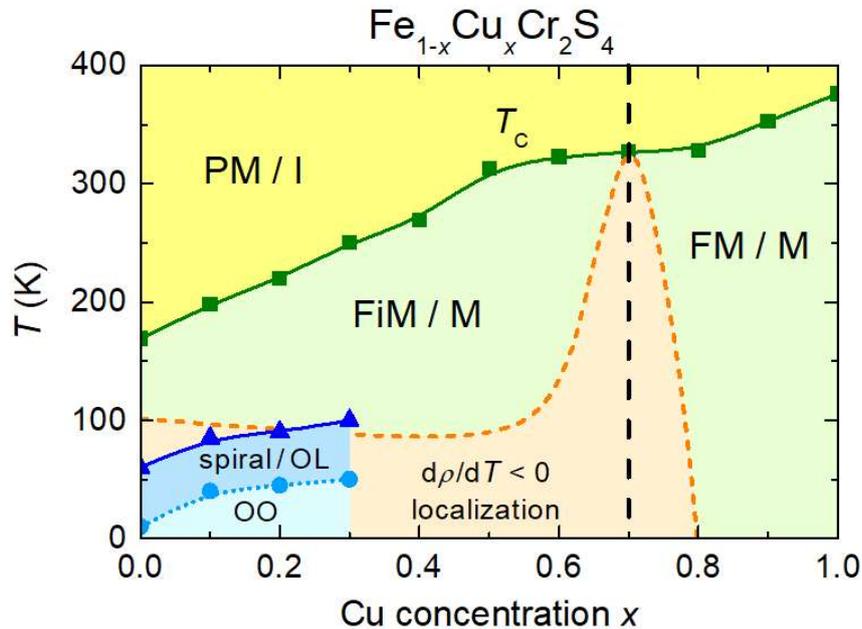

Fig. 50. Magnetic phase diagram of $Fe_{1-x}Cu_xCr_2S_4$. Characteristic temperatures are plotted vs. the Cu concentration $x$: at elevated temperatures paramagnetic (PM), ferrimagnetic (FiM), and ferromagnetic (FM) regimes are indicated. At low temperatures, and for low Cu concentrations $x$, spiral-spin order coexisting with an orbital liquid (OL) is followed by a ground state with a spin-spiral and orbital order (OO). For $x > 0.7$, the FiM phase is followed by a FM metal. For concentrations $0.3 < x < 0.7$, the low temperature magnetism in these alloys awaits final characterization.

With increasing Cu concentration $x$, the regime of the helical spin structure is followed by disordered spin configuration coexisting with short-range OO. With further increasing Cu concentrations, the FiM ordering temperature strongly increases. At low temperatures, this concentration regime so far is not well characterized. It is interesting to note that the regime exhibiting charge-carrier localization strongly extends and finally, close to x ~ 0.7, there is no metallic-like regime at all. Interestingly, beyond Cu concentrations of $x$ ~ 0.7 the system reveals metallic and almost FM behaviour. At the Cu-rich side, the metallic behaviour of the electrical resistance can be observed down to the lowest temperatures with no signs of localization and the compounds behave like conventional metals.

The only oxide spinel superconductor, $LiTi_2O_4$, was synthesized and characterized by Johnston and coworkers [385,386]. The homogeneity of the spinel phase in the ternary compounds $Li_{1+x}Ti_{2-x}O_4$ ranges from $x$ values 0 to 1/3. For all concentrations the samples reveal an almost temperature-independent Pauli spin susceptibility in the metallic high-temperature phase. From a crystallographic point of view, it seems more adequate to write the cation distribution for this compound following a normal spinel scheme with $Li(Li_xTi_{2-x})O_4$ with $0 \leq x \leq 1/3$. For $x = 0$, only titanium cations occupy the octahedral sites with a formal valence of 3.5 providing some arguments for itinerant $d$ electrons. For non-zero $x$ values a statistical distribution of Li and Ti cations at the octahedral sites is expected. Depending on the fraction of the lithium to titanium concentration, superconducting transition temperatures up to 13.7 K were found, with the highest $T_c$ values close $x$ ~ 0. Johnston [386] concluded that it is likely that the conduction electrons in $LiTi_2O_4$ reside primarily on the titanium octahedral-site sublattice of the spinel structure and hence that the title compound is a $d$ electron superconductor. To document that $LiTi_2O_4$ probably can be described in terms of a conventional BCS superconductor, Fig. 51 shows the temperature-dependent heat capacity of superconducting $LiTi_2O_4$ and non-superconducting $Li_{4/3}Ti_{5/3}O_4$, results, which were published much later for comparison with the HF vanadium compound [392]. The ordering temperature $T_c$ of this single crystal close to ideal spinel stoichiometry is 11.8 K. The $\lambda$-like anomaly at the superconducting phase-transition temperature is well pronounced and the jump of the heat capacity at the critical temperature $\Delta C_p/\gamma T_c = 1.75$, somewhat lower than the ideal BCS ratio of 2.43. Here $\gamma$ denotes the Sommerfeld coefficient characterizing the conduction electrons in the normal metallic state.

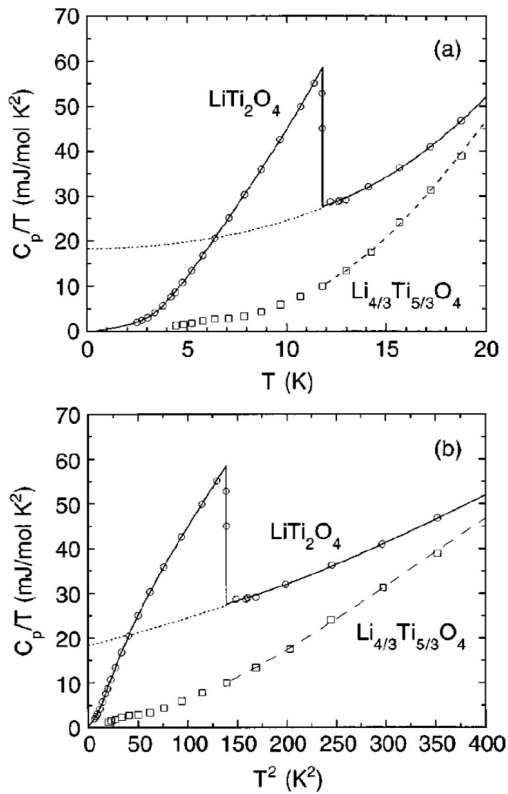

Fig. 51. Heat capacity of $LiTi_2O_4$.
Temperature dependence of the heat capacity:
a) Molar specific heat divided by temperature $C_p/T$ vs temperature $T$ of superconducting $LiTi_2O_4$ and non-superconducting $Li_{4/3}Ti_{5/3}O_4$. b) The same data as plotted $C_p/T$ vs $T^2$. The solid curves are polynomial fits to the data for $LiTi_2O_4$, with the dotted curve indicating the tentative normal-state behaviour below $T_c = 11.8$ K. The dashed line is a polynomial fit to the data for $Li_{4/3}Ti_{5/3}O_4$ above 12 K indicating the pure phonon contributions.
Reprinted figure with permission from Johnston et al. [392]. Copyright (1999) by the American Physical Society.

The spinel-type superconductors $CuRh_2S_4$ and $CuRh_2Se_4$, were synthesized and characterized by Lotgering and van Stapele [393] and where found to be metals with an almost temperature independent Pauli-like spin susceptibility. An extensive study of electrical resistivity, ac magnetic susceptibility, magnetization, specific heat, and NMR has been performed on high purity samples, identifying superconducting transitions to occur at 4.70 K in $CuRh_2S_4$ and 3.48 K in $CuRh_2Se_4$ [394]. Both compounds were characterized as type-II superconductors with values of the gap ratio, $2\Delta(0)/k_BT_C$, ranging between 3 and 4 for various types of measurements. It was argued that the lower superconducting transition temperatures of both these rhodium spinels, as compared to $LiTi_2O_4$, probably result from the significantly lower Debye temperatures that were observed in the course of this work [394].

It seems interesting to document the problems with sample purity and stoichiometry, specifically for a number of chalcogenide spinels, by mentioning a large number of contradicting reports in literature concerning the properties of $CuV_2S_4$. Originally identified as a further spinel superconductor [110], later on superconductivity was discarded in measurements on high-purity single crystals down to temperatures of 60 mK [395]. This compound gained further interest via early reports of a charge-density wave formation in a strictly 3D material [396]. However, concerning the correct temperature and nature of this phase transition there were a number of contradicting reports, including an interpretation of the phase transition in terms of a cooperative JT transition [397], which were not completely settled until today [398]. Some of these contradictory experimental reports may result from the use of samples grown by different techniques with different stoichiometry, and probably different concentrations of impurity phases.

## 3.7. Metal to insulator transition and charge order in spinels

In most spinel compounds the cations, which are located at the *A* and *B* sites in the spinel structure have well defined valences with predominantly ionic character and *d* shells forming localized magnetic moments. However, there are some exceptions: In stoichiometric $LiTi_2O_4$, the titanium ions have the formal valence 3.5 ($d^{0.5}$). This compound remains metallic down to the lowest temperature with no indications of charge ordering and even becomes superconducting below ~ 10 K. In $LiV_2O_4$, the vanadium ions at the *B* sites again have a mixed formal valence of 3.5 and hence, utilizing a strictly localized point of view for the *d* electrons, reveal a statistical distribution of $d^1$ and $d^2$ electrons. As will be discussed later, this compound again remains metallic down to the lowest temperatures, but undergoes HF formation. Magnetite, $Fe_3O_4$, which is an inverse spinel [5], should be written as $Fe^{3+}(Fe^{2+}Fe^{3+})O_4$ and reveals a statistical distribution of divalent and trivalent iron ions at the *B* sites. Because intra-atomic correlations in this compound are strong, one can assume that 50% of the *B* sites are in a $Fe^{2+}$ and 50% in a $Fe^{3+}$ configuration. Electronic configurational fluctuations into $Fe^{1+}$ or $Fe^{4+}$ are certainly suppressed and have not to be taken into account. As discussed above, this compound undergoes charge ordering and a concomitant MIT deep within the FiM state. This Verwey transition, which is still not completely unravelled, was the first example of a MIT and will be discussed in this chapter. From this introduction, it becomes clear that mixed-valence spinel compounds have many routes to handle the problem of charge disproportionation: Complete itinerancy of the *d* electrons in lithium titanate, both, partial charge localization and itinerancy in the lithium vanadate and complete CO in magnetite. There is an illuminating discussion on itinerant *d* electrons in spinel compounds by Fulde [399].

The development of models and theories of magnetism are closely related to magnetite and started more than 2500 years ago. At 585 BC Thales of Miletus stated that loadstone attracts iron because it has a soul. Since then, this material fascinated countless scientists. Magnetite is an inverse spinel and undergoes FiM ordering with a sizable spontaneous magnetization at 780 K. As detailed in the introduction, its current applications range from compass needles to magnetic recording devices, it is an important ingredient of many iron ores, and a number of animals or biorganisms use small single magnetite crystals for orientation utilizing the earth magnetic field. A further prominent role of magnetite in solid-state physics originates from the occurrence of a well-pronounced MIT at $T_V \approx 122$ K. It is named after Verwey, who suggested charge ordering of $Fe^{2+}$ and $Fe^{3+}$ ions at the *B* sites as the mechanism driving the dramatic conductivity drop at this phase-transition [30,400,401]. Verwey suggested a low-temperature phase in which $Fe^{2+}$ and $Fe^{3+}$ ions order along two families of chains. A vast number of subsequent experimental and theoretical investigations, starting in the mid-fifties and ongoing until now, including the work by Anderson [23], Mott [402], and many others, have stimulated a still unresolved dispute on the Verwey transition, trying to explain the numerous and partly mysterious aspects of this transition. Anderson [23] pointed out that under the assumption of nearest-neighbour interactions only, there will be an exceptionally large number of states with the same energy as the structure proposed by Verwey and there will be a highly degenerate ground state, typically observed in pyrochlore lattices. Mott's view of the Verwey transition, corresponding to a phase change from an electrically conducting Wigner glass ($T > T_V$) into an insulating Wigner crystal ($T < T_V$), used concepts in terms of tunnelling and variable-range hopping of small polarons [402]. The early work and a number of - at that time - unresolved questions are summarized in [403,404]. In this review, we will focus on the ac and dc conductivity as observed at this MIT and will discuss some recent and modern models and theories trying to capture the essential physics of this transition.

At elevated temperatures, the electrical conductivity is of order ~ 200 $(\Omega cm)^{-1}$ (see inset of Fig. 52) and exhibits a weak temperature dependence only. At 125 K the conductivity shows a sharp and significant drop by three orders of magnitude and exhibits semiconducting temperature characteristics on further cooling as documented in Fig. 52. Below the Verwey transition the conductivity roughly follows a thermally activated behaviour. Originally this transition has been explained in terms of pure CO of $Fe^{2+}$ and $Fe^{3+}$ ions at the *B* sites by Verwey and coworkers and was the very first illuminating example of a temperature driven MIT, stimulating an enormous amount of experimental and theoretical work. At high temperatures, the onset of FiM order at 858 K is indicated by a small dip-like minimum in the temperature dependent conductivity. Obviously magnetic ordering has no big influence on the electrical conductivity in magnetite. At elevated temperatures, above the Verwey transition electron transport between the octahedral iron sites obviously is easily possible and becomes strongly reduced in the charge-ordered regime.

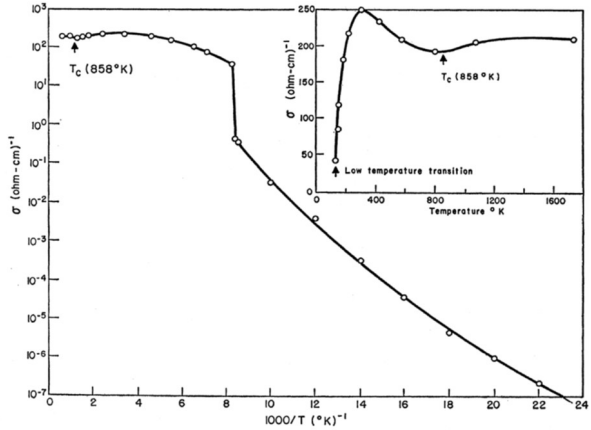

Fig.52. Electrical conductivity of $Fe_3O_4$.
Temperature dependence of the dc conductivity of a magnetite single crystal taken from ~ 1600 K down to ~ 50 K. The main frame shows the results in an Arrhenius representation; the inset shows the conductivity on linear scales. The arrows indicate phase-transition temperatures.
Reprinted figure with permission from Miles et al. [262]. Copyright (1957) by the American Physical Society.

To shed some light on the charge dynamics in magnetite when crossing the MIT transition, Pimenov et al. [405] undertook broadband measurements of the dynamic conductivity combining THz and IR techniques. The frequency dependence of the real part of the dielectric constant $\varepsilon_1$ and of the real part of the conductivity $\sigma_1$ are documented in Fig. 53 for temperatures between 100 and 130 K, for a crystal with $T_V$ = 123 K. The complex data of the frequency-dependent dielectric constant have been obtained from the transmittance and phase shift of a thin $Fe_3O_4$ plate for wavenumbers below 40 $cm^{-1}$, and via a Kramers-Kronig analysis of the reflectance of the thick sample for > 30 $cm^{-1}$. The spectral range can be separated into three distinct regimes. For wavenumbers between 70 and 700 $cm^{-1}$ phonon excitations dominate the spectra and are followed by electronic excitations at higher wavenumbers, which are not further analysed. In this regime, only minor temperature-dependent changes can be observed. For wavenumbers < 70 $cm^{-1}$, free charge carriers dominate the dielectric response, which completely change their character when crossing the Verwey transition. At the lowest temperatures, hopping conductivity with a characteristic slope $\sigma \sim v^s$, describes the real part of the conductivity. For temperatures $T > T_V$ this hopping conduction abruptly changes into a Drude-like behaviour characteristic for free charge carriers. The Drude conductivity is rather small and very narrow and can be fitted assuming a low carrier concentration density and a mean relaxation rate of ~ 40 $cm^{-1}$ corresponding to ~ 5 meV. The conductivity at low wavenumbers coincides nicely with measurements of the dc conductivity, which are indicated in the lower frame of Fig. 53. The dielectric constant (upper frame of Fig. 53) indicates a hopping term, characteristic for localized charge carriers, and turns negative at the metallic side of the transition, as expected for itinerant charge carriers. These results evidence the formation of a coherent itinerant state on the metallic side of the Verwey transition with high mobility but a low concentration of charge carriers. At low temperatures, magnetite shows the characteristic charge dynamics of semiconductors with hopping conductivity of localized charge carriers.

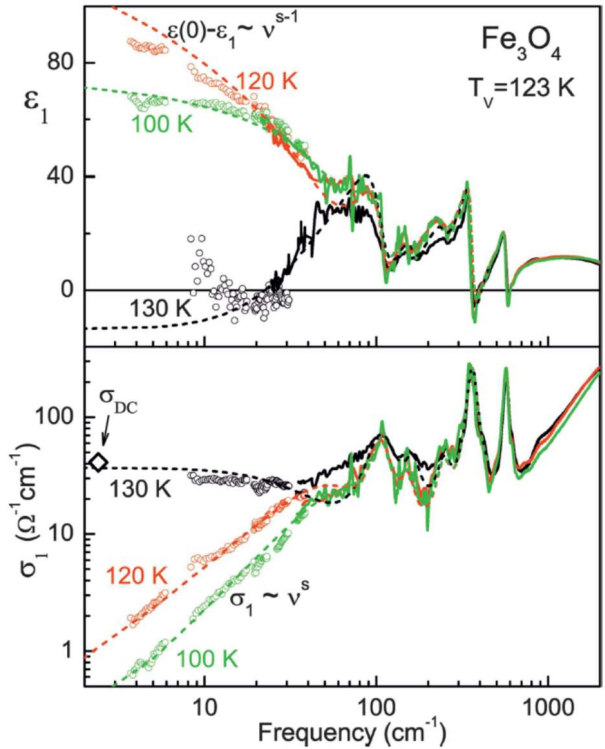

Fig. 53. Optical conductivity in $Fe_3O_4$. Frequency dependence of the real part of the dielectric constant $\varepsilon_1$ (upper panel) and of the real part of the conductivity $\sigma_1$ (lower panel) for a series of temperatures crossing the Verwey transition in magnetite at $T_V = 123$ K. Symbols below 40 $cm^{-1}$ and solid lines above 30 $cm^{-1}$ represent experimental data obtained via THz and FIR techniques, respectively; dashed lines are the results of model calculations. The dc conductivity obtained in 4-point configuration is indicated by an open diamond.
Reprinted figure with permission from Pimenov et al. [405]. Copyright (2005) by the American Physical Society.

While questions concerning the charge dynamics seem to be settled, a number of open questions remain concerning the low-temperature crystallographic structure of magnetite including the orbital degrees of freedom and the driving force for this MIT transition. The crystallographic structure of the ground state remained controversial because significant twinning of crystal domains hampers any analysis of low-temperature diffraction studies. A variety of powder-diffraction refinements and resonant x-ray studies led to different proposals of charge-ordered and bond-dimerized ground states, which are too numerous to be reviewed here. We only mention some recent progress in this field where all the relevant information can be obtained. A full low-temperature superstructure of magnetite, as determined by high-energy x-ray diffraction from a single-domain magnetite crystal was reported by Attfield and coworkers [338,339]. They impressively document that in the low-temperature charge-ordered phase, the localized $d$ electrons are distributed over three iron sites, arranged as linear molecular units, which they call 'trimerons'. In addition, this charge-order pattern, including these three-site distortions, induces substantial off-centre atomic displacements and couples the resulting large electrical polarization to the magnetization, explaining the reported observation of ME effects. Combining these microscopic structural results of a local creation of electric polarization with the results documented in Figs. 40 and 41, where dipolar relaxation and the onset of macroscopic polarization appears well below 50 K, it seems clear that this trimeron formation becomes quasi-static at low temperatures only.

Later-on it was shown that these complex trimeron molecules exist far above the MIT transition and are the driving force of the Verwey transition [406]. Spectroscopic signatures of the low-energy electronic excitations of this trimeron network were reported in Ref. [407] using terahertz spectroscopy. The use of ultrashort laser pulses allowed the authors to demonstrate the critical softening of these trimeron-derived modes and hence, to demonstrate their importance for driving the Verwey transition.

In addition to magnetite, there are two further spinel compounds with (rather poor) metallic conductivity at high temperatures, which undergo a MIT transition below room temperature, namely $CuIr_2S_4$ and $MgTi_2O_4$. However, it appears that the driving mechanism for the MIT transition in these systems is different and that CO is coupled there to spin dimerization. $CuIr_2S_4$ was successfully synthesized and characterized by Nagata et al. [408] and Furubayashi et al. [409]. At 230 K this compound reveals a first-order structural phase transition lowering the crystal symmetry from cubic to tetragonal. This structural-phase transition is accompanied by a drop of the electrical resistivity by almost three orders of magnitude and a concomitant jump in the magnetic susceptibility from Pauli-like paramagnetism at high temperatures to diamagnetism in the low-temperature phase, indicative for spin-dimerization. The diffraction profile in the insulating phase evidences a superstructure, which was not identified in these early studies [409]. In subsequent studies, the crystallographic structure in the insulating low-temperature phase has been determined by Radaelli et al. [39]. Their results indicate that $CuIr_2S_4$ undergoes a simultaneous charge-ordering and spin-dimerization transition. Remarkably, the authors found that the charge-ordered pattern consists of isomorphic octamers of $Ir_8^{3+}S_{24}$ and $Ir_8^{4+}S_{24}$, which structurally are organized as isovalent bi-capped hexagonal rings. This octamer charge pattern represents an increasing complexity compared to the variety of charge-ordered structures, which are typically based on stripes, slabs, or checkerboard patterns.

$MgTi_2O_4$ was synthesized in polycrystalline form by Isobe and Ueda [410]. In this compound, the titanium ions at the $B$ site of the spinel structure constitute a pyrochlore lattice with spin $S = ½$. The authors found a structural-phase transition close to 260 K lowering the crystal symmetry from cubic to tetragonal. These structural changes are accompanied by drastic changes of the electrical conductivity and of the spin susceptibility. The title compound transforms from a bad metal with Pauli paramagnetism at high temperatures to a spin-dimerized insulator below the MIT phase transition. In subsequent high-resolution synchrotron and neutron-powder diffraction experiments, Schmidt et al. [411] solved the low-temperature crystal structure of $MgTi_2O_4$ and found that it contains dimers with alternating short and long bonds, which form a helix-like structure. Additional band-structure calculations characterized the low-temperature structure as an orbitally ordered band insulator. The MIT transition in $MgTi_2O_4$ was characterized by IR spectroscopy, measuring the temperature-dependent opening of the charge gap when crossing the MIT [412]. The authors showed that the spectral changes across the MIT could be well understood utilizing the concept of a one-dimensional Peierls transition driven by the ordering of the titanium $d_{yz}$ and $d_{zx}$ orbitals.

The exotic superstructures observed in $CuIr_2S_4$ and $MgTi_2O_4$ are hard to understand from the almost cubic crystal structure of these spinel compounds. However, from an electronic point of view and taking the orbital degrees of freedom into account, these compounds are one-dimensional systems as discussed in detail by Khomskii and Streltsov [29]. Utilizing 1D chains as the natural building blocks of spinel structures with $t_{2g}$ electrons, these superstructures can be rather naturally explained.

### 3.8. Heavy fermion formation in spinel compounds

In the late seventies of the last century, the topic of heavy fermions (HFs) was rapidly developing in the field of solid-state physics with an incredible amount of theoretical and experimental work over the past decades. HF physics is an important sub-area in the arena of strongly correlated matter, closely linked with the physics of transition-metal oxides, high-$Tc$ superconductors and quantum criticality. Usually HFs are metallic alloys, which contain rare earth ($4f$, e.g., cerium) or actinide ions ($5f$, e.g., uranium). At high temperatures, the magnetization of these compounds follows CW laws, indicative of local-moment magnetism. However, below a characteristic temperature $T^*$ (order 10 – 100 K) they exhibit unusual low-temperature properties, like a linear specific heat, a large temperature-independent Pauli spin susceptibility, and a $T^2$-dependent electrical resistivity, with values pointing towards the existence of exceptionally large masses of the conduction electrons, in many cases as large as a few hundred up to thousand times the free-electron mass. This electronic mass enhancement results from an enormous built-up of a high quasi-particle density of states at the Fermi energy. One well-known source of this exotic behaviour is the Kondo effect, which requires ions with quasi-localized $f$ electrons of nearly integer number. The characteristic temperature $T^*$, usually called Kondo-lattice or coherence temperature, can be interpreted to result from a coherent action of these $f$ electrons establishing the anomalous low-temperature properties. Andres et al. [413] discovered the first known HF system, namely CeAl3. HF superconductivity was identified by Steglich and coworkers in CeCu$_2$Si$_2$ [414], a work that gave enormous stimulus to the field. There are numerous reviews covering theory and experiment of HF compounds. Here we reference two early reviews provided by Stewart [415] as well as by Grewe and Steglich [416]. According to the recent understanding and following a relatively arbitrary definition, at low temperatures HFs are compounds characterized by Sommerfeld coefficients $\gamma \geq 400$ mJ/($f$-atom mol K).

The normal spinel LiV$_2$O$_4$ was first synthesized by Reuter and Jaskowsky [417] and identified as a mixed-valence spinel Li(V$^{3+}$V$^{4+}$)O$_4$, where the pyrochlore lattice is statistically occupied by vanadium ions with either $d^1$ or $d^2$ electron configurations. In single crystals, Rogers et al. [384] reported the occurrence of metallic conductivity down to the lowest temperatures. Rather unexpectedly, on decreasing temperatures a crossover from localized moment magnetism to heavy Fermi-liquid behaviour was reported by Kondo et al. [418] utilizing magnetic susceptibility, heat-capacity, and NMR techniques. At the lowest temperatures, the authors found an electronic heat-capacity coefficient $\gamma \sim 420$ mJ/(mol K$^2$), which is exceptionally large for transition-metal compounds. By subtraction off the phonon-derived specific heat, as determined via measurements of the non-magnetic and non-superconducting Li$_{4/3}$Ti$_{5/3}$O$_4$, it was possible to determine the purely magnetic heat capacity at low temperatures. This allowed the authors to identify a continuous increase of the magnetic heat capacity, which was well described using a conventional $S = 1/2$ single-ion Kondo model and hence, led to the characterization of LiV$_2$O$_4$ as the first $d$-derived HF metal. In subsequent work, the structural properties were studied in detail [419] and at low temperatures, the authors identified a strongly enhanced electronic Grüneisen parameter, similar to observations in some $f$-electron HF compounds.

Subsequently, transport, magnetic susceptibility, and specific-heat measurements were reported on hydrothermally-grown LiV$_2$O$_4$ single crystals by Urano et al. [420]. Similar to the findings reported for polycrystalline samples documented above, these results also reveal the formation of a heavy-mass Fermi liquid below a coherence temperature $T^* \sim 20 – 30$ K. Fig. 54 documents the temperature dependencies of electrical resistivity, heat capacity, magnetic susceptibility, and Hall coefficient, all quantities documenting the characteristic features of HF systems below the crossover temperature. At low temperatures, the resistivity significantly decreases and shows a clear $T^2$ dependence at the lowest temperatures [inset of Fig. 54(a)]. This $T^2$ dependence signals the importance of electron-electron scattering of heavy-mass quasi particles. Characteristic for HF behaviour, $C/T$, representing the temperature dependence of the Sommerfeld coefficient, reveals a continuous increase on cooling, reaching values close to 400 mJ/(molK$^2$) [Fig. 54(b)], and the magnetic susceptibility increases towards low temperatures reaching strongly increased values [Fig. 54(c)]. On cooling, the Hall coefficient $R_H$ changes sign around 50 K and develops a strong temperature dependence with a pronounced peak at ~ 30 K, a behaviour that is similar to that typically observed in conventional HF intermetallics below their coherence temperature [Fig. 54(d)].

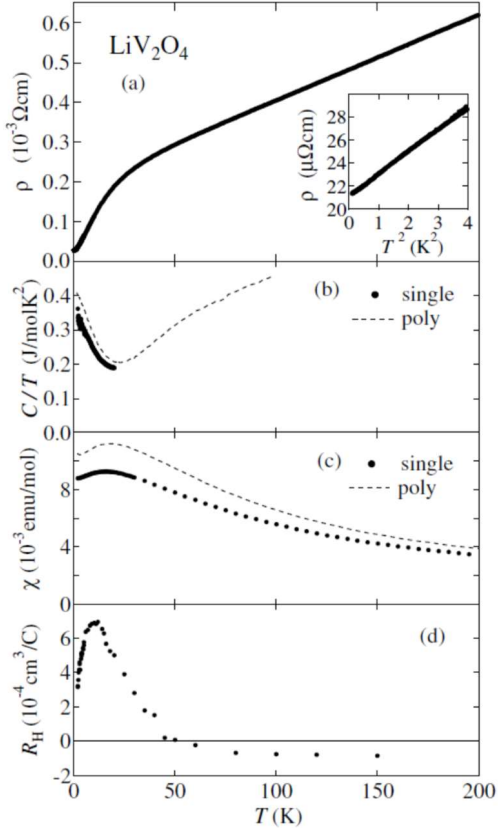

Fig. 54. HF characteristics observed in single crystalline LiV$_2$O$_4$.
Temperature dependence of various physical properties, characteristic for HF systems. (a) Electrical resistivity $\rho$, (b) specific heat $C$ plotted as $C/T$, (c) magnetic susceptibility $\chi$, and (d) Hall coefficient $R_H$. In (b) and (c), $C/T$ and $\chi$ data, as obtained from polycrystalline samples are shown for comparison. The inset of (a) shows $\rho$ vs. $T^2$ to document the HF-like $T^2$ dependence of the resistivity at the lowest temperatures.

Reprinted figure with permission from Urano et al. [420]. Copyright (2000) by the American Physical Society.

To unravel the exotic HF behaviour of this transition-metal compound a number of quasielastic and inelastic neutron scattering experiments were performed by Krimmel et al. [421], Lee et al. [422], Murani et al. [423], and Tomiyasu et al. [424]. In strongly correlated electron systems, the relaxation rate $\Gamma(Q,T)$, which can be measured as function of momentum transfer $Q$ via the generalized dynamic susceptibility in neutron-scattering experiments, gives a rough estimate of the hybridization strength of the localized spins with the conduction electrons and, at low temperatures, is a rough measure of the characteristic temperature $T^*$. All these spectroscopic experiments agree that strong spin fluctuations are built up in LiV$_2$O$_4$ at low temperatures, a characteristic signature of strongly correlated materials. The relaxation rate exhibits a weak temperature dependence only and was described by a square-root [421] or a roughly linear temperature dependence [422] with a residual value of the relaxation rate of approximately 1 - 2 meV when extrapolated towards zero temperature, again a characteristic feature of conventional HF systems with a characteristic coherence temperature of order 20 K. However, there was no agreement concerning the origin of these spin fluctuations, as originating rather from hybridization effects between conduction electrons and localized moments as observed in Kondo-type materials or being driven by strong magnetic frustration. In Fig. 55 we reproduce a typical spin-fluctuation spectrum as observed by Murani et al. [423]. Here the dynamic structure factor $S(Q,\omega)$, a direct measure of the dynamic susceptibility multiplied by a detailed-balance temperature factor, is colour-code plotted in the energy ($\hbar\omega$)

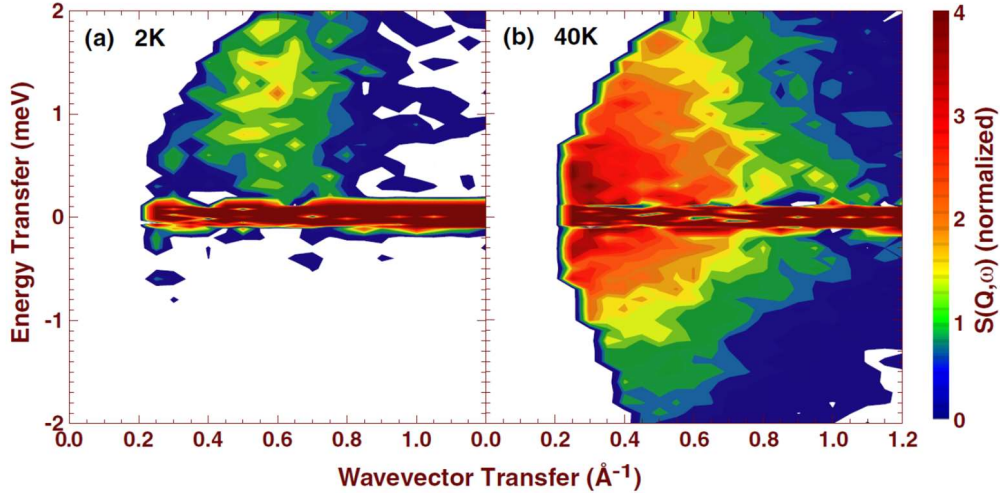

Figure 55. Neutron scattering results in LiV$_2$O$_4$. Two-dimensional contour plots of the normalized spectral distribution $S(Q, \omega)$ as observed at 2 (a) and 40 K (b). The energy transfer is plotted vs. the momentum transfer. The colour code is indicated in the bar at the left side of the figure. Note that at low temperatures the intensity is peaked at $Q$ = 0.6 Å$^{-1}$, but is shifted to low momentum and energy transfers at 40 K. Figure taken from Murani et al. [423], with permission from IOP Publishing Ltd. © 2004

vs. momentum transfer ($Q$) plane. At 2 K [Fig. (55a)], a broad intensity maximum close to momentum transfers of ~ 0.5 Å$^{-1}$ and to energy transfers ~ 1.5 meV, signals dominant AFM fluctuations. With increasing temperatures, this fluctuation spectrum is shifted towards low energies and low momentum transfer, which is nicely documented in Fig. 55(b) at 40 K. From an experimental point of view based on the PM scattering rate, LiV$_2$O$_4$ at low temperatures is dominated by AFM spin fluctuations, while at higher temperatures it has similarities with a metal close to a FM instability. This temperature-induced crossover from AFM to FM fluctuations was further elucidated by a detailed analysis of the magnetic-scattering cross section in polarized neutron studies [425]. It was proposed that the increasing importance of FM correlations may result from an increasing importance of Hund's rule coupling between itinerant and localized $d$ moments [426].

Staring with the pioneering work by Kondo et al. [418], a series of NMR experiments was published in LiV$_2$O$_4$, as well as in zinc and titanium doped samples, mainly utilizing $^7$Li and $^{51}$V line width, Knight shift, and spin-lattice relaxation [427,428,429,430,431,432]. The temperature dependence of the linewidth usually is determined by an inhomogeneous broadening due to local magnetic fields and often can be taken as measure of homogeneity and purity of the samples under investigation. The Knight shift provides a direct measure of the microscopic and local static susceptibility, while the spin-lattice relaxation rate $1/T_1$ traces the temperature dependence of the imaginary part of the dynamic susceptibility and is determined by $1/T_1 = T \chi_0 / \Gamma(T)$, where $\chi_0$ denotes the static bulk susceptibility and $\Gamma(T)$ the temperature-dependent magnetic relaxation rate at zero momentum transfer. These are the most decisive properties of the electronic environment of the NMR nuclei, being responsible for the nuclear relaxation process. In heavy-electron systems, the spin-lattice relaxation is a linear function of temperature with a highly enhanced slope or alternatively, $(TT_1)^{-1}$ is a constant with a relatively large value when compared to conventional metals with effective electron masses of order 1.

On decreasing temperatures, the spin-relaxation rate in LiV$_2$O$_4$ reveals a continuous CW-like increase, passes through a maximum close to 50 K and decreases on further cooling. The peak-like feature provides a rough measure of the characteristic temperature $T^*$ where the compound crosses from a localized-moment system into a HF state with quenched moments.

Doping experiments made clear that $LiV_2O_4$ is very close to magnetic order and Zn doping immediately induces a spin-glass like low-temperature phase [429,432,433]. All the NMR results were interpreted with reference to conventional *f*-derived HF systems, but also mentioning the importance of magnetic frustration of the mixed-valence vanadium ions located at the vortices of the pyrochlore lattice. It also has to be noted that the results obtained by Fujiwara [427] were interpreted in terms of a spin-fluctuation theory as developed by Ishigaki and Moriya [434].

A convincing comparison of relaxation rates as observed in transition-metal oxides including $LiV_2O_4$ as compared to those observed in conventional *f*-derived HF systems has been provided by Krug von Nidda et al. [435] and is reproduced in Fig. 56. In the upper frame, spin-relaxation rates $1/T_1$ as observed in $^7$Li NMR experiments in $LiV_2O_4$ are compared with those observed in conventional HF compounds $CeCu_2Si_{2-x}Ge_x$. In the lower frame, the temperature dependence of the linewidth as observed in ESR experiments is compared for the Gd doped HF compound $CeCu_2Si_2$ and the transition-metal oxides $Gd_{1-y}Sr_yTiO_3$. Depending on the electronic filling of the Ti $3d$ bands controlled by the strontium concentration $y$, the latter substitutional series undergoes a MIT on decreasing $y$ between $0.2 > y > 0.1$, realizing a Mott insulator for effective half filling at $y = 0$ ($Ti^{3+}$, $3d^1$). Close to the MIT the electronic density of states at the Fermi energy is strongly enhanced resulting in a HF-like behaviour in the metallic regime of the phase diagram. In metals, the ESR linewidth is proportional to the spin-lattice relaxation rate. The solid lines through the experimental data of the Ce compounds indicate fits by the dynamic susceptibility of the fluctuating $Ce^{3+}$ spins transferred via RKKY like interactions to the $^{63}Cu$ nuclear spins and to the $Gd^{3+}$ electron spins as described in Refs. [436,437]. For an appropriate description of the dynamic susceptibility in $Gd_{1-y}Sr_yTiO_3$ the itinerant character of the Ti electrons was taken into account, following Ishigaki and Moriya [434], who developed a theoretical description of relaxation around a magnetic instability in metals.

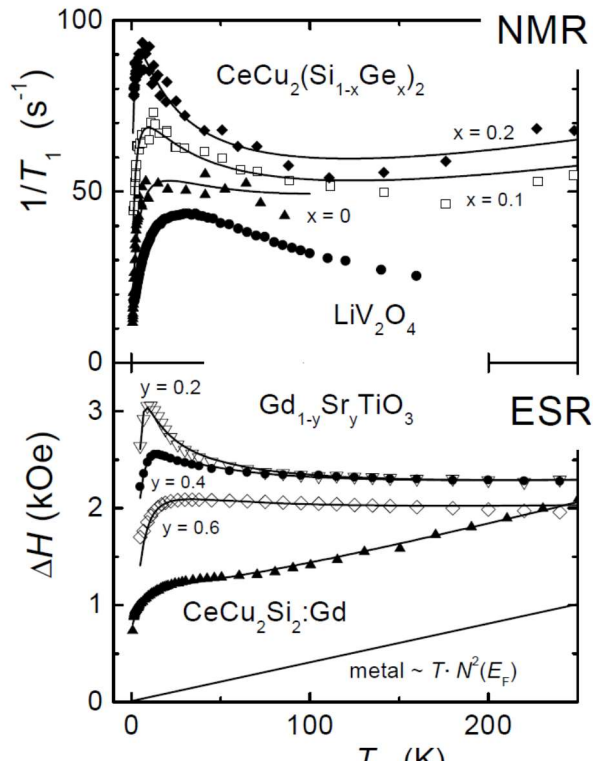

Figure 56. Temperature dependence of the relaxation rates in heavy-fermions compounds and transition-metal oxides. Upper frame: Temperature dependence of the NMR relaxation rate $1/T_1$ of $^{63}$Cu in $CeCu_2(Si_{1-x}Ge_x)_2$ compared to that obtained from $^7$Li in $LiV_2O_4$. Lower frame: Gd-ESR linewidth $\Delta H$ in $CeCu_2Si_2$ obtained after doping with 1% $Gd^{3+}$ at the Ce site. Additionally, $\Delta H(T)$ of $Gd^{3+}$ ESR in the metallic regime of $Gd_{1-y}Sr_yTiO_3$ on approaching the MIT at $y$ slightly below 0.2 is shown. The solid line at the bottom of the lower frame indicates the Korringa law as expected in a conventional metal with a regular density of electronic states. Figure taken from Krug von Nidda et al. [435]. With kind permission of The European Physical Journal (EPJ).

In all the compounds investigated, the relaxation rates exhibit an increase on cooling, pass through a maximum, and linearly decrease on further cooling towards the lowest temperatures. The low-temperature slopes of the relaxation rates correspond to an enormous Korringa relaxation due to a significantly enhanced electronic density of states at low temperatures. In the lower frame, the typical Korringa relaxation as observed in normal metals is indicated. The maximum in the relaxation rate corresponds to the crossover from itinerant HF to local-moment behaviour and the high-temperature behaviour indicates the CW-like behaviour of the magnetic susceptibility in the regime where local-moment magnetism is recovered. From a theoretical point of view, utilizing dynamical mean–field theory, the authors documented that the dynamic susceptibilities as calculated for the Hubbard model and for the periodic Anderson model qualitatively look rather similar. While the Hubbard model serves as representative microscopic model for the MIT transition, the periodic Anderson model is one of the canonical models for the heavy–fermion formation in Kondo–lattice systems. In both models a narrow quasiparticle peak appears at low temperatures and energies, resulting in strongly enhanced effective masses. It documents that on–site hybridization like in the periodic Anderson model and electronic correlations in systems close to half conduction–band filling like in the Hubbard model yield similar spectral densities and similar dynamic susceptibilities.

Over the years, numerous models and theories were developed to interpret the HF behaviour of $LiV_2O_4$. It is out of the scope of this review to adequately discuss the development of these theories. Here we provide a reference list of the most important publications in chronological order of appearance [438,439,440,441,442,443,444,445,446,447,399,448,449]. Roughly speaking, the theories can be classified into two groups, either focusing on Kondo-like aspects important for conventional HF systems or on the importance of magnetic frustration and the closeness of the title compound to magnetic order and to a MIT transition. Concerning the HF picture, in ab-initio calculations, like LDA + U schemes or similar band-structure calculations, it was assumed that with respect to the mixed valence of the vanadium ions, with an overall $d^{1.5}$ configuration, one electron is localized on the pyrochlore lattice establishing a local moment with spin S = 1/2, while the remaining 0.5 electrons are itinerant and partly fill a relatively broad band. On decreasing temperatures, local moments and itinerant charge carriers hybridize, quench the local magnetic moments and establish a low-temperature state with a strongly enhanced density of states at the Fermi energy. This ansatz is analogous to what is assumed in conventional $f$-derived HF systems, but is unusual in that both electronic species, local moments and conduction electrons, belong to the same $d$ electron manifold. References [438,441,446] are representative models preceding in this way and explaining the exotic low-temperature properties of $LiV_2O_4$. A number of models start with the dominance of strong magnetic frustration [439,442,443,444], trying to explain in this way the HF-like properties at low temperatures. Finally, triggered by photoemission-spectroscopy studies, which revealed a sharp peak-like structure at ~ 4 meV above the Fermi energy [450], the electronic properties of $LiV_2O_4$ were recalculated using realistic LDA + DMFT (Dynamical Mean-Field Theory) methods and found good agreement with the photoemission experiments [448]. According to these calculations, the physical origin of the peak structure is not a conventional hybridization, but results from the fact that this compound is very close to a Mott-Hubbard MIT transition.

## 3.9. Colossal magneto-resistance spinels

Giant magnetoresistance, which, e.g. in metallic multilayers, is caused by spin-dependent electron scattering, in the past has attracted considerable attention in the area of fundamental research, but also for device applications as magnetic recording heads. In the early nineties of the past century, so-called CMR effects were observed in Ba- [451] or Ca-doped [452] LaMnO$_3$ perovskite-derived thin films, reaching magneto-resistance (MR) values MR = [R(0) - R(H)]/R(0)] of 60% and more than 1000 %, respectively. It was speculated that these colossal MR values result from a phase transition of a high-temperature insulating AFM into a metallic and FM phase. Somewhat later, similar CMR effects were observed in FeCr$_2$S$_4$ and in Fe:CuCr$_2$S$_4$ compounds [31]. A representative result is depicted in Fig. 57. Here the temperature dependence of the magnetization $M$ (upper frame) at various external magnetic fields is shown together with the thermopower $S$ (middle frame) and the resistivity $\rho$ at different fields (lower frame, left scale). Here also the temperature dependence of the MR is shown as determined at 6 T (right scale). The upper frame reveals the weakly field dependent onset of spontaneous magnetization around the onset of FiM ordering, which for FeCr$_2$S$_4$ was discussed in detail in chapter 3.2 (see Fig. 14). The enhancement of the thermopower close to the magnetic ordering transition was recognized already much earlier [453] and was explained by the fact that at the magnetic transition temperature the charge carriers acquire additional kinetic energy, due to magnon-hole interactions. The CMR effects are documented in the lower frame of Fig. 57. The origin of this effect remains unclear: In the doped perovskite manganates, it is thought that the double-exchange mechanism plays an important role, whereby charge transport is enhanced by the magnetic alignment of neighbouring Mn ions with different valence configurations Mn$^{3+}$ and Mn$^{4+}$. Heterovalency plays no role in FeCr$_2$S$_4$ and hence, cannot explain the experimentally observed CMR effects. It can rather be associated with spin-disorder scattering.

A systematic study of the evolution of the electric resistivity and CMR effect dependent on the copper content was performed on Fe$_{1-x}$Cu$_x$Cr$_2$S$_4$ single crystals by Fritsch et al. [454].

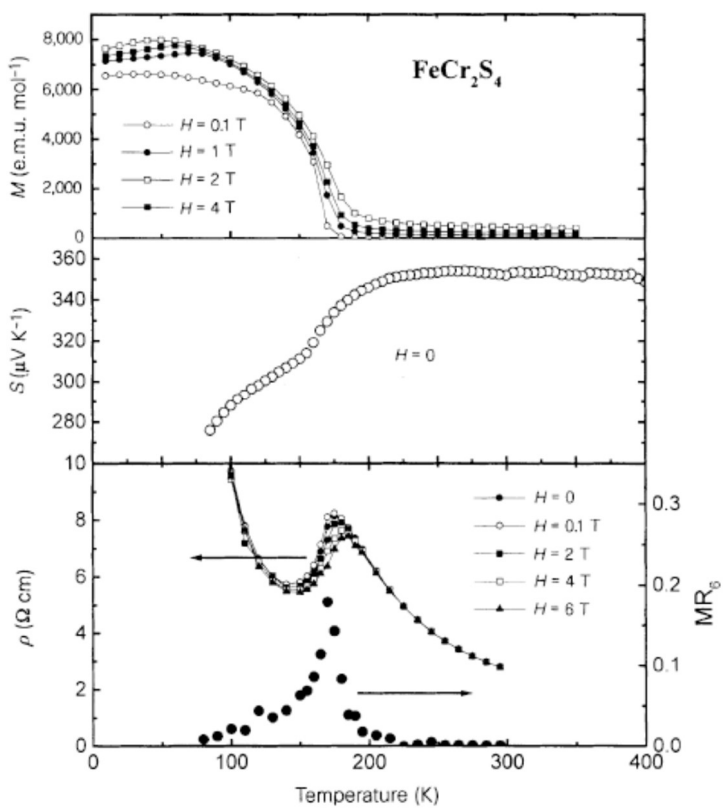

Figure 57. Colossal magneto-resistance effects in FeCr$_2$S$_4$.
Upper frame: Temperature dependent magnetization $M$ at various magnetic fields (upper frame).
Middle frame: Temperature dependence of thermopower $S$ at zero field.
Lower frame: $T$ dependence of resistivity $\rho$ at various magnetic fields (left axis) and magnetoresistance obtained at 6 T MR$_6$ = [$\rho$ (0) - $\rho$(6))/$\rho$(0)] (right axis).

Material from: Ramirez et al. [31]. Copyright © 1997, Springer Nature.

As shown in the top frame of Fig. 58 for x ≤ 0.5 the temperature dependence of the resistivity is qualitatively similar to that of x = 0, i.e. increasing on decreasing temperature in the PM regime with a peak close to $T_C$, then decreasing below $T_C$, passing a minimum, and again increasing down to lowest temperatures. While $T_C$ is monotonously increasing with $x$, as documented in Fig. 50, the resistivity minimum first drops from about 140 K for $x = 0$ down to 70 K for $x = 0.1$ and then gradually returns to 140 K for $x = 0.5$. In that study, the absolute value of the resistivity at room temperature drops from about 2 Ωcm for $x = 0$ down to 10 mΩcm for $x = 0.1$ and then increases up to 25 mΩcm for $x = 0.5$. For higher $x$ (not shown) the room-temperature resistivity passes a maximum of about 250 mΩcm at x=0.6, drops down to 1.5 mΩcm at $x = 0.7$, and remains on a level of a few mΩcm at higher x [455]. At these higher Cu contents, the anomaly at $T_C$ is less pronounced, but a slight increase of the resistivity with decreasing temperature is always observed.

In general, the conductivity in this substitutional series is explained to arise from hole doping by monovalent $Cu^+$ resulting in trivalent $Fe^{3+}$ on the A site for $x < 0.5$. Hence, electron hopping between $Fe^{2+}$ and $Fe^{3+}$ dominates the electrical-transport properties in this range of copper concentration. This hopping becomes more effective in case of magnetic order. In analogy to Zener's double exchange between ferromagnetically aligned $Mn^{3+}$ and $Mn^{4+}$ ions in CMR manganites, a triple exchange between two ferromagnetically aligned $Fe^{2+}$ and $Fe^{3+}$ ions involving the antiferromagnetically aligned $Cr^{3+}$ ions was suggested [456] in agreement with electronic band-structure calculations [457]. For $x > 0.5$, when only $Fe^{3+}$ exists, mixed valence of $Cr^{3+}$ and $Cr^{4+}$ is suggested to drive the conductivity in terms of a regular double-exchange process or alternatively hole conductivity in the sulphur $3p$ band is possible. However, a complete single-crystal study of the copper-rich side does not exist up to now, because stable single crystals have been obtained up to $x = 0.8$ so far and higher concentrations exist only in polycrystalline form.

Regarding the MR, significant effects have been observed for the iron-rich side only (Fig. 58, middle frame). Starting from high temperatures in the PM regime, one observes a positive magnetoresistance with a peak at $T_C$, as already reported for pure $FeCr_2S_4$. On further decreasing temperature and for the external magnetic field aligned along the hard <111> direction, the MR changes its sign and diverges on approaching zero temperature. But for the magnetic field applied along the easy <100> direction, the magnetoresistance remains positive, but small compared to the peak at $T_C$. This behaviour is directly connected to the increase of the magnetic anisotropy on decreasing temperature, which is documented by the concomitant increase of the magnetic anisotropy as obtained from the significant shift of the FMR fields (Fig. 58, bottom frame). This magnetic anisotropy due to the $Fe^{2+}$ Jahn-Teller ions and the related CMR effect weaken on increasing Cu+ content as the $Fe^{2+}$ content decreases. These results show that the electric conductivity strongly depends on the mutual orientation of the magnetic sublattices of iron and chromium in accordance with the suggested triple-exchange mechanism.

Note that the single crystals investigated in Refs. [454,455] were grown by chemical transport reaction using chlorine as transport agent. Later on microprobe analysis revealed small but for the conductivity non-negligible amounts of $Cl^-$ ions on the $S^{2-}$ sites. Therefore, recent, so far unpublished reinvestigations have been performed on single crystals, which were grown using bromine as transport agent, which is not built in on the sulphur sites. It turned out that qualitatively the behaviour of the bromine-grown series is comparable to that of the chlorine-grown one, while absolute values of transition temperatures and resistivity may differ and slight shifts of the phases with respect to the Cu content $x$ show up.

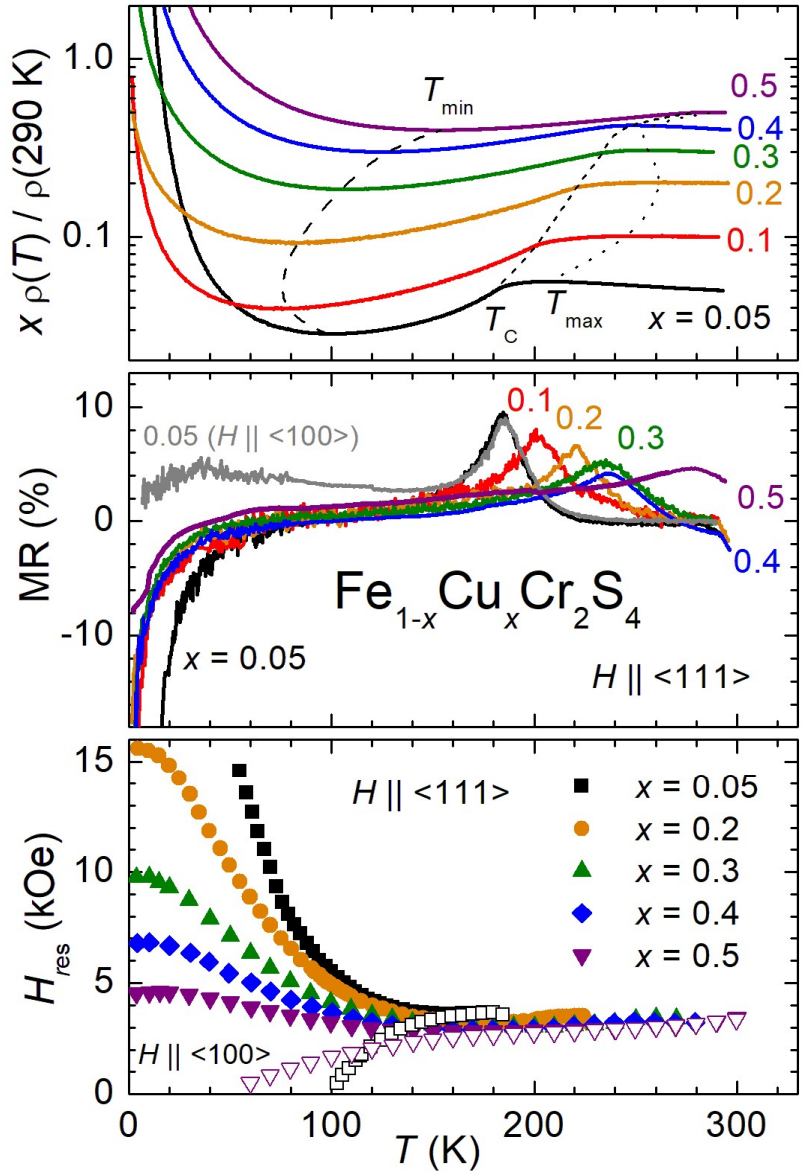

Figure 58. Temperature dependence of electric resistivity (top), magneto resistance in an external field of 50 kOe (middle), and FMR field obtained at a microwave frequency of 9.4 GHz (bottom) of $Fe_{1-x}Cu_xCr_2S_4$ single crystals ($x \leq 0.5$). The resistivity is normalized to its value at 290 K and scaled by the copper content $x$ to separate the data of different $x$. Temperatures of relative resistivity minima $T_{min}$ and maxima $T_{max}$ as well as the Curie temperature $T_C$ are indicated by dashed, dotted, and short-dash lines, respectively. For all samples, the magnetoresistance is shown for magnetic fields applied along the hard <111> direction. For x=0.05 the data for the field applied along the easy <100> direction are included. The FMR shift from its high-temperature value, is a measure for the magnetic anisotropy, closed (open) symbols are chosen for the magnetic field applied along the hard (easy) direction. The data for this figure were taken from Ref. [454].

### 3.10. Spinels under extreme conditions

3.10.1. Spinels in high magnetic fields

The enormous progress in measuring a variety of physical properties, like magnetization, sound velocity, and damping, thermal expansion, pyrocurrent or magnetocaloric effects, in high magnetic fields was documented already above. Experiments in pulsed magnetic fields are possible up to 100 T in different high-field laboratories around the world and have contributed enormously to our understanding of the complex high-field physics, including that of spinel compounds. The fractionalized magnetization plateaus in some chromium oxides, as documented in Fig. 24, or the complex spin states including the MF properties in $MnCr_2S_4$, documented in Fig. 47, are illuminating examples of the power of high-field experiments. Over the years, many more experiments of oxide and chalcogenide spinels have been published, establishing ($H,T$)-phase diagrams of outstanding complexity.

In first respect, it seemed intriguing to measure different physical properties in the chromium spinels exhibiting fractionalized magnetization plateaus. Sound velocities have been measured in $CdCr_2O_4$ [458,459] showing a half-magnetization plateau at high fields (Fig. 24). Longitudinal and transverse sound velocities were measured along the crystallographic [111] direction with the external magnetic field also parallel to [111]. A previously unknown transformation of the magnetic structure below $T_N$ was found and ascribed to a magnetic phase transition between different spin-spiral phases, which are transformed into canted phases at higher magnetic fields. At 4.2 K both sound velocities yield significant anomalies close to 30 and 60 T, exactly corresponding to the field limits of the half-magnetization plateau. The sound-velocity anomalies show clear hysteresis effects indicating a first-order character of these transitions. By applying an exchange-striction model, the authors were able to quantitatively describe the field dependence of the sound velocity below and above the magnetization plateau [459].

A detailed ($H,T$)-phase diagram up to 60 T has also been established for $CoCr_2O_4$ utilizing magnetization, sound velocity and damping in pulsed magnetic fields [460] and is reproduced in Fig. 59. At zero magnetic fields, the sequence of spin structures, collinear FiM, IC, and commensurate spin spiral was identified as function of temperature. These phases extend to higher fields, with the notable exception that the commensurate spiral phase is interrupted by a metastable two-phase region, which is followed by a high-field phase of unknown spin structure. At low temperatures and on decreasing fields, this high-field phase is stabilized down to the lowest fields, at least on the time scale of the pulsed experiments. It is interesting that, while significant anomalies were detected in the ultrasound experiments, at all temperatures the magnetization evolves continuously and evidences no sign of magnetic phase transitions, neither in passing the commensurate-to-IC phase boundary nor in entering the high-field phase regime detected in the ultrasound experiments.

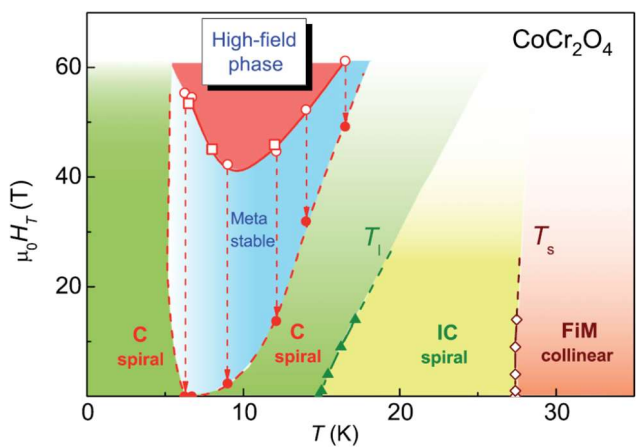

Figure 59. Schematic low-temperature ($H,T$)-phase diagram of $CoCr_2O_4$. Phase boundaries between the metastable and the helical commensurate (C) phase are shown by dashed lines. The arrows connect the boundaries observed in increasing (open symbols) and decreasing fields (solid circles). The open squares and circles refer to ultrasound data measured in pulsed fields applied along the [001] and [111] directions, respectively.
Reprinted figure with permission from Tsurkan et al. [460]. Copyright (2013) by the American Physical Society.

High-field experiments were also performed for $ZnCr_2S_4$ and $ZnCr_2Se_4$. As is documented already in Fig. 24, in both compounds the magnetization evolves continuously without any plateau up to the field-induced FM phase. Complete field polarization is established in the selenide well below 10 T, and in the sulphide beyond 40 T. A detailed $(H,T)$-phase diagram for $ZnCr_2S_4$ was measured up to 50 T and at low temperatures exhibits a sequence of magnetic phases with different ordering wave vectors in different crystallographic phases [223]. A $(H,T)$ phase diagram below 25 K and 12 T was measured for $ZnCr_2S_4$. In this compound, a field-induced FM cubic phase is established beyond 10 T. It is followed at lower fields probably by a spin-nematic phase and finally by a tetragonal phase with a spin-spiral ground state [224].

For exploration of the phase diagram of $LiCuVO_4$ up to full polarization of all copper spins at the saturation field $\mu_0 H_{sat}$ = 45 T, high-field $^{51}$V and $^{7}$Li NMR measurements contributed valuable information [461,462]. In the IC spin-modulated phase above 7.5 T the $^{51}$V NMR spectra reveal a characteristic double-horn shape for the magnetic field applied along the crystallographic *c* axis. This results from the fact that the magnetic moments are aligned parallel to the applied field ***H*** and their values alternate sinusoidally along the magnetic chains. As expected for such a modulated spin structure, the spin-spin relaxation time $T_2$ at the vanadium sites depends on the particular position of the probing $^{51}$V nucleus with respect to the magnetic copper moments. The largest $T_2$ value is observed for the vanadium nuclei next to the $Cu^{2+}$ ion with the largest ordered magnetic moment [461].

Close to saturation, a long-range ordered spin-nematic phase was theoretically predicted for the frustrated chain antiferromagnet [463]. Indeed, on approaching the saturation field, the double-horn shape of the $^{51}$V NMR spectra transforms into a single Lorentzian line around $\mu_0 H_{c3}$ = 40.5 T. Although the magnetization curve $M(H)$ shows a linear increase of magnetization in the field range $H_{c3} < H < H_{sat}$ [464], the internal field corresponding to the peak of the NMR spectra remains unchanged above 41.4 T, indicating that the moments surrounding the majority of vanadium nuclei are already saturated in this field range. Moreover, the temperature- and field-dependence of the $^{7}$Li nuclear spin-lattice relaxation rate reveals an energy gap expected for bound-magnon pairs above saturation at 41.4 T. Thus, the theoretically predicted nematic ordered phase is supposed to be realized in the narrow field range between 40.5 T and 41.4 T [462].

As discussed already in chapter 3.4.2, the most detailed work has been performed on MF $MnCr_2S_4$ [26,353]. In this compound, using a variety of techniques like magnetization experiments, ultrasound propagation and damping, thermal expansion and pyrocurrent measurements, a complete $(H,T)$-phase diagram was established, with complex spin states, which can be interpreted as symmetry analogues of spin supersolid and spin superliquid phases. The most recent version of this phase diagram has been determined by Miyata et al. [354] up to pulsed magnetic fields of 100 T and is documented in Fig. 60. At low temperatures, the time-honoured YK structure is followed by a phase where the triangular spin structure of the manganese spin system is tilted, exhibiting coexisting order of longitudinal and transverse spin components, and on further increasing fields followed by an ultra-stable magnetization plateau: Here the exchange field induced from the chromium moments is fully compensated by the external magnetic field and the manganese spins keep their intrinsic AFM spin structure, as in an *A*-site spinel without field. Beyond 50 T the plateau region is followed by a spin structure, again with coexisting transverse and longitudinal spin order of the manganese spins and finally by an inverse YK structure, which so far never has been observed before, and finally by the field-polarized FM beyond 90 T.

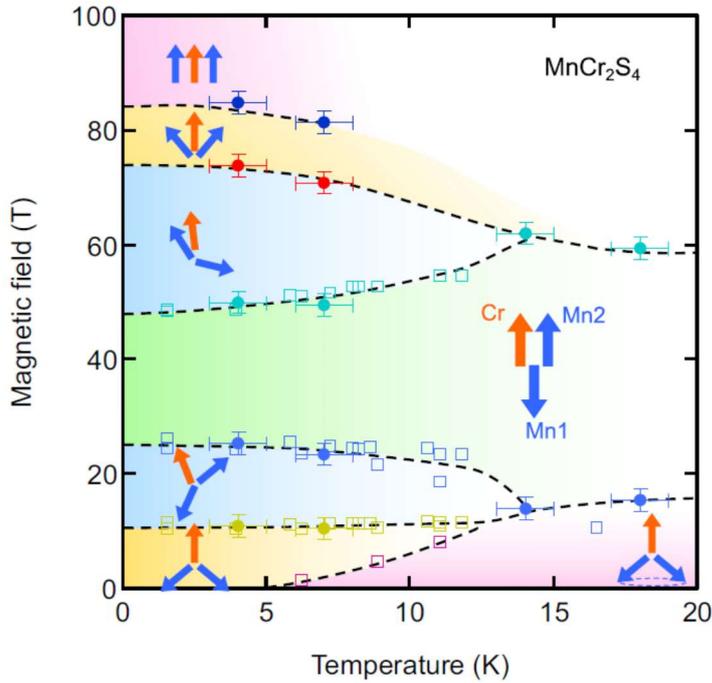

Figure 60. (*H,T*)-phase diagram of MnCr$_2$S$_4$.
The phase boundaries were obtained from the magnetization, ultrasound propagation, and magnetostriction experiments. The solid circles were obtained in ultrahigh-field experiments. Typical spin structures of chromium (orange) and manganese moments (blue) are illustrated. The Cr spins in most phases are aligned with the external field. The Mn spins show a complex, strongly field dependent order.
Reprinted figure with permission from Miyata et al. [354]. Copyright (2020) by the American Physical Society.

Miyata et al. [354] succeeded to calculate the phase boundaries and the magnetic-field dependence of magnetization, ultrasound and thermal expansion in MnCr$_2$S$_4$, assuming a minimal model with magnetic Cr-Cr, Cr-Mn, and Mn-Mn interactions, including a biquadratic term between manganese and chromium spins, utilizing classical Monte-Carlo methods. The resulting spin structures, which follow from these calculations are indicated in Fig. 60. Notably, in the intermediate phases with the tilted triangular structure of the manganese spins, which appear at the phase boundaries of the plateau phase, the chromium moments are slightly tilted off from the direction of the external field. The optimized parameters, which were found from these model calculations yield values of the leading magnetic exchange paths, $J_{Cr-Cr}$ = - 9.1 K, $J_{Mn-Cr}$= 3.1 K, and $J_{Mn-Mn}$= 3.4 K and also reproduce the two magnetic phase transition temperatures $T_C \approx$ 65 K and at $T_{YK} \approx$ 5 K The bi-quadratic exchange between manganese and chromium spins was found to be $b_{Mn-Cr}$ = 0.04 K.

As documented in Fig. 60 and discussed above, a complex tilted triangular structure with the chromium moments slightly tilted off the direction of the external magnetic field shows up just before the plateau region of the *H,T* phase diagram is reached. The existences of this spin structure, which was characterized as super-solid spin phase, was further supported by element-selective soft x-ray circular dichroism measurements on MnCr$_2$S$_4$ in external magnetic fields up to 40 T [465]. In addition, in these studies additional thermodynamic and thermo-caloric experiments established an unusual tetracritical point in the *H,T* phase diagram

3.10.2. Spinels under hydrostatic pressure

From the very beginning, research on magnetism of spinel compounds tried to unravel the dependence of different exchange interactions on lattice spacing. This was done by comparing ordering temperatures and CW temperatures of different compounds and investigating the evolution of these characteristic temperatures in solid solutions as function of lattice spacing, ion-ion separation, and bonding angles. The first detailed study in this respect was performed by Baltzer et al. [106] trying to relate the characteristic Curie or Néel and CW temperatures of the FM semiconducting chalcogenides $ACr_2X_4$ ($A$ = Cd, Hg; $X$ = S, Se) on the 90° FM Cr-$X$-Cr SE interactions and on the more distant AFM Cr-$X$-$A$-$X$-Cr exchange via three intermediate ions. In the sulphides and selenides with the larger lattice constants compared to the oxides, the direct AFM NN Cr-Cr exchange is negligible. For $CdCr_2S_4$, $CdCr_2Se_4$ and $HgCr_2Se_4$ compounds with FM ground state and metamagnetic $HgCr_2S_4$, Baltzer et al. [106] introduced a FM NN exchange $J$ and an additional AFM NNN exchange $K$ averaging over 30 similar but slightly different exchange paths and calculated the observed ordering and CW temperatures depending on the parameters $J$ and $K$. With increasing lattice spacing when going from sulphides to selenides, an increase of the value of the FM NN exchange $J$ and decrease of the AFM NNN exchange $K$ was show.

In the early work, experiments on the pressure dependence of ordering temperatures of spinel compounds were conducted to check these ideas concerning the competing FM and AFM exchange and to determine the pressure dependencies of $J$ and $K$ [466,467]. Srivastava [466] measured the pressure dependencies of the Curie temperatures $T_C$ of $CdCr_2S_4$, $CdCr_2Se_4$ and $HgCr_2Se_4$ up to 1 GPa. From the negative sign of the pressure derivative $dT_C/dp$ for these three FM compounds a decrease of the AFM exchange $K$ under pressure was concluded, however assuming a week dependence of the FM NN exchange $J$ on the lattice compression. In contrast, for $HgCr_2S_4$ a positive sign of $dT_C/dp$ was found. At that time, $HgCr_2S_4$ was treated as ferromagnet, while later studies found that this mercury-sulphide compound undergoes AFM spin-spiral ordering close to 25 K.

Later on the research moved towards pressure-dependent investigations of strongly frustrated oxides [468,469]. Jo et al. [468] studied magnetic properties under pressure up to 0.96 GPa on the geometrically frustrated $ZnCr_2O_4$ spinel. This compound shows a simultaneous AFM and structural transition at 12.5 K, which is attributed to a spin-Jahn-Teller effect that reduces the crystal symmetry from cubic to tetragonal and thereby releases magnetic frustration. The reported increase of the magnetic transition temperature under pressure, which was treated in the frame of a quantum mean-field theory. In this model, it was shown that increasing pressure leads to an increase of the NN exchange $J_1$ that characterizes interaction between the nearest spins that belong to the same tetrahedron and a decrease of the exchange $J_2$ representing next-nearest-neighbour interactions. Ueda and Ueda [469] studied two further geometrically frustrated chromium oxide spinels, namely $CdCr_2O_4$ and $HgCr_2O_4$. They found that both, Néel and CW temperatures for these compounds increase by applying pressure indicating that the nearest-neighbour direct AF interactions were enhanced through increasing overlap of $t_{2g}$ orbitals. A further important and significant observation of this work concerns the exotic field-induced transition to a half-magnetization plateau observed in high magnetic fields for these oxide spinels. The critical field of the transition into the plateau state for $HgCr_2O_4$ was found to scale with the CW temperature under pressure, implying that this transition is governed by AFM NN interactions [469].

Attempts to tune a pressure-induced MIT in a series of vanadium oxides were reported by Blanco-Canosa et al. [470]. These authors provide a systematic investigation of $AV_2O_4$ spinels, with $A$ = Mg, Zn, Mn, and Cd. The authors speculate that these vanadium spinels might be close to a MIT due to the direct overlap of the vanadium orbitals. They plot the AFM ordering temperature as function of the inverse V-V separation and locate the Cd and Mg compound in the PM and insulating regime, while the Zn and Mg vanadates come close to the regime with strong overlap enhancing the probability for the vanadium electrons to become itinerant. By systematic pressure-dependent studies of the AFM ordering temperatures, they indeed find arguments for a breakdown of localized electron behaviour within this vanadium spinel series [470]. Finally there were numerous high-pressure x-ray diffraction and Raman scattering studies up to much higher pressure of 50 GPa, mainly to detect pressure-induced magneto-structural phase transitions, also combined with possible MIT transitions [471,472,473,474,475,476,477,478]. Some of these high-pressure experiments dealt with the investigations of the ME and MF compounds.

Efthimiopulous et al. [472] found that pressurizing ME HgCr$_2$S$_4$ induces three reversible structural phase transitions. Here, the starting cubic ($Fd\bar{3}m$) phase adopts a tetragonal $I4_1/amd$ structure at 20 GPa. This structural transition is of the first order and accompanied by a 4% reduction of the cell volume. The cubic-to-tetragonal transformation appears to be concomitant with a MIT transition as was concluded from the substantial reduction of the Raman intensity under pressure. This conclusion is supported by spin-polarized LSDA+U calculations for both, cubic and tetragonal phases. These calculations also have shown that at high pressure a significant reduction of the density-of-states of the Cr $t_{2g}$ electrons at the Fermi energy takes place, indicating a tendency to suppress magnetism in the $I4_1/amd$ phase [472]. Above 26 GPa, in HgCr$_2$S$_4$ a further structural transformation into a state with orthorhombic distortions takes place. Above 36 GPa a third transition into a high-pressure phase HP3 phase was found. However, the symmetry of this phase still has to be resolved.

Under pressure, X-ray and Raman studies of the mercury-chromium-selenide spinel HgCr$_2$Se$_4$ identified two subsequent structural phase transitions [471]. The ambient-pressure ($Fd\bar{3}m$) phase was found to transform into a tetragonal $I4_1/amd$ structure above 15 GPa. Again, it was speculated that this transition is accompanied by an MIT, resulting in a vanishing Raman signal in the high-pressure phase. Further compression of HgCr$_2$Se$_4$ leads to structural disorder beyond 21 GPa. This behaviour was interpreted as precursor phenomenon of another crystalline phase with higher cationic coordinations, which cannot crystallize due to kinetic effects. The respective spin-resolved band structure calculations performed in [471] revealed significant changes in the electronic structure of HgCr$_2$Se$_4$ after the cubic-to-tetragonal transition, whereas the FM NN SE interactions are found to dominate in both structures.

The cubic ($Fd\bar{3}m$) structure of MF compound CoCr$_2$O$_4$ is stable up to 16 GPa [473]. Further compression leads to a reversible asymmetric broadening for the Bragg peaks. Detailed x-ray and Raman studies excluded non-hydrostatic conditions or cationic inversion as a possible origin of the broadening of the Bragg peaks and attributed it to a first-order phase transition into a phase with lower, probably tetragonal $I4_1/amd$ symmetry. The analysis of the evolution of the cation-oxygen bond lengths and interatomic bond angles under pressure have shown that the Co-O bond length exhibits a significant reduction compared to the Cr-O length, which remains almost uncompressible. These two bond lengths become comparable just before arriving the critical pressure of 16 Gpa. At the same time, the interatomic bond angles Cr-O-Co and Cr-O-Cr change very little with pressure. The *ab-initio* density-functional-theory calculations performed in [473] have shown that the high pressure tetragonal phase favours a FM ground state compared to the PM state in the cubic $Fd\bar{3}m$ phase. The analysis of the exchange interactions J$_{AA}$, J$_{AB}$ and J$_{BB}$ revealed that with increasing pressure the ratio J$_{BB}$/J$_{AB}$ increases significantly suggesting enhanced magnetic frustration.

High-pressure Raman and x-ray studies of MF ZnCr$_2$Se$_4$ [474] documented that the cubic $Fd\bar{3}m$ structure persists up to 15 GPa, while above 18 GP a new structure comparable to a CrMo$_2$S$_4$-type structure - a distorted variant of the monoclinic Cr$_3$S$_4$ structure - is established. This structural transformation in ZnCr$_2$Se$_4$ is reversible and of first order. In the monoclinic phase both type of cations (Zn and Cr) are in six-fold coordination with alternative layers of ZnSe$_6$ and CrSe$_6$ stacked along the *c* axis. The substantial reduction of the Raman signal above 12.5 GPa suggests that this structural transformation again is accompanied by a MIT. *Ab initio*-DFT calculations successfully reproduced the cubic-to-monoclinic structural transition. These calculations also reveal significant electronic and magnetic changes accompanying the structural modification, i.e., a significant reduction of the electronic band gap and a reduction of the magnetic moment of the Cr ions [474].

High-pessure x-ray diffraction and Raman studies up to 42 GPa were conducted on CdCr$_2$Se$_4$, which is governed by FM exchange, and resolved three structural phase transitions [475]. The cubic structure at ambient pressure transforms at ~11 GPa into a tetragonal $I4_1/amd$ structure and undergoes a further orthorhombic distortion at ~15 GPa. It seems that significant structural disorder is induced beyond 25 GPa. Comparing these results with pressure results from other chromium sulphide and selenide spinels, the authors found a clear correlation between the cubic-tetragonal transition pressures and the next-nearest-neighbour magnetic exchange interactions and concluded that the cubic-to-tetragonal transitions in these systems are governed by magnetic exchange alone [475]. These conclusions were further corroborated by high-pressure x-ray diffraction studies on $MnCr_2O_4$ and $NiCr_2O_4$ spinels [478], providing experimental evidence that the Cr-Cr magnetic exchange interactions alone suffice to account for the high-pressure behaviour of these systems.

High-pressure x-ray diffraction and Raman spectroscopy studies of $ZnCr_2S_4$ spinel found a sequence of structural transformations from cubic $Fd\bar{3}m$ to tetragonal $I4_1/amd$ structure at 21 GPa, followed by transition to orthorhombic $Imma$ state above 33 GPa [476]. Interestingly, similar structural transformations take place at ambient pressure on decreasing temperature from 300 K (in the PM state) to 16 K and 8 K into two different magnetic states with complex AFM spin ordering [178]. This supports the above conclusion that the structural cubic-to-tetragonal transformation mainly is driven by the variation of magnetic exchange. A careful inspection of the XRD and Raman data revealed an additional isostructural transformation in $ZnCr_2S_4$ between 9 and 14 GPa, which was speculated to be induced by changes of the electronic state of the Cr ions. The sequence of structural phase transitions as observed in a number of chromium sulphide and selenide spinels up to a hydrostatic pressure of ~ 40 GPa is shown in Fig. 61 [476]. In all compounds with the exception of $ZnCr_2Se_4$, the cubic phase as observed at room temperature and ambient pressure is followed by a tetragonal phase. There are convincing arguments that this structural phase transition is driven by magnetic exchange.

A second example of cubic-to-monoclinic structural transformation under pressure was reported for $CuCr_2Se_4$ [477]. At ambient pressure this compound exhibits FM behaviour and metallic conductivity. The cubic $Fd\bar{3}m$ structure of $CuCr_2Se_4$ is stable up to 8 GPa. Above this pressure, several new Bragg reflections appear in the XRD pattern, however this transformation becomes complete only at 20 GPa. The contrasting pressure behaviour of Cu and Zn based selenides, as compared to sulphide and oxide spinels was attributed to steric effects, which are proposed to be the driving force of the cubic-to-monoclinic transformation.

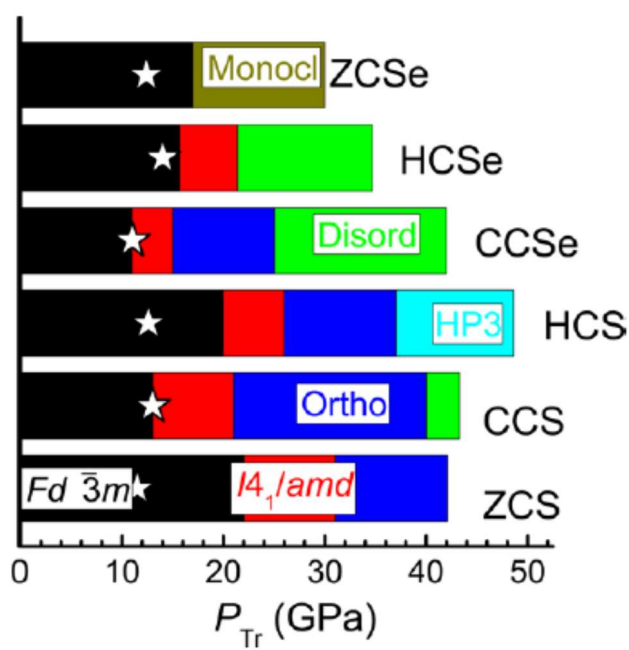

Figure 61. Overview of the pressure-induced structural phase transitions in several Cr-based spinels at ambient temperature. The stability of each phase is indicated by a different color: black for the starting cubic $Fd\bar{3}m$ phase, red for tetragonal $I4_1/amd$, blue for orthorhombic, green for disorder, cyan for an unidentified high-pressure phase (HP3), and yellow for the $CrMo_2S_4$-type. The open star symbols denote the onset of the Raman intensity drop. Abbreviations: ZCS = $ZnCr_2S_4$, CCS = $CdCr_2S_4$, HCS = $HgCr_2S_4$, CCSe = $CdCr_2Se_4$, HCSe = $HgCr_2Se_4$, and ZCSe = $ZnCr_2Se_4$.
Reprinted with permission from Efthimiopoulos et al. [476]. Copyright (2017) American Chemical Society.

## 4. Summary and Outlook

This review summarizes the research on spinel compounds, mainly focusing on recent studies of the magnetic and polar properties of this class of compounds. In addition to detailed discussions of the complex magnetic ground states, including Jahn-Teller and orbital-ordering effects, and of the many open questions concerning the occurrence of ferroelectricity and multiferroicity, we also have mentioned a number of interesting and outstanding features, like the observation of heavy-fermion behaviour, superconductivity in spinels, as well as the occurrence of CMR and metal-to-insulator transitions. Finally, we have commented ongoing attempts to measure and analyse high magnetic-field data, as well as high-pressure results. This review clearly documents the outstanding importance of spinels in developing models and theories for magnetic compounds, and specifically the significance of magnetic frustration in these compounds.

The research on magnetism of spinel compounds, performed early on after World War II, mainly focused on many aspects of magnetic semiconductors characterized by relatively high resistivities with concomitant high magnetic ordering temperatures and high saturation magnetization useful for a variety of applications. With the main focus on technical relevance and possible applications from millimeter wave integrated circuitry to magnetic recording, we also provided a short survey of the relevant modern field of spinel thin films and spinel-based heterostructures. Also in the field of multiferroics, well-designed nanostructures, i.e. regular arrays of magnetic materials embedded in a FE, were synthesized and characterized.

In parallel, the basic principles of magnetic exchange and magnetic order, as well as the importance of crystal-electric-field effects and orbital-ordering phenomena, were tested and exemplified on a variety of spinel compounds: The occurrence of ferrimagnetism and of the triangular YK phase were experimentally verified in spinel compounds. In addition, since decades the Verwey transition, a metal-to-insulator like transition observed in magnetite, is in the focus of basic research. Spinels in modern research of magnetism are highly interesting because of frustration effects: The $A$ sites in the spinels constitute a bipartite diamond lattice, and are responsible for unconventional magnetic ground states, which strongly depend on the ratio of the exchange interactions within and between the two sublattices. $B$-site ions are located at the vertices of a pyrochlore lattice, one of the strongest contenders of frustration in 3D. If $A$ and $B$ sites in the lattice are occupied by magnetic ions, again, competing magnetic exchange interactions are responsible for a variety of complex spin configurations. The complexity of many spinel compounds is further enhanced by additional orbital-ordering and charge-ordering phenomena and in many cases, a strong spin-lattice coupling. Spinels belong to a class of transition-metal compounds with an extremely strong entanglement of spin, charge, orbital, and lattice degrees of freedom, with a variety of emerging complex phenomena: spin liquids, spin-spiral liquids, spin-orbital liquids, orbital glasses, spin-driven Jahn-Teller effect, the occurrence of metal-to-insulator transition, heavy-fermion behaviour, fractionalized magnetization plateaus, and colossal magnetoresistance, to name a few, are illuminating examples and are discussed in detail in the present review.

Many of the phenomena reported during the last years are far from being completely understood: The microscopic origin of multiferroicity in a variety of spinels is not finally unravelled. Whether some of the spinel compounds indeed exhibit a tendency to off-centring of the $B$-site ions is not clear and far from being well established. The existence of spin-supersolid and spin-superliquid phases, which are claimed to exist for some chromium spinels at high magnetic fields, has not been finally verified experimentally and the correspondence of these phases to supersolids and superliquids is not settled from a theoretical point of view. Thus, many open questions remain and have to be resolved in the forthcoming years. It is clear that highly improved sample-growing techniques, which allow the growth of single crystals large enough, even for inelastic neutron scattering studies, and the continuous and significant improvement of various experimental techniques operating in a broad range of external parameters, will strongly advance the field. Specifically, X-ray absorption techniques and X-ray magnetic circular dichroism experiments, which can nowadays be performed at high external magnetic fields will provide microscopic views on spin structures and moment orientations and should help to finally solve a number of open questions.

**Acknowledgements**

Research on spinel compounds was in the focus of the experimental work performed in Experimental Physics V at the Center of Electronic Correlations and Magnetism at the University of Augsburg. This research was carried out by the authors of the present review over the past 20 years. However, this research has benefitted a lot from contributions of a number of people, either being former members of the group or researchers with close and long-standing collaborations over the years. We are especially grateful to Manuel Brando, Norbert Büttgen, Ilias Efthimiopoulos, Viorel Felea, Veronika Fritsch, Axel Günther, Joachim Hemberger, Christian Kant, Alexander Krimmel, Stephan Krohns, Franz Mayr, Atsuhiko Miyata, Andrei Pimenov, Lilian Prodan, Thorsten Rudolf, Alexander Ruff, Michael Schmidt, Florian Schrettle, Yurii Skourski, Natalia Tristan, Zhe Wang, and Sebastian Widmann, who have worked with us over the years and helped to make this research exciting and successful. In addition, we are indebted to our colleagues Leon Balents, Mike Kalvius, Peter Lemmens, Jochen Litterst, Christian Rüegg, Yuejian Wang, Jochen Wosnitza, Oksana Zaharko, and Sergei Zherlitsyn for fruitful collaborations and illuminating discussions.

Over the years this research was funded by the Bundesministerium für Forschung und Technolgie (BMFT) via VDI/EKM, FKZ 13N6917, as well as by the Deutsche Forschungsgemeinschaft (DFG) via the Sonderforschungsbereich SFB 484 "Kooperative Phänomene im Festkörper: Metall-Isolator-Übergänge und Ordnung mikroskopischer Freiheitsgrade", via the Transregional Research Collaboration TRR 80 "From Electronic Correlations to Functionality", and via the Research Unit FOR 960 "Quantum Phase Transitions".